\documentclass[12pt]{article}
\usepackage{epsfig}
\def\bq{\begin{equation}}
\def\eq{\end{equation}}
\def\bqa{\begin{eqnarray}}
\def\eqa{\end{eqnarray}}
\def\bqb{\begin{eqnarray*}}
\def\eqb{\end{eqnarray*}}

\def\pr#1#2#3{ Phys. Rev. ${\bf{#1}}$ (#2) #3}

\def\pl#1#2#3{ Phys. Lett. ${\bf{#1}}$ (#2) #3}

\def\np#1#2#3{ Nucl. Phys. ${\bf{#1}}$ (#2) #3}
\def\zp#1#2#3{ Z. Phys. ${\bf{#1}}$ (#2) #3}

\def\ie{{\it i.e.\/}}
\def\eg{{\it e.g.\/}}

\global\nulldelimiterspace = 0pt
 
\def\roughly#1{\mathrel{\raise.3ex
    \hbox{$#1$\kern-.75em\lower1ex\hbox{$\sim$}}}}

\def\gsim{\roughly>}

\def\L{ {\cal L }}
\def\O{ {\cal O }}
\def\A{ {\cal A }}

\def\R{ {\cal R }}

\def\mwd{M_W^2}

\def\mt{m_t}
\def\mtd{m_t^2}

\def\pvec{\overrightarrow p}

\def\numero{ \begin{tabular}{c}
     
           {\bf UNIVERSITE MONTPELLIER  II}\\[0.2cm]
           {\bf Laboratoire de Physique Math\'ematique}\\[1cm]
           {\bf  PM/96-38}\\

            \end{tabular}  }
\def\numeroa{ \begin{tabular}{c}
     
           {\bf ARISTOTLE UNIVERSITY}\\
           {\bf of THESSALONIKI}\\[0.2cm]
            {\bf Department of Theoretical Physics}\\[0.6cm]
            {\bf THES-TP 96/12 }

            \end{tabular}  }

\def\hepplace{ \begin{tabular}{c}
            {\bf November 1996}\\
            \end{tabular}  }

\begin{document}
\thispagestyle{empty}
\begin{minipage}[b]{18cm}
\vspace*{-2cm}
\hspace*{-1.7cm}
\epsfig{file=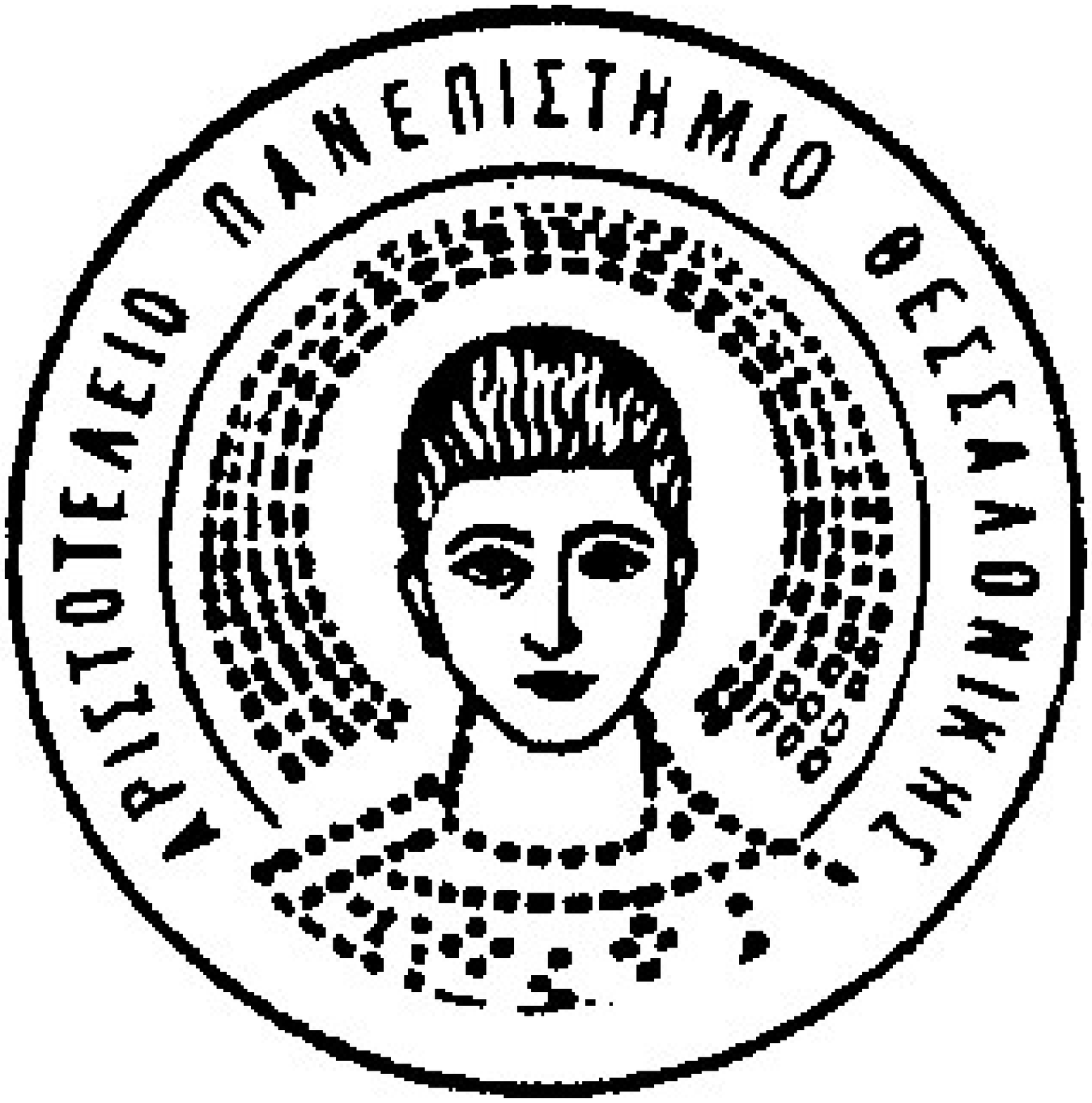,height=3.5cm}
 \hspace{7cm}
\epsfig{file=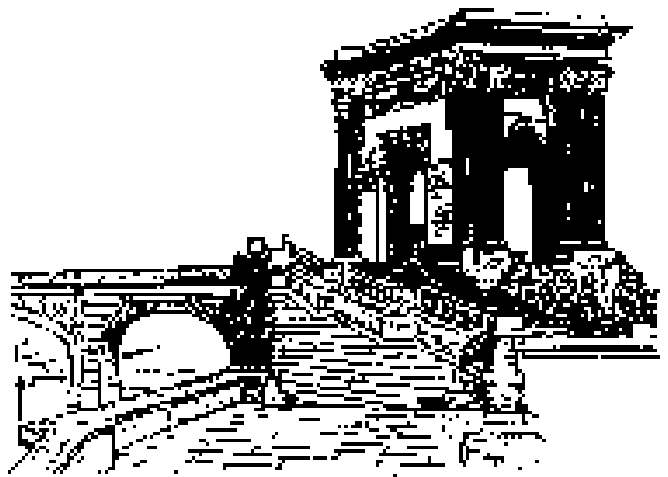,height=3.5cm}\\
\hspace*{-3.8cm}\numeroa\hspace{3.cm} \numero\\

\vspace*{-0.6cm}\hspace{4.cm} \hepplace 
\end{minipage}
\vspace*{2cm}

\hspace*{-0.5cm}
\begin{center}
{\Large {\bf  A Model-independent Description of\hspace{2.2cm}\null}}\\
\vspace{0.2cm}
{\Large {\bf \hspace{0.5cm} New Physics effects
in $e^+e^-\to t\bar t$ }}\footnote{Partially 
supported by the EC contract CHRX-CT94-0579.}\hspace{2.2cm}\null 
\hspace*{-0.5cm}
\vspace{1.cm} \\{\large G.J.
Gounaris$^a$, J. Layssac$^b$ and F.M. Renard$^b$}\hspace{2.2cm}\null
 \vspace {0.5cm} \\
$^a$Department of Theoretical Physics, University of Thessaloniki,
\hspace{2.2cm}\null \\
Gr-54006, Thessaloniki, Greece.\\ \vspace{0.2cm} $^b$ Physique
Math\'{e}matique et Th\'{e}orique, UPRES-A 5032\hspace{2.2cm}\null\\
Universit\'{e} Montpellier
II,  F-34095 Montpellier Cedex 5.\hspace{2.2cm}\null\\[1cm]
{\bf Abstract}\hspace{2.2cm}\null
\end{center}
\hspace*{-1.2cm}
\begin{minipage}[b]{16cm}
We study the potential of a future $e^+e^-$ collider
for the search of anomalous
$\gamma t\bar t$ and $Zt\bar t$ couplings,
assuming that CP-invariance holds. This is
done in a model-independent way, considering that all six
possible couplings do appear. Two
experimental situations are envisaged, with and without
$e^{\pm}$ beam polarization. Observability limits in
the form of domains in the 6-dimensional parameter space are
established. Illustrations for specific
constrained models are also presented
and implications for new physics searches
are discussed.
\end{minipage} 

\setcounter{footnote}{0} 
\clearpage
\newpage 
  
\hoffset=-1.46truecm
\voffset=-2.8truecm
\textwidth 16cm
\textheight 22cm
\setlength{\topmargin}{1.5cm}

\section{Introduction} 
The properties of the recently discovered top quark \cite{tmass}
has been the
subject of many speculations \cite{treview}.  Theoretical motivations
for them deal with the problems of
the scalar sector associated to mass generation and the 
high value of the top mass, close to the value
of the electroweak scale. Some experimental
hints  also came from
possible anomalies observed in the heavy
quark production at LEP1/SLC \cite{texp}.\par
 
The phenomenological description of non-standard top 
quark properties mostly rely on the
effective lagrangian method which is the proper way to
describe New Physics (NP) in case all new degrees of freedom are too
heavy to be directly produced in the various Colliders \cite{Leff}.
This effective lagrangian is supposed to be
derived from a more fundamental interaction after integrating 
out the heavy NP degrees of freedom. It describes the 
residual NP effects in terms of operators involving the 
$Z$ ,$W$, $\gamma$, $t$, $b$ and gluon fields,
 with the option of including \cite{GRVbb, GKR, unitop} or not 
the Higgs boson as a fundamental particle \cite{tchiral}.\par

This way one predicts anomalous properties
for the top quark; \ie\@ departures from the Standard Model 
(SM) couplings that could be revealed by studies of the 
production and decay modes. Assuming that NP is CP invariant,
the set of operators used to construct the effective lagrangian,
although restricted by dimensional and
gauge symmetry considerations,
is rather large, which means a large number of
unknown couplings.
An underlying theory \cite{Arzt, tdyn, dynbos}
would certainly relate these couplings to more fundamental
parameters. But nowadays such
relations are lacking so that studies of NP effects predicted 
by this description usually proceed by taking
each operator one by one and ignoring possible correlations.\par
 
In this paper we take a somewhat different attitude. Assuming
that  NP is CP invariant and that its effects on the 
process $e^-e^+ \to t\bar t$  
only arise from modifications of the $\gamma t \bar t$ and $Z
t \bar t$ vertices, we present a fully model independent 
analysis of  this process, keeping as free parameters all six 
vector, axial and tensor couplings describing  these vertices.  
We establish observability limits in this 6-dimensional
parameter space spanned by the departures from
SM values of these couplings. This
allows us to discuss discovery limits for 
NP contributions of any structure. We then
also give a few simple illustrations
for constrained models where
a certain number of relations are imposed among the couplings, 
leaving for example, only four, three, two or even one 
free parameters.\par
 
We illustrate in full detail the case of 1TeV $e^+e^-$ collider
and briefly mention how the results change for the cases of  
0.5 or 2~TeV colliders.
With the foreseen luminosity one expects more 
than $10^{4}$ $t\bar t$ events
in this high energy region \cite{eelin}, so that even after taking into
account detection efficiencies \cite{effic},
the basic accuracy in the determination of the NP couplings is still
at the few percent level. If the  $\gamma t \bar t$ and
$Zt \bar t$ couplings are assumed to be generated from  a
specific operator then we can translate the 
sensitivity limits on the above couplings, to bounds on the 
corresponding NP scale \cite{unitop, uniY}.\par
 
In establishing the observability limits we emphasize the 
special role of a set of angular asymmetries which, 
to first order in the NP couplings, are independent of the 
top decay properties and depend only on the 
structure of the relative magnitudes of the spin density matrix 
elements of the produced $t$-quark. This allows us to separate
the anomalous effects in the production
process that we want to study in the present paper, 
from those the top decay amplitude, which either modify the 
$tWb$ vertex or open new decay channels involving \eg\@ 
Higgs bosons. We find that a study of the top spin
density matrix
allows to construct 4 different asymmetries when the 
$e^{\pm}$ beams are unpolarized, and 7 additional ones 
when the $e^{\pm}$ beams are
longitudinally polarized. This  way we obtain
model-independent constraints on  two  $\gamma
t\bar t$ and one $Zt\bar t$ couplings which do not receive 
any SM contribution at tree level; (\ie\@ on $d_2^\gamma$, 
$d_3^\gamma$ and $d_3^Z$ below). For the determination of the 
~remaining 3 couplings, one more information is needed which
should determine their overall scale. For this
we use the magnitude of an integrated top quark density
matrix element, like \eg\@  the $t\bar t$ production cross section.
Such a density matrix element though is of course also sensitive
to uncertainties on the top quark decay width and branching ratios. 
We quantitatively discuss these effects, as well as the
implications on the measurements of the aforementioned two sets
of couplings for the general 6-parameter case and also 
for the restricted 1-, 2-, 3- and 4- parameter cases.\par
 
Finally we discuss the implications that our results 
could have on the study of the structure of the 
underlying NP which can most ~generally be described in terms of a
set of $dim=6$ operators \cite{Leff, unitop}.  \par

The organization of the paper is the following. In Section 2,
we present the amplitudes, cross sections and asymmetries, 
and  establish their
~dependencies on the NP couplings. Section 3 is devoted to
the derivation of the constraints on the NP the parameters and 
to an estimate of the experimental accuracies. Applications are
then made in Section 4 to the general 6-parameter cases, as well
as the 4-, 3-, 2- and 1- parameter cases. 
Observability domains are given
in Figures and Tables. The final discussion is
given in Section 5. Appendix A collects the coefficients of the NP
couplings controlling the sensitivity of each observable.

\section{Observables}
The general CP-conserving structure of the $\gamma t\bar t$ and
$Zt\bar t$ vertices is written as\footnote{The definition of
$d^V_3$ in (\ref{eq:dgz}) differs from the one in \cite{GKR} by
a factor of $1/\mt$.} \cite{GKR}
\begin{equation}  \label{eq:dgz}
-i \epsilon_\mu^V J^{\mu}_V = -i e_V \epsilon_\mu^V \bar
u_t(p)[\gamma^{\mu}d^V_1(q^2)+\gamma^{\mu}\gamma^5d^V_2(q^2)
+(p-p^{\prime})^{\mu}d^V_3(q^2)/m_t] v_{\bar t}(p^{\prime}) \ ,
\end{equation}
where $\epsilon_\mu^V$ is the polarization of the vector boson
$V=\gamma, Z$. The
outgoing momenta $(p,~p^{\prime})$ refer to $(t,~\bar t) $ 
respectively and
satisfy $q\equiv p+p^{\prime}$. The normalizations are
determined by $e_{\gamma}\equiv e$ and $e_Z\equiv e/(2s_Wc_W)$,
while $d^V_i$ are in general $q^2$ dependent form factors. 
Non-vanishing contributions to the these couplings from 
\underline{SM at tree level} only arise for
\begin{equation} \label{eq:dgzSM}
d^{\gamma, SM0}_1 = {\frac{2}{3}} \ \ , \ \ d^{Z, SM0}_1=
g_{Vt}={\frac{1}{2}}%
-{\frac{4}{3}}s^2_W\ \ , \ \ d^{Z, SM0}_2=-g_{At}=-\,
 {\frac{1}{2}} \ \ . 
\end{equation}
\underline{Departures from the SM} values are then defined as:
\bq 
\bar d^V_j \equiv d^V_j-d^{V,SM}_j \ \ \ , \ \label{eq:dvbar}
\eq
for $V=\gamma ,Z$. 
The $e^-e^+\to t\bar t$ helicity amplitude is ~written as  
$F_{\lambda,\tau,\tau^{\prime}}$, where
$\lambda\equiv\lambda(e^-)=-\lambda^{\prime}(e+)=\pm 1/2$ 
denote the $e^-$, $e^+$ helicities,
while $\tau$ and $\tau^{\prime}$ represent respectively the 
$t$ and $\bar t$ helicities.
Using the couplings defined in (\ref{eq:dgz}),
we obtain
\begin{eqnarray}  \label{eq:fltt}
F_{\lambda,\tau,\tau^{\prime}}&=&\sum_{V=\gamma,Z} 2\lambda e^2\sqrt{s}
(A_V-2\lambda B_V)\Bigg \{ d^V_1 [2m_t \sin\theta
\delta_{\tau \tau^{\prime}}
+\sqrt{s} \cos\theta(\tau^{\prime}-\tau) -2\lambda\sqrt{s}%
\delta_{\tau,-\tau^{\prime}}]  \nonumber \\
&&-d^V_2 2|\overrightarrow p|
[\cos\theta\delta_{\tau,-\tau^{\prime}}+2
\lambda(\tau-\tau^{\prime})] -d^V_3
{4|\overrightarrow p|^2\over \mt} \sin\theta
\delta_{\tau \tau^{\prime}}\Bigg \} \ \ , \
\end{eqnarray}
\noindent
where $A_V$, $B_V$ describe the contribution of the vector and
axial $Vee$ vertex. In this paper we assume that the
$Vee$ vertices are standard, which
implies that\footnote{An NP contribution to the 
$Vee$ vertex of the type suggested by the Class 3 operators of
\cite{unitop} can be also described by the present formalism, 
by introducing vector and axial $\gamma ee$ ~form factors through  
$A_\gamma=- d_{1e}^\gamma /s$ and $B_\gamma=- d_{2e}^\gamma /s$,
and correspondingly modifying also $g_{Ve}$, $g_{Ae}$.}   
\bq
A_{\gamma} =-\frac{1}{s} \ \ , \ \ A_Z=\frac{g_{Ve}}{ 4s^2_Wc^2_W
D_Z}\ \ , \ \
 B_{\gamma} = 0 \ \ , \ \  
B_Z=\frac{g_{Ae}}{ 4s^2_Wc^2_W D_Z}\  \ \ ,\label{eq:AVBV} \  
\eq
\bq
g_{Ve}=-\frac{1}{2}+2s^2_W \ \ , \ \ \ g_{Ae}=-1/2 \ \ \ , 
\eq
and $D_Z=s-M^2_Z+iM_Z\Gamma_Z$, $s\equiv q^2$. 
In (\ref{eq:fltt}), $\theta$ is the $(e^-,t)$
scattering angle in the $(e^-,e^+)$ c.m. frame. 
Because of CP invariance, the amplitude in (\ref{eq:fltt}) 
satisfies \cite{Chang}
\bq
F_{\lambda,\tau,\tau^{\prime}}=F_{\lambda,-\tau^\prime ,-\tau}
\ , \label{eq:CPrelation}
\eq
and is normalized
so that the unpolarized $e^-e^+\to t\bar t$ differential cross 
section is given by
\begin{equation}
{\frac{d\sigma(e^-e^+\to t\bar t)}{~d\cos\theta}}= 
{\frac{3 \beta_t}{128\pi s}}\sum_{\lambda, \tau, 
\tau^{\prime}} |F_{\lambda, \tau, \tau^{\prime}}|^2 \
\ ,
\end{equation}
where $\beta_t=(1-{\frac{4m^2_t}{s}})^{1/2}$ and the colour 
factor has been included. \par
 
The density matrix for top-quark production is
\begin{equation}
\rho^{L,R}_{\tau_1 \tau_2}=\sum_{ \tau^{\prime}} F_{\lambda, \tau_1,
\tau^{\prime}}F^*_{\lambda, \tau_2, \tau^{\prime}} \ \ ,
\end{equation}
where $L,R$ correspond respectively to 
$\lambda\equiv\lambda(e^-)=-\lambda^{\prime}(e+)=\mp 1/2$. 
All density  matrix elements are 
real, so long as the imaginary parts of the $d^V_j(q^2)$ are
neglected. Because of (\ref{eq:CPrelation}), there are only six 
independent such elements, namely 
$\rho^{L,R}_{++}$, $\rho^{L,R}_{--}$ and $\rho^{L,R}_{+-}=
\rho^{L,R}_{-+}$, which can be measured through the top
production and decay distributions.   
Unpolarized $e^\mp$ beams allow to measure only the three $(L+R)$
density matrix elements, whereas the three $(L-R)$ ones require 
longitudinal $e^{\pm}$ beam polarization.\par
 
The general expression of the differential cross section for $e^+e^-\to
t\bar t$ with $t\to bW \to b l \nu_l $ and longitudinally
polarized $[L (R)~ ~e^-]$
and $[R(L)~ ~e^+]$ beams, is written (compare eq.(B6,B7) of 
\cite{GKR}) as
\newpage
\begin{equation}
\frac{d\sigma^{L, R}}{d\cos\theta d\varphi_1 d\cos\vartheta_1 d\psi_1
d\cos\theta_l} =\frac{9\beta_t \Gamma(t\to bW)_{SM} 
Br(W \to l \nu)}{(8\pi)^3 s(2\mwd+\mtd) \Gamma_t } 
\rho^{L, R}_{\tau_1 \tau_2}\cdot {\cal R }_{\tau_1 \tau_2} \ ,
\label{eq:dsigma}
\end{equation}
where $\Gamma_t$ is the total top width, $\Gamma (t\to bW)_{SM}$ 
describes the SM contribution
to $t \to bW$ and the $Wl\nu_l$ vertex is taken to be standard
\cite{unitop}.\par

In (\ref{eq:dsigma}) the definition  
\begin{eqnarray}
\label{eq:rhoR}
\rho^{L, R}_{\tau_1 \tau_2} \cdot {\cal R }_{\tau_1 \tau_2}&=
&\frac{1}{2}~
(\rho_{++}+\rho_{--})^{L, R}( {\cal R }_{++}+ {\cal R }_{--})
\nonumber \\
&+&\frac{1}{2} (\rho_{++}-\rho_{--})^{L, R}( {\cal R }_{++}- {\cal R }_{--})
+~ \rho^{L, R}_{+-}( {\cal R }_{+-}+ {\cal R }_{-+})\ \ 
\end{eqnarray}
is used, where the  $ {\cal R}_{\tau_1\tau_2}$ factors 
are elements of
the top decay matrix  introduced in ~eq.(B9-B11) of
ref.\cite{GKR} and include any possible NP contribution to the 
$t\to bW$ decay. The quantities $\R$ and 
the differential cross section in (\ref{eq:dsigma}),  
depend on the three Euler angles 
$(\varphi_1, \vartheta_1, \psi_1)$ determining 
the $t\to bW \to b l\nu_l$ decay plane in the rest
frame of the top, as well as on an additional angle
$\theta_l$ describing the decay distribution of $l$
within the top decay plane.
In \cite{GKR} it has been shown how
appropriate averages over the Euler angles allow to
project out quantities proportional to each of
the three different $\rho$-factors
given in (\ref{eq:rhoR}),
multiplied by top-decay functions depending on the 
$t\to Wb$ decay couplings. These averages allow the ~construction
of three types of observables and are done as follows:
\begin{itemize}
\item 
Type A arises by projecting out the first term in (\ref{eq:rhoR})
including 
$(\rho_{++}+\rho_{--})^{L, R}$, which is ~achieved  by integrating
eq.(\ref{eq:dsigma}) 
over $d\varphi_1 d\cos\vartheta_1 d\psi_1  d\cos\theta_l$.
These observables measure the structure of the top production
differential cross section, when we sum over the $t$-polarizations.
\item
Type $H$, arising from the second term  
$(\rho_{++}-\rho_{--})^{L, R}$, is   
related to the top
quark helicity. It is extracted though an 
integration of eq.(\ref{eq:dsigma})
using the projector 
\bq
\label{eq:PH}
P_H = \cos\psi_1 + r \sin\psi_1\ \ \ , 
\eq
where the parameter $r$ is
a priori free. It is chosen   
\bq
\label{eq:r}
 r\equiv {3\pi m_t M_W\over4(m^2_t-2M^2_W)} \ \ ,
\eq
so  that to maximize the statistical significance of the results, by
optimizing to the angular dependence 
of the SM distribution.
\item
Type $T$, arising from $\rho^{L, R}_{+-}$, is
related to the top quark transverse
polarization. It is obtained by integrating (\ref{eq:dsigma}) 
using the optimized projector,
\bq
\label{eq:PT}
P_T = \cos\psi_1 \sin\varphi_1 \cos\vartheta_1- \sin\psi_1 
\sin\varphi_1 +
r (\sin\psi_1 \cos\varphi_1 \cos\vartheta_1 + \cos\psi_1
\sin\varphi_1 )\ \ .
\eq
\end{itemize}\par

In each of these cases, various asymmetries with
respect to the top production angle $\theta$ may be ~constructed, 
which allows us 
to get rid of the top-decay couplings contained in the $\R$
factors in (\ref{eq:rhoR}). Thus, these asymmetries depend
only on the six different $\rho$ elements  
appearing in (\ref{eq:rhoR}). 
To present them, let us first single out more precisely the contents
of the three types of $\rho$ terms in (\ref{eq:rhoR}). 
Using \cite{GKR}, the $\rho^{L\pm R}$ top density matrix
elements are expressed,
in terms of the $\gamma t\bar t$ and $Z t\bar t$ couplings
(compare (\ref{eq:dgz}))
\begin{equation} \label{eq:dLR}
d^L_i=d^{\gamma}_i+{\frac{1-2s^2_W}{4s^2_Wc^2_W}}\chi d^Z_i \ \ , \ \
d^R_i=d^{\gamma}_i-{\frac{\chi}{2c^2_W}}d^Z_i \ ,
\end{equation}
where
$\chi\equiv s/(s-M^2_Z)$ and the Z width is neglected for $s=q^2
>4m^2_t$. We thus have
\begin{eqnarray}
(\rho_{++}+\rho_{--})^{L\pm R} & = & 2 e^4\left [\sin^2\theta
 \left({\frac{4 m^2_t}{s}}\right ) A^{L\pm R}_1
+(1+\cos^2\theta)A^{L\pm R}_2 -  4\beta_t \cos\theta
  A^{L\pm R}_3 \right ] ,\ \ \ \label{eq:rhotrace} \\[0.1cm]
(\rho_{++}-\rho_{--})^{L\pm R}& = & 4 e^4[(1+\cos^2\theta) 
~\beta_t  B^{L\pm R}_1
-  \cos\theta B^{L\pm R}_2] \ , \ \label{eq:rho++m--} \\[0.1cm]
 \rho^{L\pm R}_{+-}& = & e^4\left ( \frac{4\mt}{\sqrt{s}} \right )
\sin\theta \left[  C^{L\pm R}_1 - \cos\theta ~\beta_t 
C^{L\pm R}_2 \right ]
 \ \ , \label{eq:rho+-}
\eqa
where
\bqa
A^{L\pm R}_1&=&\left [d^{L}_1-{2|\pvec|^2\over \mtd} d^{L}_3
\right ]^2 \pm \left [d^{R}_1-{2|\pvec|^2\over \mtd} d^{R}_3
\right ]^2 \ \ , \label{eq:ALpmR1}\\ 
B^{L\pm R}_2= A^{L\mp R}_2&=&
\left [(d^{L}_1)^2+\beta_t^2 (d^{L}_2)^2  \right ] \mp
\left [(d^{R}_1)^2+\beta_t^2 (d^{R}_2)^2  \right ]
\ \ , \\
B^{L\pm R}_1= A^{L\mp R}_3 &=&
 d^{L}_1d^{L}_2 \pm  d^{R}_1d^{R}_2 
\ \ \ , \\
C^{L\pm R}_1&=&
d^{L}_1
\Big [d^{L}_1 - \frac{2|\pvec|^2}{\mtd} d^{L}_3 \Big ]
\mp d^{R}_1
\Big [d^{R}_1 - \frac{2|\pvec|^2}{\mtd} d^{R}_3 \Big ]
\ \ , \\
C^{L\pm R}_2&=&
d^{L}_2 \Big [d^{L}_1 - \frac{2|\pvec|^2}{\mtd} d^{L}_3 \Big ]
\pm
d^{R}_2
\Big [d^{R}_1 - \frac{2|\pvec|^2}{\mtd} d^{R}_3 \Big ]
\ \  \ . \label{eq:CLpmR2} 
\eqa
Introducing in analogy to (\ref{eq:dvbar}),
the NP contributions $\bar d^L_i$, $\bar d^R_i$
to the couplings defined in (\ref{eq:dLR}), and expanding to
first order in them, we get
\bqa
A^{L\pm R}_1&=& (d^L_{1SM})^2\pm(d^R_{1SM})^2  \nonumber \\
&&+2d^L_{1SM}\left ({\bar d}^L_1-{2|\pvec|^2\over \mtd}
{\bar d}^L_3 \right )\pm
2d^R_{1SM}\left ({\bar d}^R_1-{2|\pvec|^2\over \mtd}
{\bar d}^R_3\right ) \ , \label{eq:ALpmR1a} \\[0.2cm]
B^{L\pm R}_2= A^{L\mp R}_2 &= &
(d^L_{1SM})^2 \mp (d^R_{1SM})^2
+\beta_t^2 [(d^L_{2SM})^2 \mp (d^R_{2SM})^2]\nonumber \\
&&+2d^L_{1SM}{\bar d}^L_1\mp 2d^R_{1SM}{\bar d}^R_1+
 2\beta_t^2 [d^L_{2SM}{\bar d}^L_2\mp d^R_{2SM}{\bar d}^R_2]
\ , \\[0.1cm]
B^{L\pm R}_1= A^{L\mp R}_3 & = &
 d^L_{1SM}d^L_{2SM}\pm d^R_{1SM}d^R_{2SM} \nonumber \\
&&+d^L_{1SM}{\bar d}^L_2\pm d^R_{1SM}{\bar d}^R_2
+d^L_{2SM}{\bar d}^L_1\pm d^R_{2SM}{\bar d}^R_1
\ , \\[0.1cm]
C^{L\pm R}_1&=& (d^L_{1SM})^2\mp(d^R_{1SM})^2  \nonumber \\
&&+2d^L_{1SM}{\bar d}^L_1\mp 2d^R_{1SM}{\bar d}^R_1
-{2|\pvec|^2\over \mtd}[d^L_{1SM}{\bar d}^L_3 \mp d^R_{1SM}{\bar d}^R_3]
\ , \\[0.1cm]
C^{L\pm R}_2&=& d^L_{1SM}d^L_{2SM}\pm d^R_{1SM}d^R_{2SM}
-{2|\pvec|^2\over \mtd}[d^L_{2SM} \bar d^L_3 \pm d^R_{2SM}
{\bar d}^R_3]
\nonumber \\
&+& d^L_{1SM}{\bar d}^L_2\pm d^R_{1SM}{\bar d}^R_2
+d^L_{2SM}{\bar d}^L_1\pm d^R_{2SM}{\bar d}^R_1
 \ .\label{eq:CLpmR2a}
\eqa\par

The angular dependence 
of each of these $\rho$ elements in 
(\ref{eq:rhotrace}-\ref{eq:rho+-}) is determined by linear 
combinations of terms of the form 
$(1+\cos^2\theta)$, $\sin^2\theta$, $\cos\theta$,
$\sin\theta$ and $\sin\theta \cos\theta$. 
The aforementioned asymmetries just measure the relative ratios
of the coefficients of these terms, in each of the three
$\rho$ elements ~mentioned above.  
Thus, from the 
$(\rho_{++}+\rho_{--})^{L\pm R}$ elements in
(\ref{eq:rhotrace}), involving six $A_j^{L\pm R}$ terms
as coefficients of the 
$\sin^2\theta$, $(1+\cos^2\theta)$ and $\cos\theta$ angular 
dependencies, we construct five ratios not depending 
on the top decay properties. 
>From the $(\rho_{++}-\rho_{--})^{L\pm R}$ elements
in (\ref{eq:rho++m--}), containing four $B_j^{L\pm R}$ terms
associated to the
$(1+\cos^2\theta)$ and $\cos\theta$ dependencies, we can construct
three such ratios. Finally, from the $\rho^{L\pm R}_{+-}$  
elements in (\ref{eq:rho+-}) and its four 
$C_j^{L\pm R}$ terms associated to the
$\sin\theta$ and $\sin\theta \cos\theta$ angular dependencies,
another three ratios are possible.
Altogether we thus have 11 ratios that can be measured. This is
achieved  by constructing 4 asymmetries for
unpolarized beams and another 7 for polarized ones.
They are asymmetries of $A$-, $H$- and $T$-type mentioned above
\cite{GKR}. Below, we enumerate them:\par 

\vspace{0.2cm}

\noindent 
\underline{Unpolarized asymmetries}\\
For unpolarized beams there are two A-type asymmetries possible
which test  
$(\rho_{++}+\rho_{--})^{L+R}$. The forward-backward asymmetry
\begin{equation}
\label{eq:AFB}
A_{FB}=-\left ({3\beta_t\over2}\right)
{A^{L+R}_3\over A^{L+R}_2+{2m^2_t\over s}
A^{L+R}_1} \ \ , \
\end{equation}
and the edge-central one 
\begin{equation}
A_{EC}=\left ({3\over16}\right )
{ A^{L+R}_2-{4m^2_t\over s}A^{L+R}_1\over
A^{L+R}_2+{2m^2_t\over s}A^{L+R}_1} \ \ . \
\end{equation}
Moreover, there is 
the forward-backward asymmetry for $(\rho_{++}-\rho_{--})^{L+R}$
\begin{equation}
H_{FB}=-\left ({3\over8\beta_t}\right )
{B^{L+R}_2\over B^{L+R}_1} \ \ , \
\end{equation}
and the forward-backward asymmetry for $(\rho_{+-})^{L+R}$
\begin{equation}
T_{FB}=-\left ({4\beta_t\over3\pi}\right )
{C^{L+R}_2\over C^{L+R}_1} \ \ . 
\end{equation}

\noindent
\underline{Polarized asymmetries}\\
For polarized beams there is the left-right asymmetry 
for the total cross section which tests
$(\rho_{++}+\rho_{--})^{L,R}$ and gives
\begin{equation}
A_{LR}={ {2m^2_t\over s} A^{L-R}_1+A^{L-R}_2\over
{2m^2_t\over s}A^{L+R}_1+A^{L+R}_2}\ \ , 
\end{equation}
the forward-backward asymmetry of $(\rho_{++}+\rho_{--})^{L-R}$
\begin{equation}
A^{pol}_{FB}=-\left ({3\beta_t\over2}\right)
{A^{L-R}_3\over A^{L-R}_2+{2m^2_t\over s}A^{L-R}_1} \ \ ,
\end{equation}
the edge-central asymmetry of $(\rho_{++}+\rho_{--})^{L-R}$
\begin{equation}
A^{pol}_{EC}=\left ({3\over16}\right )
{ A^{L-R}_2-{4m^2_t\over s}A^{L-R}_1\over
A^{L-R}_2+{2m^2_t\over s}A^{L-R}_1} \ \ .
\end{equation}
In ~addition there are two H-type asymmetries which test the 
left-right asymmetry of $(\rho_{++}-\rho_{--})^{L,R}$ integrated
over the $t$ production angle $\theta$
\begin{equation}
H_{LR}={B^{L-R}_1\over B^{L+R}_1} \ \ ,
\end{equation}
and the forward-backward asymmetry of $(\rho_{++}-\rho_{--})^{L-R}$
\begin{equation}
H^{pol}_{FB}=-\left ({3\over8\beta_t}\right )
{B^{L-R}_2\over B^{L-R}_1} \ \ ,
\end{equation}
and another two T-type asymmetries testing
the left-right asymmetry of the integrated $(\rho_{+-})^{L,R}$
\begin{equation}
T_{LR}={C^{L-R}_1\over C^{L+R}_1} \ \ ,
\end{equation}
and the forward-backward asymmetry of $(\rho_{+-})^{L-R}$
\begin{equation}
\label{eq:TFBpol}
T^{pol}_{FB}=-\left ({4\beta_t\over3\pi}\right )
{C^{L-R}_2\over C^{L-R}_1}\ \ .
\end{equation}
 
To these 11 asymmetries
which are independent of any NP effects in the top decay 
couplings, we now add the three quantities measuring the 
overall magnitude of the unpolarized  $\rho^{L+R}$ elements 
in (\ref{eq:rhoR}), integrated over $\theta$. These elements, 
multiplied by $Br(t \to bW \to bl \nu_l)$, are obtained from
the event distribution (\ref{eq:dsigma}), either 
by simply integrating over the Euler angles of the top decay
plane, or by also using 
 the projectors mentioned (\ref{eq:PH}, \ref{eq:PT}) 
\cite{GKR}.\par

 For determining the overall
magnitude of the $\rho$'s, the  knowledge of 
$Br(t \to bW \to b l \nu_l)$ is needed. Thus, integrating  
(\ref{eq:dsigma}) and dividing the result by  
$Br(t \to bW \to bl \nu_l)$, we get
the unpolarized $e^-e^+ \to t\bar t$ total
cross section 
\bq
\label{eq:sigmat}
\sigma_t=\sigma(e^-e^+ \to t \bar t)=
\frac{\int  d \sigma^{L+R}}{Br(t \to bW \to bl \nu_l)}=
{2\pi\alpha^2\beta_t\over s}
\left [{2m^2_t\over s}A^{L+R}_1+ A^{L+R}_2 \right ] \ \ ,
\eq
determined by $(\rho_{++}+\rho_{--})^{L+R}$ in (\ref{eq:rhotrace}).
The quantity $\sigma_t$ is of course a type A observable,
according to the above ~classification. We also have the type H
observable ~determined by $(\rho_{++}-\rho_{--})^{L+R}$ 
in (\ref{eq:rho++m--}) and defined by 
\bq
\label{eq:Ht}
H_t=\frac{\int P_H d \sigma^{L+R}}{Br(t \to bW \to bl \nu_l)}= 
\ {\pi^2 \beta_t^2\alpha^2(m^2_t-2M^2_W)(1+r^2)
\over2 s(m^2_t+2M^2_W)}~B^{L+R}_1 \ \  ,
\eq
in terms of the projector in
(\ref{eq:PH}) and $d\sigma^{L+R}$ from (\ref{eq:dsigma}).
Finally  $(\rho_{+-})^{L+R}$ in (\ref{eq:rho+-}) 
~determines $T_t$ through
\bq
\label{eq:Tt}
T_t=\frac{\int P_T d \sigma^{L+R}}{Br(t \to bW \to bl \nu_l)}
~=~ {\pi^2\beta_t\alpha^2m_t(m^2_t-2M^2_W)(1+r^2)\over
2s^{3/2}(m^2_t+2M^2_W)}~C^{L+R}_1  \ \ ,
\eq
\noindent
using the projector in (\ref{eq:PT}) and $r$ given in eq.(\ref{eq:r}).
The effect on these quantities, of the top decay branching ratio 
and of the detection efficiencies,  will be taken into account 
in Section 3.\par

So in the whole, we have the 14 observables in 
(\ref{eq:AFB}-\ref{eq:Tt}) corresponding to the measurements 
of the coefficients of the various angular terms appearing in
the $\rho$ elements. Eleven of these observables, 
defined in (\ref{eq:AFB}-\ref{eq:TFBpol}),
are ratios of such coefficients, which are therefore independent
of the top-decay parameters involving 
the 10 combinations of couplings defined in
(\ref{eq:ALpmR1}-\ref{eq:CLpmR2}) or
(\ref{eq:ALpmR1a}-\ref{eq:CLpmR2a}) .\par

\section{Experimental accuracies and constraints}

The analysis then proceeds by writing for each of the 14
observables $\A^i$ in (\ref{eq:AFB}- \ref{eq:Tt}), 
the departures from the SM prediction
at first order in NP parameters, as 
\begin{equation}
\delta\A^i\equiv \A^i-\A^{i,SM}=\sum_{j=1,6} K^i_j \bar d_j \ \ .
\end{equation}
The coefficients $K^i_j$, deduced from the expansions
(\ref{eq:ALpmR1a}-\ref{eq:CLpmR2a}), are given in Appendix
A.\par
 
A priori it could seem sufficient to just use the 11 asymmetries
for a determination of the six NP couplings, since the
asymmetries are experimentally preferable quantities being 
independent of the overall normalization of the data. 
But this is actually not true, since 
asymmetries are only ~sensitive to the relative ratios
of the $d_j^Z,~d_j^\gamma$ ($j=1-3$) couplings.
Thus, asymmetries cannot impose important constraints on
couplings ~receiving ~non-vanishing ~values in SM at tree level.
Additional information is needed to fix the overall 
normalization of these
couplings. This can either be done by using an extra 
measurement sensitive to overall coupling normalization, 
or by imposing an appropriate ~non-homogeneous constraint which
will force  some of the ~couplings that have non vanishing SM
values, to retain their magnitudes even in the presence of NP.
\begin{itemize} 
\item
Thus we can use the measurements of the total $e^+e^-\to t\bar
t$ cross section $\sigma_t$,  and/or any 
of the two other combinations of the integrated $\rho$ density matrix
elements expressed through the quantities 
$H_t$ or $T_t$; compare (\ref{eq:sigmat}-\ref{eq:Tt}).
As mentioned
previously, these quantities are sensitive to the top quark
decay width. To take this into account we 
consider two extreme cases for the uncertainties of 
$\sigma_t$, $H_t$ or $T_t$. In the first case these
uncertainties are taken to be $\sim 2\%$, which is of the order 
of magnitude of possible  NP effects in $t\to Wb$ and the 
$\gamma t\bar t$ or $Zt\bar t$ couplings; while in the second
case we take  the uncertainties to be of order of $20\%$,
which is the type of
uncertainty expected for the experimental measurement of the top
decay width\cite{effic}.
\item 
Alternatively, we impose a (non-homogeneous) constraint on the
$d_j^V$, like \eg\@ forcing some of them to have their SM
values. For
example this is what happens in the 3- or 4-parameter cases
presented below. 
\end{itemize}

The observability limits are now obtained by the following procedure.
For each observable $\A^i$, we estimate an experimental uncertainty
$\delta\A^i_{exp}$. Following \cite{effic}, we
assume an overall reduction of the number of events by a factor
0.18 due to branching ratios, reconstruction of events,
efficiencies and detector acceptance. We then apply statistical
considerations to the projected events, in the three types 
A, H and T observables defined above. The results are given in Table 1.\par

Assuming that the measurements
coincide with the SM expectations and demanding that for an NP
effect to be observable, the statistical  uncertainty
should be smaller than the effect expected due to the NP
couplings $\bar d_j^V$ (compare
\ref{eq:ALpmR1a}-\ref{eq:CLpmR2a}), we write for each observable
the  inequality 
\bq
|\sum_{j=1}^n K^i_j \bar{d}_j|~~ \geq ~~{\delta\A^i_{exp}\over\A^i_{SM}}
\ \ , \eq
where $n\leq 6$ NP is the number of the NP couplings considered.
We then combine quadratically all such information coming from the $l$
available observables. This
gives at one standard deviation 
the observability domain which is outside the ellipsoid surface
\bq
\sum_{i=1}^l~~|\sum_{j=1}^n [K^i_j \bar{d}_j].
[{\delta\A^i_{exp}\over\A^i_{SM}}]^{-1}~|^2 ~~ = ~~1
\eq\par

\section{Applications}
\subsection{6-parameter case}
 
We first consider the most general case
with 6 free parameters $\bar{d}^{\gamma}_i$,
$\bar{d}^Z_i$, $i=1,2,3$.
Results are collected in Table 2 for a 0.5, 1, 2~TeV collider and
illustrated in Fig.1 for the 1~TeV case.
Remember that only the three parameters $d^{\gamma}_1$, $d^Z_1$, $d^Z_2$
receive SM tree level contributions.
We start by considering the
constraints due to all 11 asymmetries, assuming that polarized
$e^{\pm}$ beams
are available.
As expected, only the
3 pure NP couplings $\bar{d}^{\gamma}_2$, $\bar{d}^{\gamma}_3$,
$\bar{d}^Z_3$, (for which there is no SM analog), 
are then strictly constrained, as one can see in Fig.1(d,e,f)
and in Table 2. For these three couplings, the observability
limit is around
\bq
|\bar{d}_j|~~\gsim~~0.01 \ \ .
\eq

Among the other
3 couplings,  $\bar{d}^{\gamma}_1$, $\bar{d}^Z_1$, $\bar{d}^Z_2$,
one constraint is missing and this is the origin of the band
in Figs.1a,c,e relating pairs of these parameters. The
width of the band is of a few percent.\par
 
We have then considered how a
measurement of $\sigma_t$, $H_t$ and $T_t$ with an uncertainty of 20\%
or 2\% on the decay width would limit these bands. In fact only
one of these quantities would be sufficient to limit the bands, and
it turns out that $\sigma_t$ is the most efficient, since
the statistical weight associated to it is the largest. In the
following results all three informations are statistically 
included in the ellipsoid. 
This is
also shown in Fig.1a,c,e where one sees that the band is
transformed into ellipses whose sizes are of the
order of $\pm0.025$ to $0.05$. In addition in Table 2 and in
Fig.1(g,i,k) one sees, as expected, that this
additional information does not much improve the determination
of the  set of three pure NP couplings which are already
severely constrained by the asymmetries.\par

If no $e^{\pm}$ beam polarization is available, we have only 4
asymmetries at our disposal, and it is necessary to add the
information coming from $\sigma_t$, $H_t$, $T_t$
in order to constrain the system of 6 free parameters.
The results are shown in Table 3 and in Fig.1(b,d,f,h,j,l). 
The constraints on the pure
NP couplings $\bar{d}^{\gamma}_2$, $\bar{d}^{\gamma}_3$,
$\bar{d}^Z_3$, are now a factor 2 less
stringent; \ie\@ they lie at the 0.02 level. The other three
couplings get also more freedom by a factor 2 and are now
allowed to reach the 0.08 level.\par

All our figures correspond to the case of a 1 TeV
collider. Expectations
for 0.5 TeV and 2 TeV can be compared in Tables 2-7. The change
is not dramatic. The most
notable energy dependence concerns the tensor
couplings $d_3^V$. For the other ones the sensitivity is much milder.
Remember that we assumed that the luminosity grows like $s$ so that the
number of events is roughly the same at all three considered energies.
So only the intrinsic s dependence of the NP couplings shows up in
this comparison.\par

\subsection{4-parameter case}
We here consider the case of only 4 free parameters. We impose
the photon NP couplings to be of a pure
$\sigma_{\mu\nu}q^{\nu}$ type with no axial coupling, 
that means
\bq
\bar{d}^{\gamma}_1=-2\bar{d}^{\gamma}_3 \ \ , 
\ \ \bar{d}^{\gamma}_2=0\ \ ,
\eq
\noindent
so that the 4 free parameters are taken as
$\bar{d}^{\gamma}_3$ and $\bar{d}^Z_i$ ($i=1,2,3$).
This constraint is satisfied by the tree level NP contributions of
all operators considered
in \cite{GKR}. In the notations of \cite{GKR}, these 
operators are $\O_{t2}$, $\O_{Dt}$, $\O_{tW\Phi}$ and
$\O_{tB\Phi}$.\par 
 
We have considered both polarized (11 asymmetries)
and unpolarized (4 asymmetries) cases, illustrated in
Fig.2 (a,c,e,g,i,k) and (b,d,f,h,j,l), respectively. 
We observe that a meaningful constraint
is already obtained in the unpolarized case at the 0.1 level
without $\sigma_t$,
$H_t$ or $T_t$ measurements. Furthermore, in this  case
the ~additional constraint due to
a measurement of $\sigma_t$ at a 20 \% accuracy, reduces the
allowed domain by a factor 5, while another factor 2 would be
obtained if the accuracy is at 2\%.
Having polarized observables at our disposal 
improves notably the constraints, reaching the 0.01 level, even
when only asymmetries are used; compare Fig.2 and 
Table 4.\par
 
\subsection{3-parameter case}
A 3-parameter case is obtained from the 4-parameter one 
by imposing that the NP
contribution to the $Zt\bar t$ couplings only involves
the $\sigma^{\mu\nu}q_{\nu}$ and the right-handed
$\gamma^{\mu}(1+\gamma^5)$ couplings. This is suggested by the
tree level contributions of the dynamical models studied in 
\cite{tdyn}. Indeed, it has been found that in the dynamical
models studied in \cite{tdyn}, that only
three operators can be generated which contribute at ~tree level
to the $Zt\bar t$ vertex. These operators, named $\O_{t2}$,  
$\O_{tW\Phi}$ and $\O_{tB\Phi}$ in \cite{tdyn},
lead precisely to the aforementioned types of couplings.\par

We first show in Fig.3 the ellipsoid containing the invisible domain at
1 TeV in the 3-parameter space 
($\bar{d}^{\gamma}_3$, $\bar{d}^{Z}_2$, $\bar{d}^{Z}_3$),  
for the polarized case. When projected on the
three axes one obtains the ellipses shown in Fig.4(a,c,e). The
comparison with the
unpolarized case is done with Fig.4(b,d,f).
Results for these NP couplings, collected in Table 5, 
appear to be of similar type as those of the
4-parameter case, with the constraints reduced by roughly a factor 2.
In some cases these constraints reach the 0.005 level. \par

\subsection{2-parameter case}
We have also analyzed the 2-parameter case
suggested by the chiral description. In this case NP contribute
only to the vector and axial couplings of the Z. 
The effective Lagrangian is written as \cite{tchiral}
\bq
\L=-{e\over4s_W c_W}\bar{\Psi}_t[\kappa^{NC}_L\gamma^{\mu}(1-\gamma^5)
+\kappa^{NC}_R\gamma^{\mu}(1+\gamma^5)]\Psi_t Z_{\mu}\  \ ,
\eq
which gives the NP contributions in our notations as:
\bq
\bar{d}^Z_1={1\over2}[\kappa^{NC}_L+\kappa^{NC}_R-(1-{8\over3}s^2_W)]
\ \ , 
\eq
\bq
\bar{d}^Z_2=-~{1\over2}[\kappa^{NC}_L-\kappa^{NC}_R-1]\ .
\eq
 
The corresponding constraints are shown in Fig.5 and Table 6. One
can appreciate the role of the polarization (an improvement from
the 0.1 to the 0.01--0.02 level). Also in the unpolarized case
the measurement of the cross section allows to reach the 
few 0.01 level, but with polarization this level is already
obtained with asymmetries only.\par

\subsection{1-parameter case}
Finally we reconsider the simplest case in which all effective
operators listed in \cite{GKR} are taken one by one. We want to
appreciate the role of the different observables in obtaining 
the constraints on the
associated coupling constants. The expressions of the $\bar{d}_j$
couplings have been established in \cite{GKR}. Results are given in
Table 7 for the polarized and the unpolarized case. Corresponding
to each coupling we give in parenthesis the value of the NP
scale obtained from the
unitarity relations established in \cite{unitop}. The values are in the
order of magnitude expected from the rough analysis made in 
\cite{GKR}. Polarization generally increases the visibility domain by
a factor 2. Operators contributing at tree level lead obviously to
the largest effects and in some cases NP scales of the order of 50
TeV can be reached.\par

\section{Conclusions}
In this paper we have studied the observability of NP effects on
the $\gamma t\bar t$ and $Z t\bar t$ couplings in a model
independent way through the process $e^-e^+ \to t\bar t$. 
We have considered the rather general situation involving the 6
anomalous CP-conserving couplings.\par

We have first tried an analysis which would not be affected by
the unknown top quark decay couplings. For that purpose we have
used 4 unpolarized and 7 polarized asymmetries, constructed from the
top quark density matrix elements, which are independent of the
decay couplings. We have obtained severe
constraints on the set of the 3 pure NP couplings; 
\ie\@ those that receive no SM contributions. The observability 
limits  in this set of NP couplings ($d_2^{\gamma}$, 
$d_3^{\gamma}$ and $d_3^Z$) vary from 0.01
to 0.02, independently of any top decay uncertainties.\par

 For the other three couplings $d_1^{\gamma}$, 
$d_1^Z$ and $d_2^Z$, which receive tree level 
SM contributions,  asymmetries are not sufficient to 
constrain them and one additional information is needed. 
We have thus added the
informations coming from the integrated top density matrix
elements and ~discussed the influence of the uncertainty 
affecting the top decay width.
Varying it from $2\%$ to $20\%$, the observability limits for
this set of couplings lie in the range 0.01 to 0.05.\par

The improvement brought by $e^{\pm}$ beam
polarization corresponds to roughly a reduction of the invisible
domain by a factor 2.\par

 We have also considered 3 different $e^+e^-$
collider energies, 0.5, 1 and 2 TeV. The order of magnitude 
of the observability limits for 
$\bar{d}_1$ and $\bar{d}_2$, do not depend strongly on the
energy. On the contrary, for the tensor coupling $\bar{d}_3$, 
the sensitivity increases by one order of magnitude, as the 
energy increases from 0.5 to 2TeV.\par

We have also considered more specific cases of NP models in
which the number of free parameters is reduced. In these cases
asymmetries alone allow to constrain all parameters, because
there exist relations between the two aforementioned sets of 
NP couplings. In such cases, it is possible to get strong constraints 
that are independent of the top decay couplings. 
We have made illustrations for the case of 4-parameter,
3-parameter, 2-parameter and 1-parameter models, 
suggested by the effective
lagrangian descriptions. Visibility domains are obviously
now increased, reaching in some cases the few permille level.
As compared to present indirect constraints from LEP/SLC (which
are at the $10\%$ level in the 2-parameter case) \cite{Lang}, and
to the constraints expected on the charged current couplings
$Wtb$ at LHC (also at the $10\%$ level) \cite{tchiral}, this
represents an important improvement.\par

Translating to NP scales through the unitarity
relations \cite{unitop, uniY}, we find that the percent level
in the sensitivity to these couplings 
(precise values depending on the considered
operators), corresponds to NP scales in the 10 TeV region.\par

Finaly we should state that our analysis is of theoretical 
nature and should
certainly be adapted to specific experimental and detection
configurations. We believe though, that it has the advantage of 
pointing out
the merit of each type of observable, and 
of specifying the results which do not depend on assumptions
about the possibly unknown top decay couplings.

\newpage

\begin{center}
{\bf Table 1: Expected accuracy on observables}\\
\vspace{0.5cm}
\begin{tabular}{|c|c|c|c|}
\hline
\multicolumn{1}{|c|}{}&
\multicolumn{3}{|c|}{ Integrated
$\rho$-elements}
\\[0.1cm] \hline
\multicolumn{1}{|c|}{Norm. accuracy}&
\multicolumn{1}{|c|}{$\delta\sigma_t/\sigma_t$} & 
\multicolumn{1}{|c|}{$\delta H_t/H_t$} & 
\multicolumn{1}{|c|}{$\delta T_t/T_t$}
\\[0.1cm] \hline
$\delta\Gamma_t/\Gamma_t=0.02$&
$0.028$& $0.19$ & $0.29$
 \\[0.1cm] \hline
$\delta\Gamma_t/\Gamma_t=0.20$&
$0.20$ & 0.27& 0.35\\
[0.1cm] \hline
\end{tabular}
\end{center}
 
\vspace{0.3cm}
 
\begin{center}
\begin{tabular}{|c|c|c|c||c|c|c|c|c|c|c|}
\hline
\multicolumn{4}{|c||}{ Unpolarized Asymmetries}&
\multicolumn{7}{|c|}{ Polarized Asymmetries}
\\[0.1cm]\hline
\multicolumn{1}{|c|}{$\delta A_{FB}$} & \multicolumn{1}{|c|}{
$\delta A_{EC}$} & \multicolumn{1}{|c|}{$\delta H_{FB}$}
&\multicolumn{1}{|c||}{$\delta T_{FB}$}&
\multicolumn{1}{|c|}{$\delta A^{(A)}_{LR}$}
& \multicolumn{1}{|c|}{$\delta A^{(pol)}_{FB}$}
& \multicolumn{1}{|c|}{$\delta A^{(pol)}_{EC}$}
 & \multicolumn{1}{|c|}{
$\delta A^{(B)}_{LR}$} & \multicolumn{1}{|c|}{$\delta H^{(pol)}_{FB}$}
&\multicolumn{1}{|c|}{$\delta A^{(C)}_{LR}$}
&\multicolumn{1}{|c|}{$\delta T^{pol}_{FB}$}
\\[0.1cm] \hline
$0.017$&0.020& $0.23$&0.31&
$0.019$&0.050& $0.058$&0.75&0.053&0.82&0.11
 \\[0.1cm] \hline
\end{tabular}
\end{center}

\vspace{1cm}
\begin{center}
{\bf Table 2: Sensitivity limits for 6 free parameters in the
polarized case}\\
(For each energy the three lines correspond to constraints
obtained, without ($\sigma_t$, $H_t$, $T_t$), or with 
$\sigma_t$, $H_t$, $T_t$, assuming a normalization uncertainty 
of $2\%$ or $20\%$)\\
\vspace{0.3cm} 
\begin{tabular}{|c|c||c|c|c|c|c|c|}
\hline 
\multicolumn{1}{|c|}{\null}&
\multicolumn{1}{|c||}{\null} & 
\multicolumn{1}{|c|}{\null} &
\multicolumn{1}{|c|}{\null} &
\multicolumn{1}{|c|}{\null} &
\multicolumn{1}{|c|}{\null} &
\multicolumn{1}{|c|}{\null} &
\multicolumn{1}{|c|}{\null} \\[-0.4cm]
\multicolumn{1}{|c|}{$\sqrt{s}$}&
\multicolumn{1}{|c||}{$\delta\Gamma_t/\Gamma_t$} & 
\multicolumn{1}{|c|}{$\bar d^{\gamma}_1$} &
\multicolumn{1}{|c|}{$\bar d^{\gamma}_2$} &
\multicolumn{1}{|c|}{$\bar d^{\gamma}_3$} &
\multicolumn{1}{|c|}{$\bar d^{Z}_1$} &
\multicolumn{1}{|c|}{$\bar d^{Z}_2$} &
\multicolumn{1}{|c|}{$\bar d^{Z}_3$} \\[0.1cm] \hline
0.5&{no $\sigma_t$,$H_t$,$T_t$ } & --- & 0.020 & 0.049 & --- & --- &
0.050 \\[0.1cm] \hline
0.5&${20\%}$ &0.055  & 0.019 &0.043  & 0.032 &0.049
& 0.044 \\[0.1cm] \hline
0.5&${2\% }$ & 0.023 &0.018  &0.043  &0.026  &0.031
 & 0.039 \\[0.1cm] \hline
1&{no $\sigma_t$,$H_t$,$T_t$ } & --- & 0.014 & 0.016 & --- & --- &
0.010 \\[0.1cm] \hline
1&${20\%}$ & 0.056 & 0.013 &0.014  &0.025 &0.044
&0.008  \\[0.1cm] \hline
1&${2\% }$ & 0.019  & 0.013 & 0.014 & 0.018 & 0.022
 & 0.007 \\[0.1cm] \hline
2&{no $\sigma_t$,$H_t$,$T_t$ } & --- & 0.013 & 0.012 & --- & --- &
0.006 \\[0.1cm] \hline
2&${20\%}$ &0.056  & 0.012 & 0.008 &0.023  &0.044
&0.003  \\[0.1cm] \hline
2&${2\% }$ &0.016  &0.011  & 0.008 & 0.016 &0.020
 &0.003  \\[0.1cm] \hline
\end{tabular}
\end{center}

\newpage

\vspace{1cm}
\begin{center}
{\bf Table 3: Sensitivity limits for 6 free parameters 
in the unpolarized 
case}\\
(with $\sigma_t$, $H_t$, $T_t$ and a $2\%$ or $20\%$ normalization
uncertainty)\\
\vspace{0.3cm} 

\begin{tabular}{|c|c||c|c|c|c|c|c|}
\hline
\multicolumn{1}{|c|}{\null} &
\multicolumn{1}{|c||}{\null} & 
\multicolumn{1}{|c|}{\null} &
\multicolumn{1}{|c|}{\null} &
\multicolumn{1}{|c|}{\null} &
\multicolumn{1}{|c|}{\null} &
\multicolumn{1}{|c|}{\null} &
\multicolumn{1}{|c|}{\null} \\[-0.4cm]
\multicolumn{1}{|c|}{$\sqrt{s}$} &
\multicolumn{1}{|c||}{$\delta\Gamma_t/\Gamma_t$} & 
\multicolumn{1}{|c|}{$\bar d^{\gamma}_1$} &
\multicolumn{1}{|c|}{$\bar d^{\gamma}_2$} &
\multicolumn{1}{|c|}{$\bar d^{\gamma}_3$} &
\multicolumn{1}{|c|}{$\bar d^{Z}_1$} &
\multicolumn{1}{|c|}{$\bar d^{Z}_2$} &
\multicolumn{1}{|c|}{$\bar d^{Z}_3$} \\[0.1cm] \hline
0.5&${20\%}$ &0.082  & 0.032 &0.056  & 0.103 &0.063
& 0.109 \\[0.1cm] \hline
0.5&${2\% }$ & 0.034 &0.031  &0.055  &0.093  &0.041
 & 0.096 \\[0.1cm] \hline
1&${20\%}$ & 0.081 & 0.024 &0.021  &0.081 &0.057
&0.016  \\[0.1cm] \hline
1&${2\% }$ & 0.031  & 0.023 & 0.021 & 0.067 & 0.031
 & 0.014 \\[0.1cm] \hline
2&${20\%}$ &0.081  & 0.023 & 0.017 &0.078  &0.056
&0.006  \\[0.1cm] \hline
2&${2\% }$ &0.030  &0.022  & 0.017 & 0.063 &0.030
 &0.006  \\[0.1cm] \hline
\end{tabular}
\end{center}

 \vspace{1cm}
\begin{center}
{\bf Table 4: Sensitivity limits with 4 free parameters}\\
(same captions as in Table 2)

\vspace{0.3cm}
\begin{tabular}{|c|c||c|c|c|c||c|c|c|c|} \hline
\multicolumn{1}{|c|}{$\sqrt{s}$}&
\multicolumn{1}{|c||}{$\delta\Gamma_t/\Gamma_t$}&
\multicolumn{4}{|c||}{ Polarized}&
\multicolumn{4}{|c|}{Unpolarized} \\[0.1cm]\hline
\multicolumn{1}{|c|}{} &
\multicolumn{1}{|c||}{}&
\multicolumn{1}{|c|}{} &
\multicolumn{1}{|c|}{} &
\multicolumn{1}{|c|}{} &
\multicolumn{1}{|c||}{} &
\multicolumn{1}{|c|}{} &
\multicolumn{1}{|c|}{} &
\multicolumn{1}{|c|}{} &
\multicolumn{1}{|c|}{} \\[-0.4cm]
\multicolumn{1}{|c|}{} &
\multicolumn{1}{|c||}{}&
\multicolumn{1}{|c|}{$\bar d^{\gamma}_3$} &
\multicolumn{1}{|c|}{$\bar d^{Z}_1$} &
\multicolumn{1}{|c|}{$\bar d^{Z}_2$} &
\multicolumn{1}{|c||}{$\bar d^{Z}_3$} &
\multicolumn{1}{|c|}{$\bar d^{\gamma}_3$} &
\multicolumn{1}{|c|}{$\bar d^{Z}_1$} &
\multicolumn{1}{|c|}{$\bar d^{Z}_2$} &
\multicolumn{1}{|c|}{$\bar d^{Z}_3$} \\[0.1cm] \hline
0.5&no $\sigma_t$,$H_t$,$T_t$ &0.042  &0.040  &0.089  &0.042&0.140&0.391&0.235&0.693
 \\[0.1cm] \hline
0.5&${20\%}$ &0.018  &0.026  &0.043  &0.038&0.006&0.057&0.029&0.070
 \\[0.1cm] \hline
0.5&${2\% }$ &0.004  &0.020  &0.025  &0.033&0.006&0.070&0.029&0.071
 \\[0.1cm] \hline
1&no $\sigma_t$,$H_t$,$T_t$ &0.015  &0.020  &0.046 
&0.010&0.030&0.302&0.111&0.123
 \\[0.1cm] \hline
1&${20\%}$ &0.010  &0.017  &0.032  &0.007&0.015&0.065&0.041&0.013
 \\[0.1cm] \hline
1&${2\% }$ &0.004  &0.014  &0.017  &0.006&0.005&0.048&0.020&0.010
 \\[0.1cm] \hline
2&no $\sigma_t$,$H_t$,$T_t$ &0.012  &0.016  &0.041  &0.005&0.018&0.291&0.104&0.051
 \\[0.1cm] \hline
2&${20\%}$ &0.007  &0.014  &0.026  &0.003&0.004&0.038&0.018&0.004
 \\[0.1cm] \hline
2&${2\% }$ &0.003  &0.012  &0.016  &0.003&0.005&0.044&0.018&0.004
 \\[0.1cm] \hline
\end{tabular}
\end{center}

\vspace{1cm}

\newpage
 
\vspace{1cm}
\begin{center}
{\bf Table 5: Sensitivity limits with 3 free parameters}\\
(same captions as in Table 2)

\vspace{0.3cm}
\begin{tabular}{|c|c||c|c|c||c|c|c|}  \hline
\multicolumn{1}{|c|}{$\sqrt{s} $} &
\multicolumn{1}{|c||}{$\delta\Gamma_t/\Gamma_t$} &
\multicolumn{3}{|c||}{ Polarized}&
\multicolumn{3}{|c|}{Unpolarized} \\[0.1cm]\hline
\multicolumn{1}{|c|}{} &
\multicolumn{1}{|c||}{}&
\multicolumn{1}{|c|}{} &
\multicolumn{1}{|c|}{} &
\multicolumn{1}{|c||}{} &
\multicolumn{1}{|c|}{} &
\multicolumn{1}{|c|}{} &
\multicolumn{1}{|c|}{} \\[-0.4cm]
\multicolumn{1}{|c|}{} &
\multicolumn{1}{|c||}{}&
\multicolumn{1}{|c|}{$\bar d^{\gamma}_3$} &
\multicolumn{1}{|c|}{$\bar d^{Z}_2$} &
\multicolumn{1}{|c||}{$\bar d^{Z}_3$} &
\multicolumn{1}{|c|}{$\bar d^{\gamma}_3$} &
\multicolumn{1}{|c|}{$\bar d^{Z}_2$} &
\multicolumn{1}{|c|}{$\bar d^{Z}_3$} \\[0.1cm] \hline
0.5&no $\sigma_t$,$H_t$,$T_t$   &0.026  &0.055&0.029&0.048&0.104&0.220
 \\[0.1cm] \hline
0.5&${20\%}$ &0.015  &0.035&0.017&0.025&0.053&0.021
 \\[0.1cm] \hline
0.5 &${2\% }$ &0.004  &0.024 & 0.011 & 0.005 & 0.029 & 0.015
 \\[0.1cm] \hline
1&no $\sigma_t$,$H_t$,$T_t$  &0.009&0.023&0.009&0.023&0.055&0.106
 \\[0.1cm] \hline
1&${20\%}$  &0.007&0.017&0.006&0.015&0.038&0.008
 \\[0.1cm] \hline
1&${2\% }$ &  0.004&0.014&0.005&0.004&0.019&0.007
 \\[0.1cm] \hline
2&no $\sigma_t$,$H_t$,$T_t$ &0.007  &0.015&0.005&0.016&0.043&0.050
 \\[0.1cm] \hline
2&${20\%}$ &0.005  &0.011&0.003&0.038&0.018&0.004
 \\[0.1cm] \hline
2&${2\% }$ &  0.003&0.010&0.002&0.013&0.034&0.004
 \\[0.1cm] \hline
\end{tabular}
\end{center}

\vspace{1cm}
\begin{center}
{\bf Table 6: Sensitivity limits with 2 free parameters}\\
(same captions as in Table 2)

\vspace{0.3cm}
\begin{tabular}{|c|c||c|c||c|c|}
\hline
\multicolumn{1}{|c|}{$\sqrt{s}$}&
\multicolumn{1}{|c||}{$\delta\Gamma_t/\Gamma_t$}&
\multicolumn{2}{|c||}{ Polarized}&
\multicolumn{2}{|c|}{Unpolarized} \\[0.1cm]\hline
\multicolumn{1}{|c|}{}
&\multicolumn{1}{|c||}{}&
\multicolumn{1}{|c|}{} &
\multicolumn{1}{|c||}{} &
\multicolumn{1}{|c|}{} &
\multicolumn{1}{|c|}{} \\[-0.4cm]
\multicolumn{1}{|c|}{}
&\multicolumn{1}{|c||}{}&
\multicolumn{1}{|c|}{$\bar d^{Z}_1$} &
\multicolumn{1}{|c||}{$\bar d^{Z}_2$} &
\multicolumn{1}{|c|}{$\bar d^{Z}_1$} &
\multicolumn{1}{|c|}{$\bar d^{Z}_2$} 
 \\[0.1cm] \hline
0.5&no $\sigma_t$,$H_t$,$T_t$ &0.012  &0.027 &0.373  &0.090
 \\[0.1cm] \hline
0.5&${20\%}$ &0.012  &0.027 &0.042  &0.032 
 \\[0.1cm] \hline
0.5&${2\% }$ &0.010  &0.025 &0.026  &0.029 
 \\[0.1cm] \hline
1&no $\sigma_t$,$H_t$,$T_t$ &0.012  &0.020 &0.286  &0.074 
 \\[0.1cm] \hline
1&${20\%}$ &0.012  &0.020  &0.045  &0.024
\\[0.1cm] \hline
1&${2\% }$ &0.011  &0.017  &0.030  &0.020
 \\[0.1cm] \hline
2&no $\sigma_t$,$H_t$,$T_t$ &0.013  &0.019  &0.276  &0.074
 \\[0.1cm] \hline
2&${20\%}$ &0.012  &0.019  &0.052  &0.024
 \\[0.1cm] \hline
2&${2\% }$ &0.011  &0.016 &0.032  &0.018 
 \\[0.1cm] \hline
\end{tabular}
\end{center}

\newpage

\noindent
\begin{center}
{\bf Table 7: Sensitivity limits for 1 free parameter,
polarized/unpolarized cases}\\
(Limits correspond to the couplings associated to each operator,
the corresponding value of the
NP scale is indicated below.
The column "other constraints" refers to LEP1/SLC 
(a) from $\epsilon_i$,~~~ (b) from $R_b$ ~~~and (c) to limits 
expected from LEP2.\\

\vspace{0.3cm}
\begin{tabular}{|c|c|c|c|c|} \hline
\multicolumn{1}{|c|}{Operator} &
  \multicolumn{1}{|c|}{$\sqrt{s}=0.5$ TeV} &
   \multicolumn{1}{|c|}{$\sqrt{s}=1$ TeV} &
     \multicolumn{1}{|c|}{$\sqrt{s}=2$ TeV} &
       \multicolumn{1}{|c|}{other constraints}
          \\[0.1cm] \hline
  $\O_{qt}$ & 0.53/1.62& 0.26/0.53 &
0.18/0.31& $-0.14\pm0.07^{(b)}$\\
   & (0.98)/(0.56)& (1.41)/(0.99) &
(1.68)/(1.29)& \\[0.1cm]\hline
 
  $\O^{(8)}_{qt}$ & 0.10/0.30 & 0.049/0.099
&0.034/0.057 &$-0.027\pm0.013^{(b)}$
\\
  & (2.47)/(1.42) & (3.55)/(2.49)
&(4.24)/(3.26) & 
\\[0.1cm]\hline

  $\O_{tt}$ & 0.064/0.11 & 0.017/0.039 &
0.010/0.026 & ----- \\

  &(3.00)/(2.32) & (5.87)/(3.85) &
(7.46)/(4.70) & \\[0.1cm]\hline
 
  $\O_{tb}$ & 0.14/0.36 & 0.043/0.11 &
0.027/0.071 & $-0.13\pm0.06^{(b)}$\\

  & (2.33)/(1.46) & (4.24)/(2.64) &
(5.30)/(3.30) & \\[0.1cm]\hline
 
$\O_{t2}$ & 0.010/0.023& 0.0090/0.018 &
0.0089/0.017 &$0.01^{(a)}$; $0.14\pm0.07^{(b)}$ \\

 & (11.57)/(7.60)& (12.18)/(8.62) &
(12.24)/(8.75) &  \\[0.1cm]\hline
 
  $\O_{Dt}$ &0.039/0.093&
0.011/0.018&0.0052/0.0071&$0.03^{(a)}$; $-0.06\pm0.03^{(b)}$ \\

   &(2.84)/(1.85)&
(5.27)/(4.15)&(7.84)/(6.68)& \\[0.1cm]\hline

$\O_{tW\Phi}$ & 0.0010/0.0021 &
0.00067/0.0010& 0.00043/0.00056 &$0.014^{(a)}$
\\

 & (42.67)/(29.97) &
(53.14)/(42.73)& (66.16)/(58.10) &
\\[0.1cm]\hline

  $\O_{tB\Phi}$ & 0.0011/0.0027 &
0.00079/0.0015 &0.00060/0.0012 & $0.013^{(a)}$ \\

 & (41.63)/(26.52) &
(48.82)/(36.11) &(55.89)/(39.80) & \\
[0.1cm]\hline
 
$\O_{tG\Phi}$ & 0.027/0.029 & 0.023/0.025
 & 0.045/0.047 & ----- \\

 & (7.86)/(7.30) & (9.08)/(8.54)
 & (4.71)/(4.52) & \\[0.1cm]\hline
 
$\O_W$ & 0.065/0.13 & 0.021/0.045 &
0.014/0.030 & $0.1^{(c)}$ \\

 & (1.37)/(0.95) & (2.38)/(1.65) &
(2.95)/(2.02) & \\[0.1cm]\hline
 
$\O_{W\Phi}$ & 0.11/0.22 & 0.036/0.075 &
0.023/0.050 & $0.1^{(c)}$ \\

& (1.35)/(0.94) & (2.35)/(1.63) &
(2.91)/(1.99) &  \\[0.1cm]\hline
 
$\O_{B\Phi}$ & 0.071/0.14 & 0.020/0.043 &
0.012/0.028 & $0.1^{(c)}$ \\

& ((2.98)/(2.09) & (5.66)/(3.81) &
(7.14)/(4.76) &  \\[0.1cm]\hline

$\O_{WW}$ & 0.28/0.56 & 0.29/0.45 &
0.51/0.66 & $0.015^{(c)}$ \\

& (1.56)/(1.10) & (1.52)/(1.22) &
(1.15)/(1.01) &  \\[0.1cm]\hline
 
$\O_{BB}$ & 0.32/0.78 & 0.37/0.69
&0.77/1.53  & $0.05^{(c)}$ \\

 & (1.99)/(1.27) & (1.83)/(1.35)
&(1.27)/(0.91)  &  \\[0.1cm]\hline

$\O_{\Phi 2}$ & 0.57/0.68 &  0.68/0.81 & 1.74/2.08& $0.01^{(c)}$  \\
\hline

\end{tabular}
\end{center}

\newpage

\renewcommand{\theequation}{A.\arabic{equation}}
\setcounter{equation}{0}
\setcounter{section}{0}

{\large {\bf Appendix A}}\\

\vspace{0.3cm}

We give here the explicit expressions of the coefficients
determining the NP effects of each anomalous coupling on the
various observables $\A^i$ that we consider. These expressions are
given in terms of $\bar {d}^{L,R}_j$ couplings defined through 
(\ref{eq:dLR}) by
\bq
\bar d^{L,R}_j = d^{L,R}_j - d^{L,R, SM}_j \ \ ,
\eq
in the form
\bq
\delta\A^i = \sum_{j=1,3} [K^{L}_j \bar{d}^{L}_j+ K^{R}_j \bar{d}^{R}_j]
\ \ .
\eq 
They are obtained from
the expressions (\ref{eq:AFB}-\ref{eq:Tt}) of the observables and
the expansions to first order in NP couplings of the combinations 
(\ref{eq:ALpmR1a}-\ref{eq:CLpmR2a}).
 Thus,
using the definitions
\bqa
a_1 \equiv A^{L+R}_{2SM}+{2m^2_t\over s}A^{L+R}_{1SM}  & , &
a_2=A^{L+R}_{2SM}-{4m^2_t\over s}A^{L+R}_{1SM}\ , \\     
b_1=A^{L-R}_{2SM}+{2m^2_t\over s}A^{L-R}_{1SM} & , &
b_2=A^{L-R}_{2SM}-{4m^2_t\over s}A^{L-R}_{1SM} \ ,
\eqa
\noindent
and the combinations $A^{L\pm R}_{iSM}$,
$B^{L\pm R}_{iSM}$ and $C^{L\pm R}_{iSM}$ computed from 
(\ref{eq:ALpmR1a}-\ref{eq:CLpmR2a})
with
\bq
d^L_{1SM}={2\over3}+{(1-2s^2_W)(1-{8\over3}s^2_W)\over8s^2_Wc^2_W}\chi
\ \ \ \ \ \ \ d^R_{1SM}={2\over3}-{1-{8\over3}s^2_W\over4c^2_W}\chi
\eq
\bq
d^L_{2SM}=-~{1-2s^2_W\over8s^2_Wc^2_W}\chi \ \ \ \ \ \ \ 
d^R_{2SM}={\chi\over4c^2_W}
\eq
we get\\

\noindent
\underline{a) Integrated observables}\\

\vspace*{0.3cm}
\noindent
$\sigma_t$
\bqa
K^L_1={2d^L_{1SM}\over a_1}(1+{2m^2_t\over s})  & , &
K^R_1={2d^R_{1SM}\over a_1}(1+{2m^2_t\over s})\ , \nonumber \\
K^L_2=2\beta^2_t~{d^L_{2SM}\over a_1} & , &
K^R_2=2\beta^2_t~{d^R_{2SM}\over a_1}\ , \nonumber \\
K^L_3=-2 \beta^2_t~{d^L_{1SM}\over a_1} & , &
K^R_3=-2 \beta^2_t~{d^R_{1SM}\over a_1}\ ,
\eqa\\

\noindent
$H_t$
\bqa
K^L_1= {d^L_{2SM}\over B^{L+R}_{1SM}} & , &
K^R_1= {d^R_{2SM}\over B^{L+R}_{1SM}}\ , \nonumber \\
K^L_2= {d^L_{1SM}\over B^{L+R}_{1SM}} & , &
K^R_2= {d^R_{1SM}\over B^{L+R}_{1SM}}\ ,
\eqa\\

\noindent
$T_t$
\bqa
K^L_1={2d^L_{1SM}\over C^{L+R}_{1SM}} & , &
K^R_1=-~{2d^R_{1SM}\over C^{L+R}_{1SM}}\ , \nonumber \\
K^L_3=-\left ({2|\pvec|^2\over m^2_t}\right )
{d^L_{1SM}\over C^{L+R}_{1SM} } & , &
K^R_3=\left ({2|\pvec|^2\over m^2_t}\right )
{d^R_{1SM}\over C^{L+R}_{1SM} }\ ,
\eqa\\

\noindent
\underline{b) Unpolarized asymmetries}\\

\vspace*{0.3cm}
\noindent
$A_{FB}$
\bqa
K^L_1={d^L_{2SM}\over A^{L+R}_{3SM}}-
{2d^L_{1SM}\over a_1}(1+{2m^2_t\over s}) & , & 
K^R_1=-{d^R_{2SM}\over A^{L+R}_{3SM}}-
{2d^R_{1SM}\over a_1}(1+{2m^2_t\over s}) \ , \nonumber \\
K^L_2={d^L_{1SM}\over A^{L+R}_{3SM}}-2\beta^2_t~{d^L_{2SM}\over
a_1} & , &
K^R_2=-~{d^R_{1SM}\over A^{L+R}_{3SM}}-2\beta^2_t~{d^R_{2SM}\over a_1}
\ , \nonumber \\
K^L_3=2 \beta^2_t~ {d^L_{1SM}\over a_1} & , &
K^R_3=2 \beta^2_t~ {d^R_{1SM}\over a_1}\ ,
\eqa\\

\noindent
$A_{EC}$
\bqa
K^L_1={2d^L_{1SM}\over a_2 }(1-{4m^2_t\over s})-
{2d^L_{1SM}\over a_1}(1+{2m^2_t\over s}) & , &
K^L_2=2\beta^2_t d^L_{2SM}({1\over a_2}-{1\over a_1})
\ , \nonumber \\
K^R_1={2d^R_{1SM}\over a_2 }(1-{4m^2_t\over s})-
{2d^R_{1SM}\over a_1}(1+{2m^2_t\over s})
 & , & 
K^R_2=2\beta^2_t d^R_{2SM}({1\over a_2}-{1\over a_1})\ ,
\nonumber \\
K^L_3=4\beta^2_t d^L_{1SM}({1\over a_2}+{1\over2 a_1}) & , & 
K^R_3=4\beta^2_t d^R_{1SM}({1\over a_2}+{1\over2 a_1})\ , 
\eqa\\

\noindent
$H_{FB}$
\bqa
K^L_1={2d^L_{1SM}\over B^{L+R}_{2SM} }-
{d^L_{2SM}\over B^{L+R}_{1SM}}  & , &
K^R_1=-{2d^R_{1SM}\over B^{L+R}_{2SM} }-{d^R_{2SM}\over B^{L+R}_{1SM}}
\ , \nonumber \\
K^L_2=2\beta^2_t{ d^L_{2SM}\over B^{L+R}_{2SM}}-
{d^L_{1SM}\over B^{L+R}_{1SM}} & , &
K^R_2=-2\beta^2_t{ d^R_{2SM}\over B^{L+R}_{2SM}}-
{d^R_{1SM}\over B^{L+R}_{1SM}} \ , 
\eqa\\

\noindent
$T_{FB}$
\bqa
K^L_1={d^L_{2SM}\over C^{L+R}_{2SM} }-
{2d^L_{1SM}\over C^{L+R}_{1SM}} & , & 
K^R_1={d^R_{2SM}\over C^{L+R}_{2SM} }+{2d^R_{1SM}\over C^{L+R}_{1SM}}
\ , \nonumber \\
K^L_2={d^L_{1SM}\over C^{L+R}_{2SM} } & , & 
K^R_2={d^R_{1SM}\over C^{L+R}_{2SM} }
\ , \nonumber \\
K^L_3=-~{2|\pvec|^2\over m^2_t}\left [{d^L_{2SM}\over C^{L+R}_{2SM} }-
{d^L_{1SM}\over C^{L+R}_1}\right ] & , & 
K^R_3=-~{2|\pvec|^2\over m^2_t}\left [{d^R_{2SM}\over C^{L+R}_{2SM} }+
{d^R_{1SM}\over C^{L+R}_1}\right ]\ ,
\eqa\\

\noindent
\underline{c) Polarized asymmetries}\\

\vspace*{0.3cm}
\noindent
$A_{LR}$
\bqa
K^L_1=2 d^L_{1SM}(1+{2m^2_t\over s})({1\over b_1}-{1\over a_1})
& , &
K^R_1=-2 d^R_{1SM}(1+{2m^2_t\over s})({1\over b_1}+{1\over a_1})
\ , \nonumber \\
K^L_2=2\beta^2_t d^L_{2SM}({1\over b_1}-{1\over a_1}) 
& , &
K^R_2=-2\beta^2_t d^R_{2SM}({1\over b_1}+{1\over a_1})
\ , \nonumber \\
K^L_3=-2\beta^2_t d^L_{1SM}({1\over b_1}-{1\over a_1}) 
& , &
K^R_2=2\beta^2_t d^R_{1SM}({1\over b_1}+{1\over a_1})
\ ,
\eqa\\

\noindent
$A^{(pol)}_{FB}$
\bqa
K^L_1={ d^L_{2SM}\over A^{L-R}_{3SM}}-
{2d^L_{1SM}\over b_1}(1+{2m^2_t\over s}) & , &  
K^R_1={ d^R_{2SM}\over A^{L-R}_{3SM}}+
{2d^R_{1SM}\over b_1}(1+{2m^2_t\over s}) 
\ , \nonumber \\
K^L_2= {d^L_{1SM}\over A^{L-R}_{3SM}}-
2\beta^2_t { d^L_{2SM}\over b_1} & , & 
K^R_2= {d^R_{1SM}\over A^{L-R}_{3SM}}+
2\beta^2_t { d^R_{2SM}\over b_1}
\ ,  \nonumber \\
K^L_3=2\beta^2_t { d^L_{1SM}\over b_1} & , &  
K^R_3=-2\beta^2_t { d^R_{1SM}\over b_1}\ ,
\eqa\\

\noindent
$A^{(pol)}_{EC}$
\bqa
K^L_1={2d^L_{1SM}\over b_2 }(1-{4m^2_t\over s})-
{2d^L_{1SM}\over b_1}(1+{2m^2_t\over s}) & , &  
K^L_2=2\beta^2_t d^L_{2SM}({1\over b_2}-{1\over b_1})
\ , \nonumber \\
K^R_1=-~{2d^R_{1SM}\over b_2 }(1-{4m^2_t\over s})+
~{2d^R_{1SM}\over b_1}(1+{2m^2_t\over s})
& , &
K^R_2=-2\beta^2_t d^R_{2SM}({1\over b_2}-{1\over b_1})
\ , \nonumber \\
K^L_3=4\beta^2_t d^L_{1SM}({1\over b_2}+{1\over2 b_1}) 
& , & 
K^R_3=-4\beta^2_t d^R_{1SM}({1\over b_2}+{1\over2 b_1})
,  
\eqa\\

\noindent
$H_{LR}$
\bqa
K^L_1= d^L_{2SM}({1\over A^{L+R}_{3SM}}-{1\over A^{L-R}_{3SM}})
& , &  
K^R_1=- d^R_{2SM}({1\over A^{L+R}_{3SM}}+{1\over A^{L-R}_{3SM}})
\ , \nonumber \\
K^L_2= d^L_{1SM}({1\over A^{L+R}_{3SM}}-{1\over A^{L-R}_{3SM}})
& , & 
K^R_1=- d^R_{1SM}({1\over A^{L+R}_{3SM}}+{1\over A^{L-R}_{3SM}})
\ , 
\eqa \\

\noindent
$H^{(pol)}_{FB}$
\bqa
K^L_1={2 d^L_{1SM}\over A^{L+R}_{2SM}}-
{ d^L_{2SM}\over A^{L+R}_{3SM}} & , &
K^R_1={2 d^R_{1SM}\over A^{L+R}_{2SM}}+
{ d^R_{2SM}\over A^{L+R}_{3SM}}
\ , \nonumber \\
K^L_2={2\beta^2_t d^L_{2SM}\over A^{L+R}_{2SM}}-
{ d^L_{1SM}\over A^{L+R}_{3SM}} & , &  
K^R_2={2\beta^2_t d^R_{2SM}\over A^{L+R}_{2SM}}+
{ d^R_{1SM}\over A^{L+R}_{3SM}}\ ,
\eqa\\

\noindent
$T_{LR}$
\bqa
K^L_1=2 d^L_{1SM}({1\over C^{L-R}_{1SM}}-{1\over C^{L+R}_{1SM}})
& , &  
K^L_3=-\left ({2|\pvec|^2\over m^2_t}\right ) 
d^L_{1SM}({1\over C^{L-R}_{1SM}}-{1 \over C^{L+R}_{1SM}})
 , \nonumber \\
K^R_1=2 d^R_{1SM}({1\over C^{L-R}_{1SM}}+{1\over C^{L+R}_{1SM}})
 & , &  
K^R_3=-\left ({2|\pvec|^2\over m^2_t}\right ) 
d^R_{1SM}({1\over C^{L-R}_{1SM}}+{1 \over C^{L+R}_{1SM}})
 ,
\eqa\\

\noindent
$T^{(pol)}_{FB}$
\bqa
K^L_1={ d^L_{2SM}\over C^{L-R}_{2SM}}-
{2 d^L_{1SM}\over C^{L-R}_{1SM}} & , &  
K^R_1=-{ d^R_{2SM}\over C^{L-R}_{2SM}} -
{2 d^R_{1SM}\over C^{L-R}_{1SM}}
\ , \nonumber \\
K^L_2={ d^L_{1SM}\over C^{L-R}_{2SM}} & , &  
K^R_2=-{ d^R_{1SM}\over C^{L-R}_{2SM}}
\ , \nonumber \\
K^L_3=-~{2|\pvec|^2\over m^2_t}\left ({ d^L_{2SM}\over C^{L-R}_{2SM}}-
{ d^L_{1SM}\over C^{L-R}_{1SM}} \right) & , &
K^R_3={2|\pvec|^2\over m^2_t} \left ({ d^R_{2SM}\over C^{L-R}_{2SM}}+
{ d^R_{1SM}\over C^{L-R}_{1SM}}\right ).
\eqa\\
\noindent
The coefficients for the $\bar{d}^{\gamma}_j$ 
and $\bar{d}^Z_j$ couplings
are easily obtained from relations (\ref{eq:dLR}). 
\bq
K^{\gamma}_j = K^L_j + K^R_j \ \ \ \ 
K^{Z}_j ={\frac{1-2s^2_W}{4s^2_Wc^2_W}}\chi K^L_j 
-{\frac{\chi}{2c^2_W}} K^R_j \ \ \ \ \ \ \ \ 
\eq
\noindent
This allows
to write:

\bq
\delta\A^i = \sum_{j=1,3} [K^{\gamma}_j \bar{d}^{\gamma}_j
+ K^{Z}_j \bar{d}^{Z}_j]
\ \ .
\eq 

In Table A1 we give for illustration the values of these 
$K^{\gamma,Z}_j$ coefficients
at 1 TeV. With this table one can appreciate the
role of each observable, non polarized asymmetry, polarized
asymmetry and integrated $\rho$ elements (taken with a 2\%
normalization uncertainty), in establishing the limits 
for each coupling.

\vspace{1cm}
\begin{center}
{\bf Table A1: Numerical values of $K^{\gamma,Z}_j$ at 1 TeV.}

\vspace{0.3cm}
\begin{tabular}{|c|c|c|c|c|c|c|}  \hline
\multicolumn{1}{|c|}{$\A^i$}&
\multicolumn{1}{|c|}{$\bar d^{\gamma}_1$} &
\multicolumn{1}{|c|}{$\bar d^{\gamma}_2$ }&
\multicolumn{1}{|c|}{$\bar d^{\gamma}_3$} &
\multicolumn{1}{|c|}{$\bar d^{Z}_1$ }&
\multicolumn{1}{|c|}{$\bar d^{Z}_2$} &
\multicolumn{1}{|c|}{$\bar d^{Z}_3$ }
 \\[0.1cm] \hline
$A_{FB}$& -28.98& -15.86&  63.50& -9.74& -42.37& 12.17\\ \hline
$A_{EC}$& -0.68& -0.10&  52.07& -0.13& -0.95&  9.98\\ \hline
$H_{FB}$& -5.14&-39.36& 0.00& -5.62& -9.01&  0.00\\ \hline
$T_{FB}$& -0.50& -4.81&  2.23& -0.68& -0.92& -7.76\\ \hline\hline
$A^{(A)}_{LR}$& -17.51& -51.63& 14.47&  79.91&  7.24& -66.07\\ \hline
$A^{(pol)}_{FB}$& -8.42& -65.45& 9.86& -9.33& -14.80&  35.38\\ \hline
$A^{(pol)}_{EC}$& -0.06& -1.51&  10.28& -0.21& -0.16&  36.89\\ \hline
$A^{(B)}_{LR}$& -5.26& -49.29& 0.00& 18.34&  0.00&  0.00\\ \hline
$H^{(pol)}_{FB}$& -15.61&  -8.73&  0.00& -5.31& -22.85&  0.00\\ \hline
$A^{(C)}_{LR}$&  4.38& 0.00& -31.39& -15.27&  0.00& 109.39\\ \hline
$T^{(pol)}_{FB}$& -2.44& -0.98& -0.88& -0.69& -3.52& 3.05\\ \hline\hline
$\sigma_t$&  82.36& -2.62& -68.10& 15.78& -24.98& -13.05\\ \hline
$H_t$& 2.13& -55.49&  0.00& 20.35& -10.63&  0.00\\ \hline
$T_t$&  5.06&  0.00& -36.26& 18.16& 0.00& -130.11
 \\[0.1cm] \hline
\end{tabular}
\end{center}

\newpage

\begin{center}

{\large \bf Figure captions}
\end{center}
\vspace{0.5cm}

{\bf Fig.1} Observability limits in the 6-parameter case;
with polarized beams (a) (c) (e) (g) (i) (k), with unpolarized beams
(b) (d) (f) (h) (j) (l); from asymmetries alone (- - - -), from asymmetries
and $\sigma_t$, $H_t$, $T_t$ with a normalization uncertainty of
2\% (........), 20\% (----------).\\

{\bf Fig.2} Observability limits in the 4-parameter case; same captions
as in Fig.1.\\

{\bf Fig.3} The observability ellipsoid in the 3-parameter case with
polarized beams.\\

{\bf Fig.4} Observability limits in the 3-parameter case; 
with polarized beams (a)(c)(e), with unpolarized beams
(b)(d)(f); same captions.\\

{\bf Fig.5} Observability limits in the 2-parameter case; 
with polarized beams (a), with unpolarized beams
(b); same captions.\\

\newpage
\def\x{$\bar d^\gamma_1$}
\def\y{$\bar d^\gamma_2$}
\def\z{$\bar d^\gamma_3$}
\def\u{$\bar d^Z_1$}
\def\v{$\bar d^Z_2$}
\def\w{$\bar d^Z_3 $}
\vspace*{-2.cm}
\[
\epsfig{file=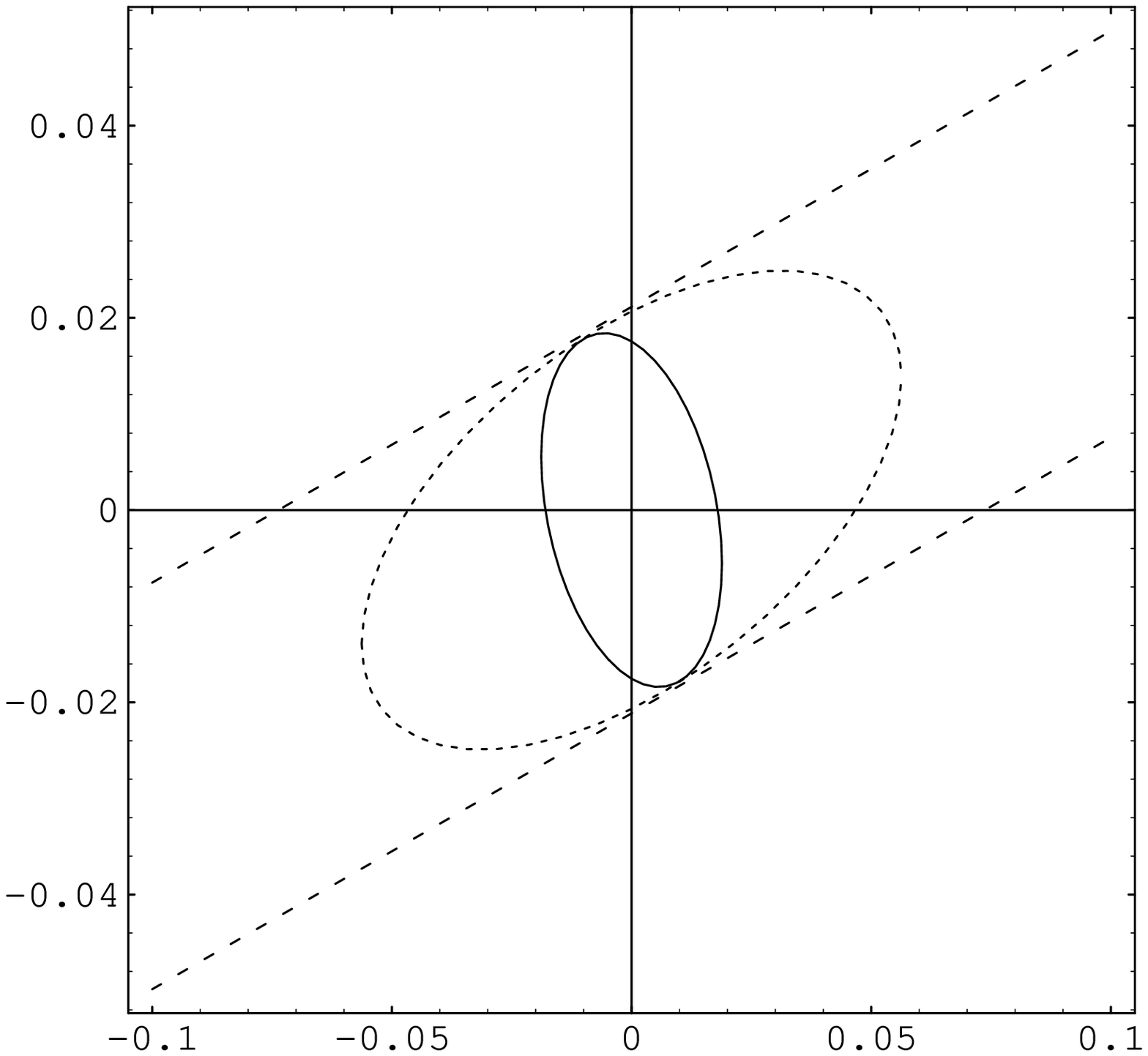,height=9cm}\hspace{2cm} 
\epsfig{file=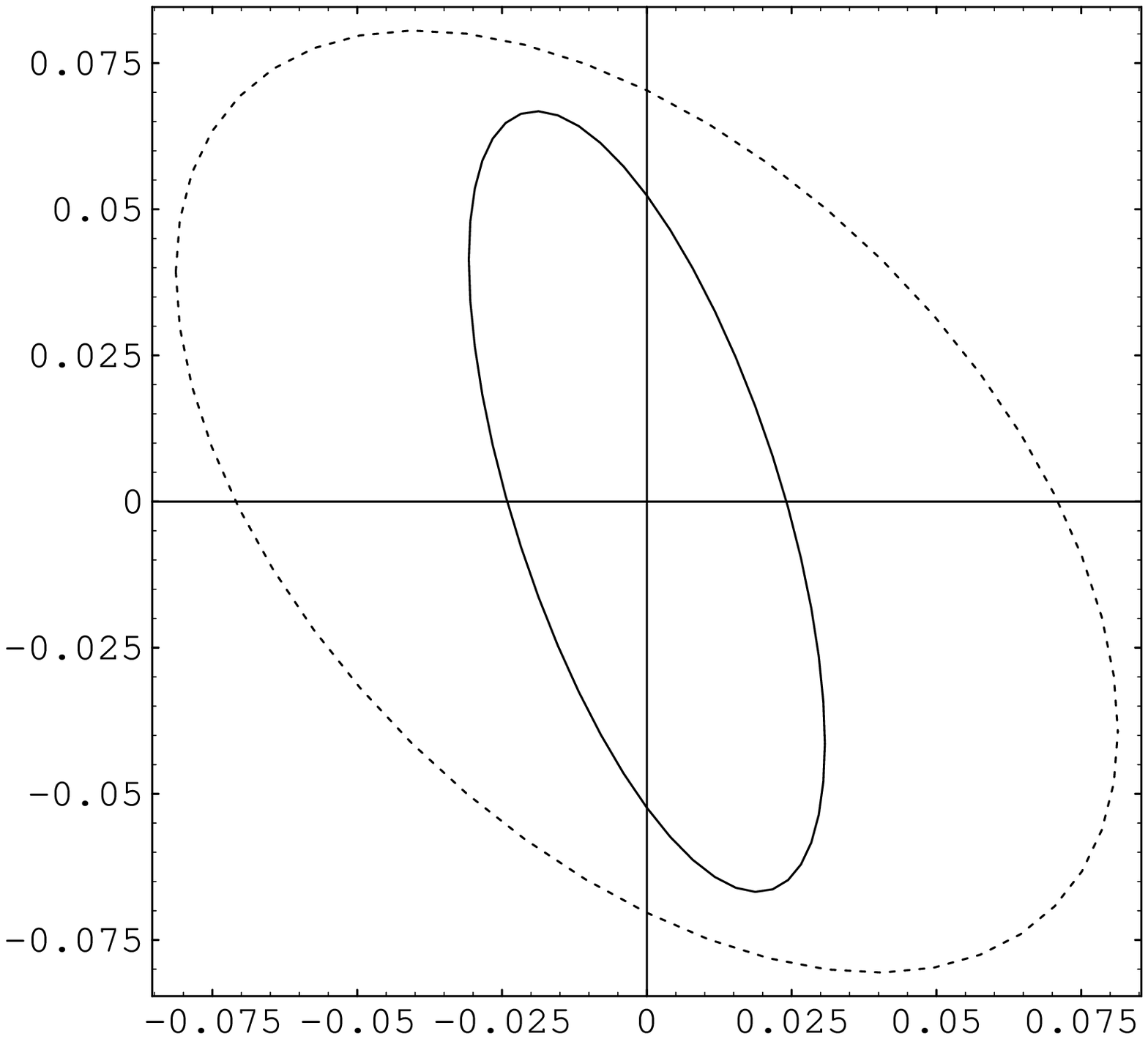,height=9cm}
\]
\vspace{-6.cm}\null\\
\hspace*{-1cm} \u \hspace{8cm} \u \\[3.2cm]
\hspace*{6cm} \x \hspace{8cm}  \x
\\
\hspace*{3.2cm} (a) \hspace{7.8cm}  (b)
%
\[
\epsfig{file=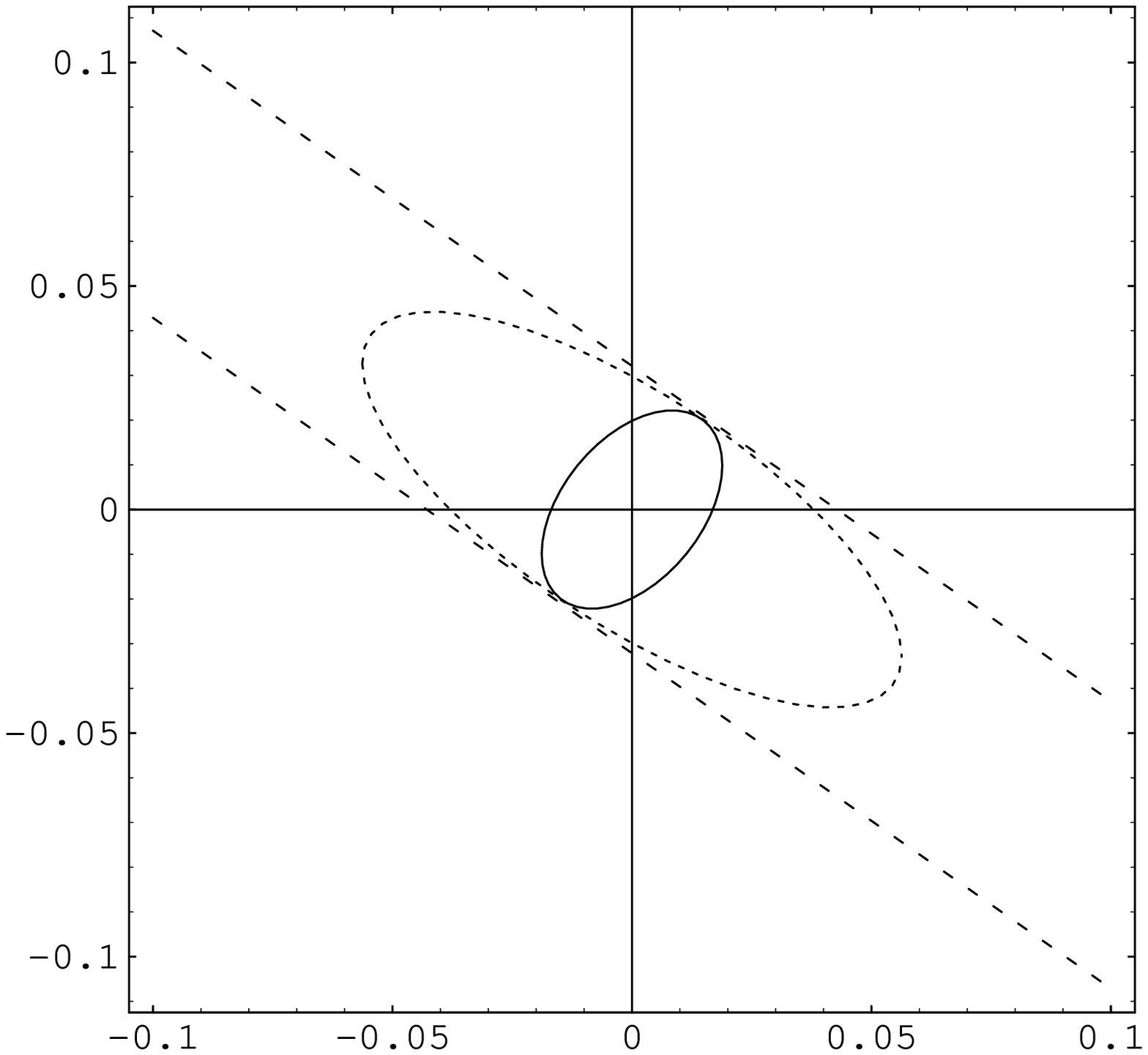,height=9cm}\hspace{2cm} 
\epsfig{file=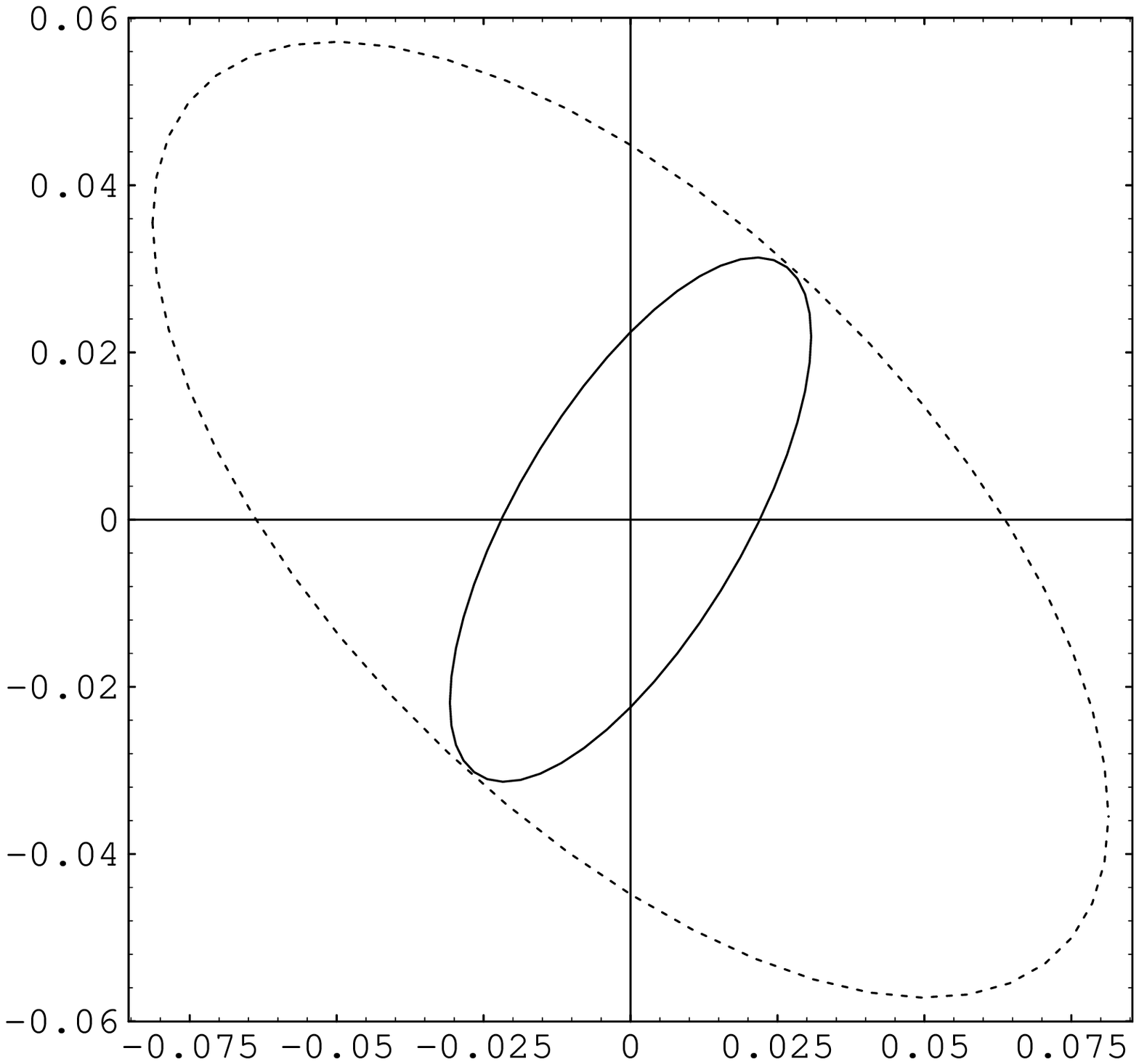,height=9cm}
\]
\vspace{-6.cm}\null\\
\hspace*{-1cm} \v \hspace{8cm} \v \\[3.2cm]
\hspace*{6cm} \x \hspace{8cm}  \x
\\
\hspace*{3.2cm} (c) \hspace{7.8cm}  (d)
\\[1.2cm]
\centerline{Fig 1}
\newpage
\vspace*{-2.cm}
\[
\epsfig{file=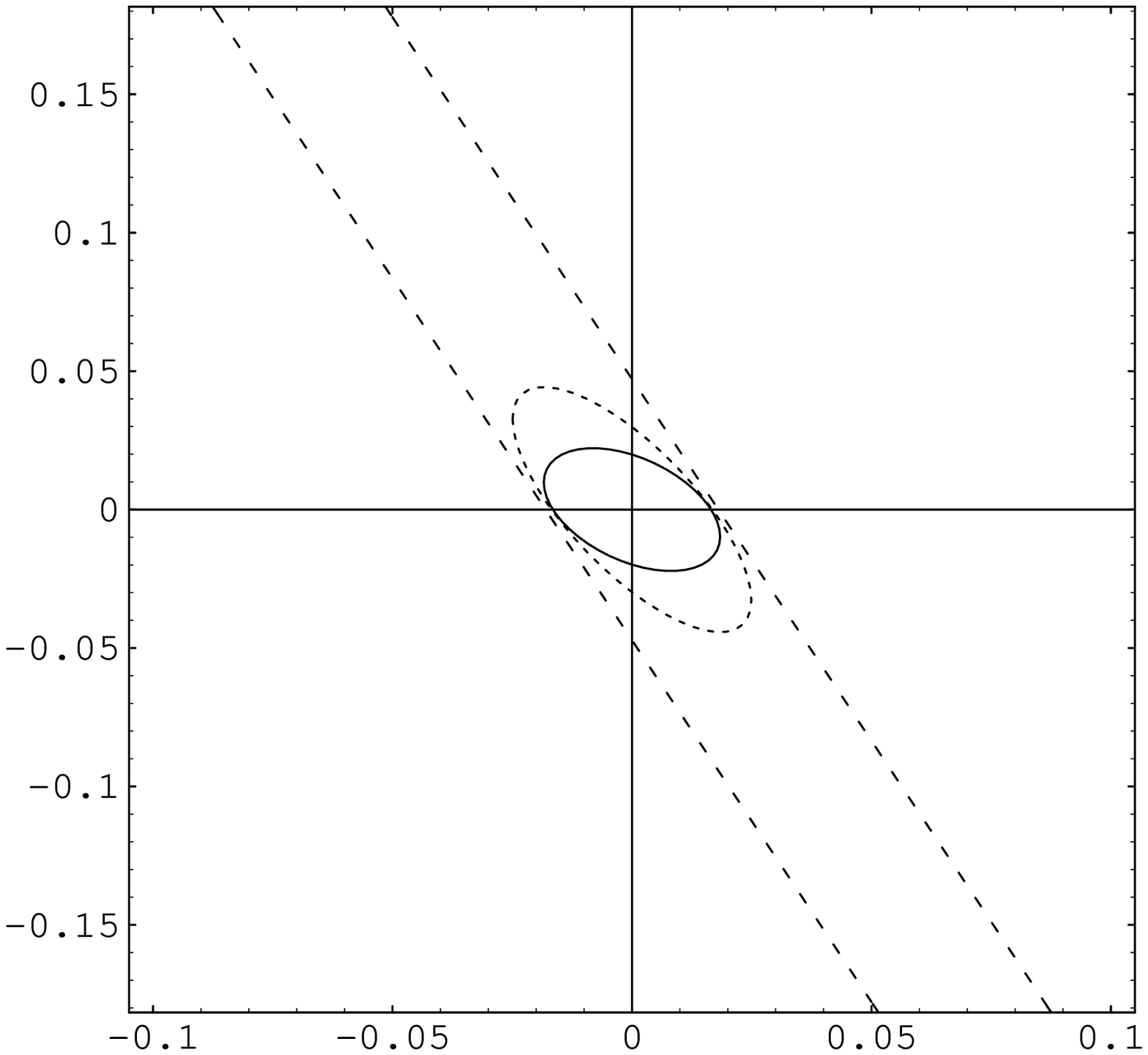,height=9cm}\hspace{2cm} 
\epsfig{file=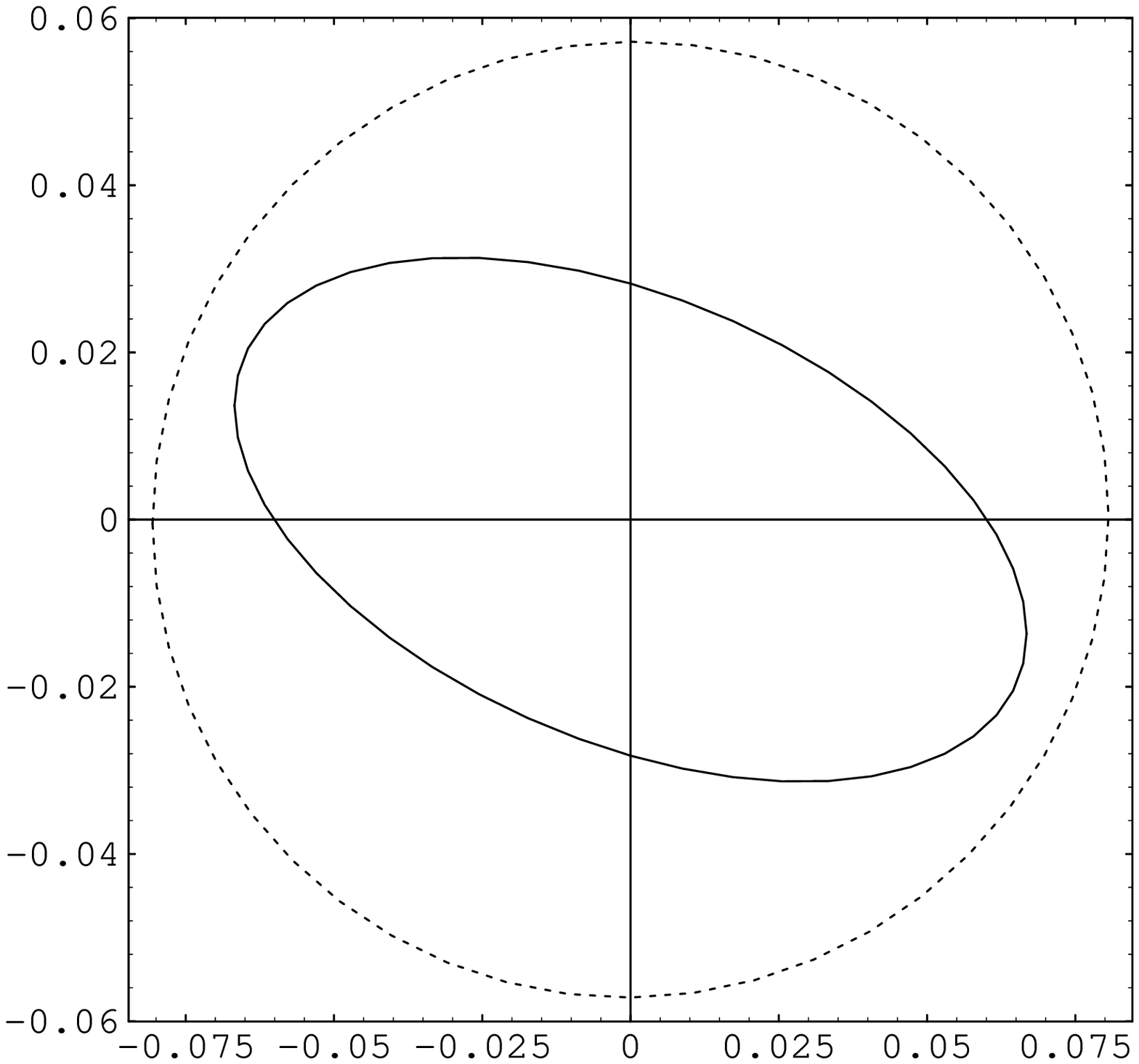,height=9cm}
\]
\vspace{-6.cm}\null\\
\hspace*{-1cm} \v \hspace{8cm} \v \\[3.2cm]
\hspace*{6cm} \u \hspace{8cm}  \u
\\
\hspace*{3.2cm} (e) \hspace{7.8cm}  (f)
%
\[
\epsfig{file=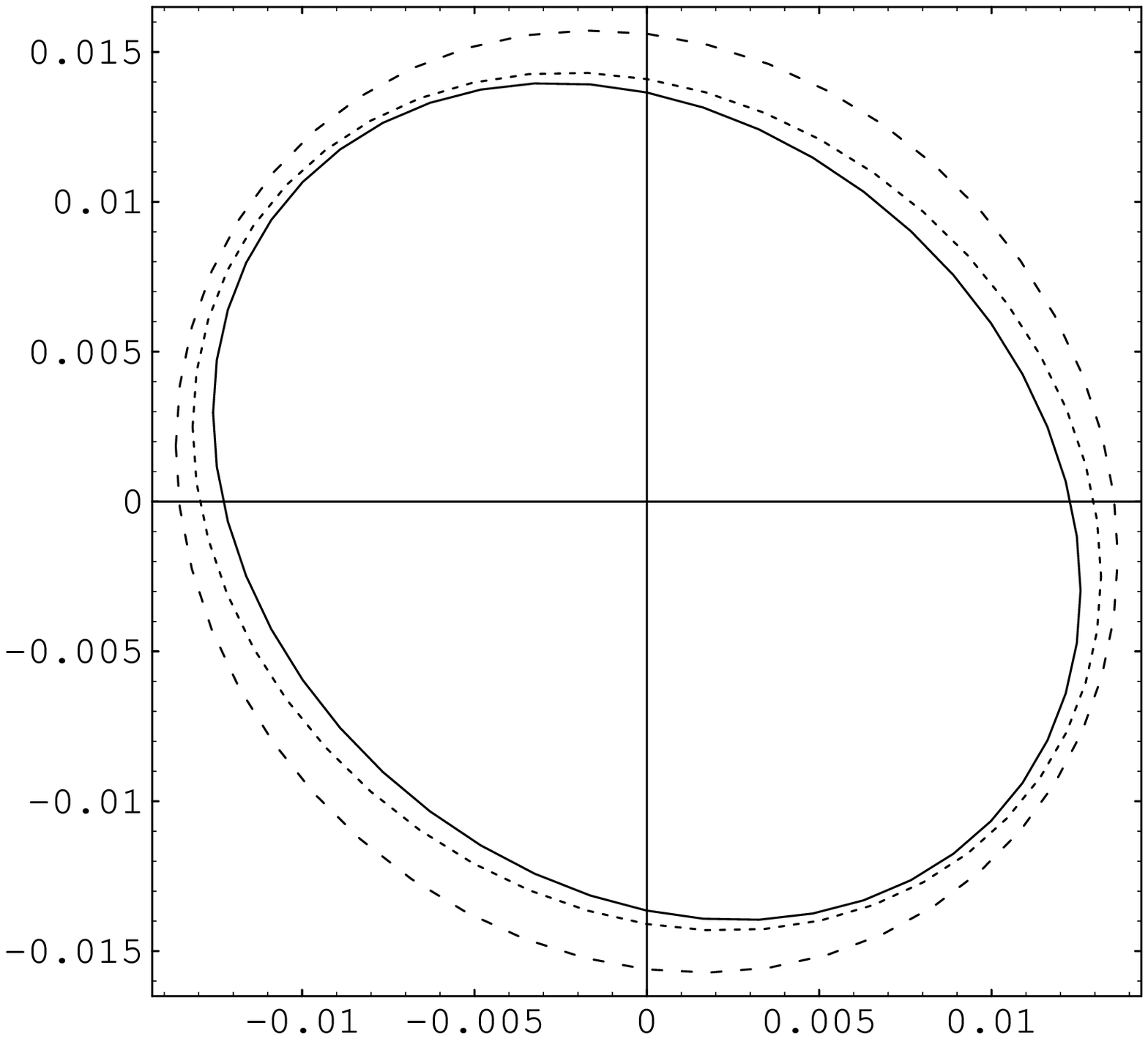,height=9cm}\hspace{2cm} 
\epsfig{file=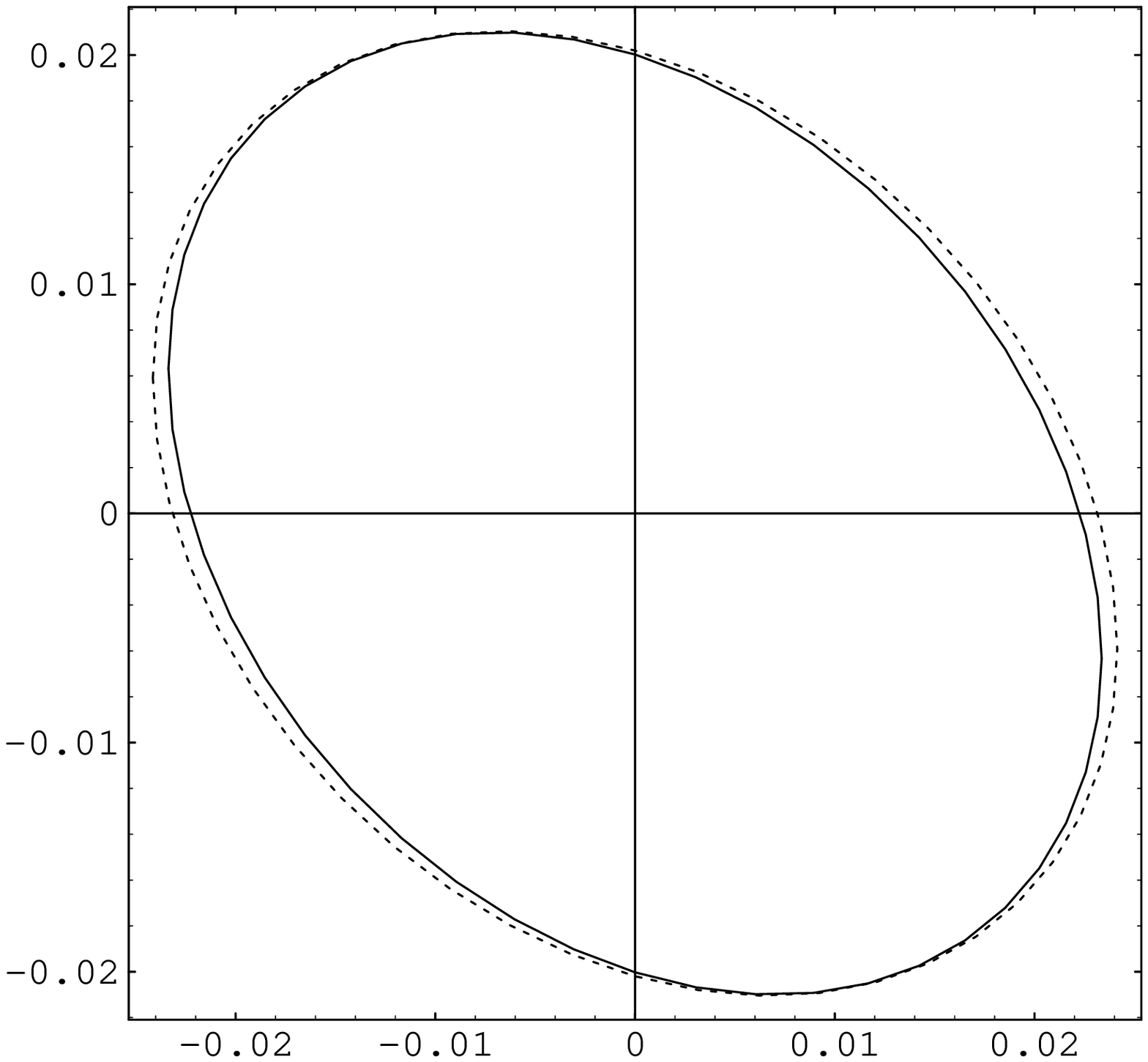,height=9cm}
\]
\vspace{-6.cm}\null\\
\hspace*{-1cm} \z \hspace{8cm} \z \\[3.2cm]
\hspace*{6cm} \y \hspace{8cm}  \y
\\
\hspace*{3.2cm} (g) \hspace{7.8cm}  (h)
\\[1.2cm]
\centerline{Fig 1}
\newpage
\vspace*{-2.cm}
\[
\epsfig{file=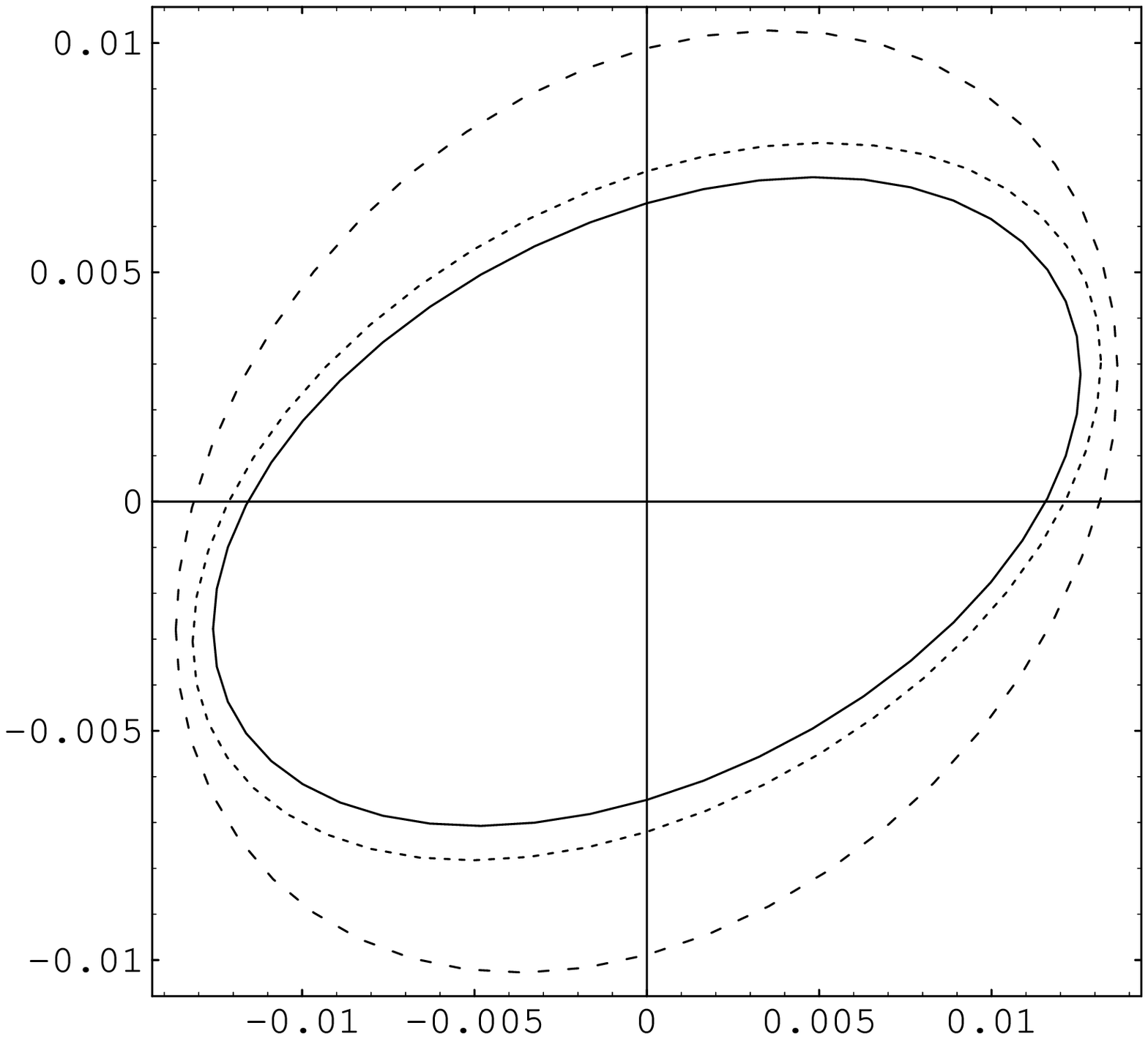,height=9cm}\hspace{2cm} 
\epsfig{file=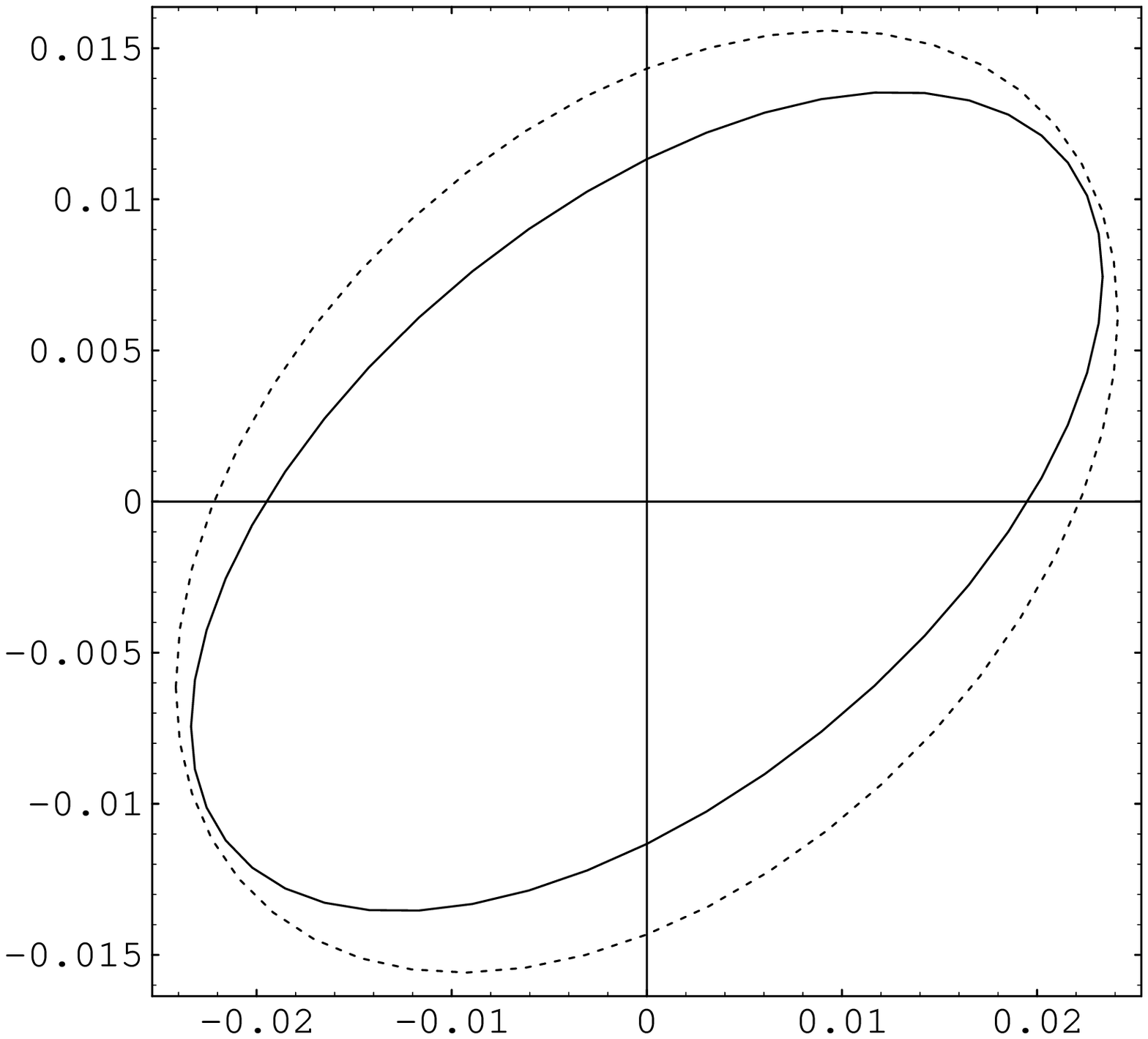,height=9cm}
\]
\vspace{-6.cm}\null\\
\hspace*{-1cm} \w \hspace{8cm} \w \\[3.2cm]
\hspace*{6cm} \y \hspace{8cm}  \y
\\
\hspace*{3.2cm} (i) \hspace{7.8cm}  (j)
%
\[
\epsfig{file=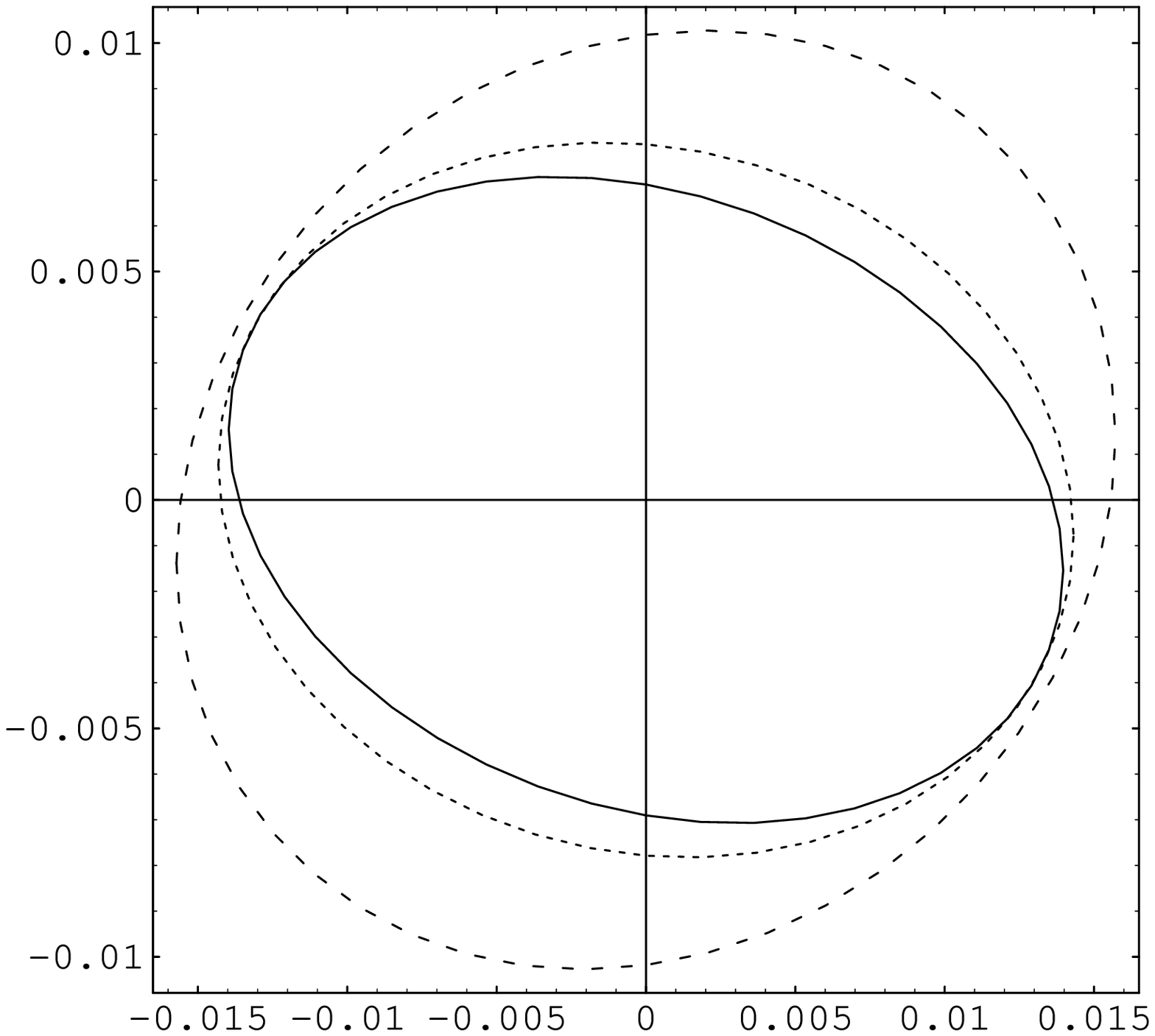,height=9cm}\hspace{2cm} 
\epsfig{file=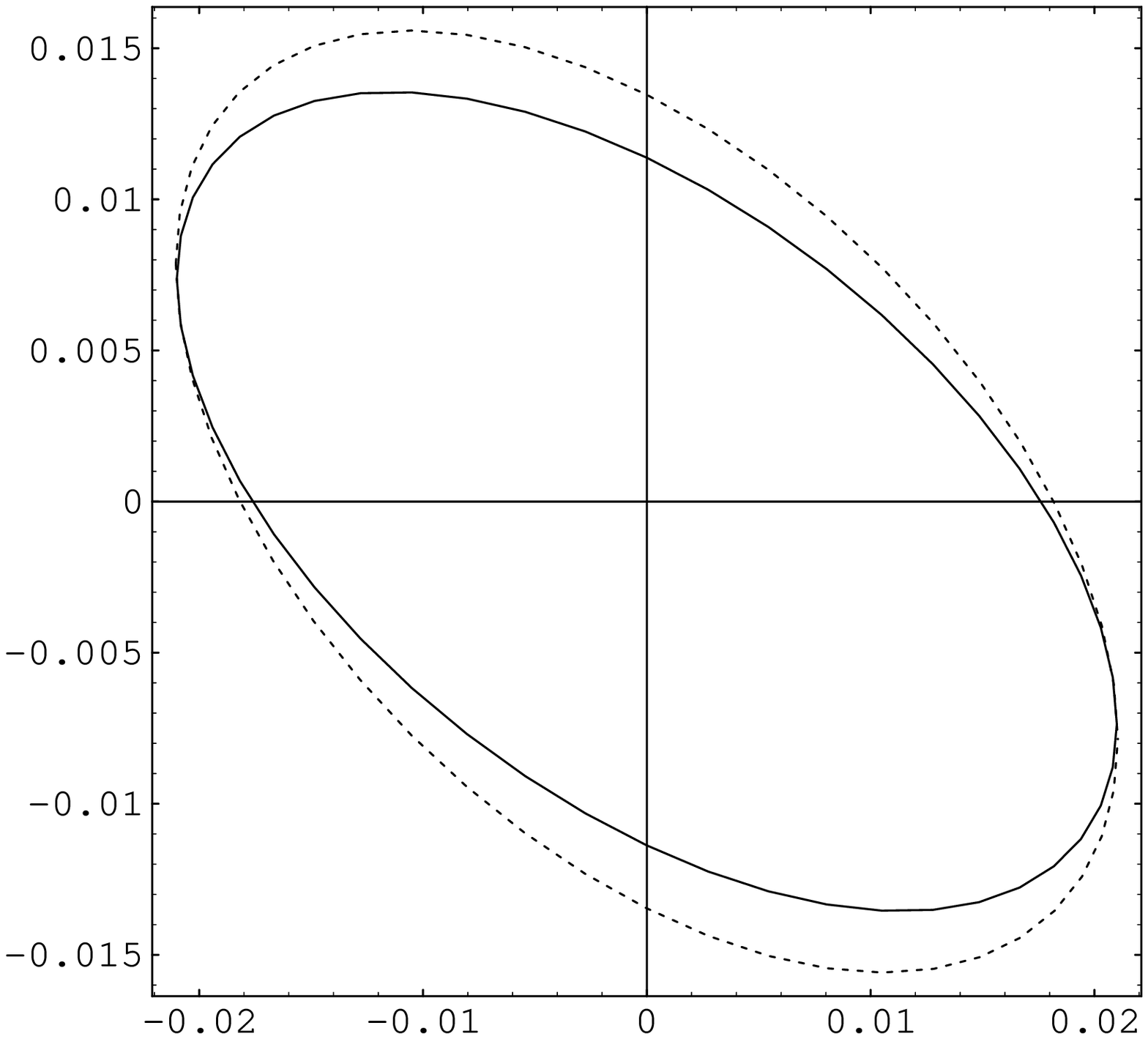,height=9cm}
\]
\vspace{-6.cm}\null\\
\hspace*{-1cm} \w \hspace{8cm} \w \\[3.2cm]
\hspace*{6cm} \z \hspace{8cm}  \z
\\
\hspace*{3.2cm} (k) \hspace{7.8cm}  (l)
\\[1.2cm]
\centerline{Fig 1}
\newpage
\vspace*{-2.cm}
\[
\epsfig{file=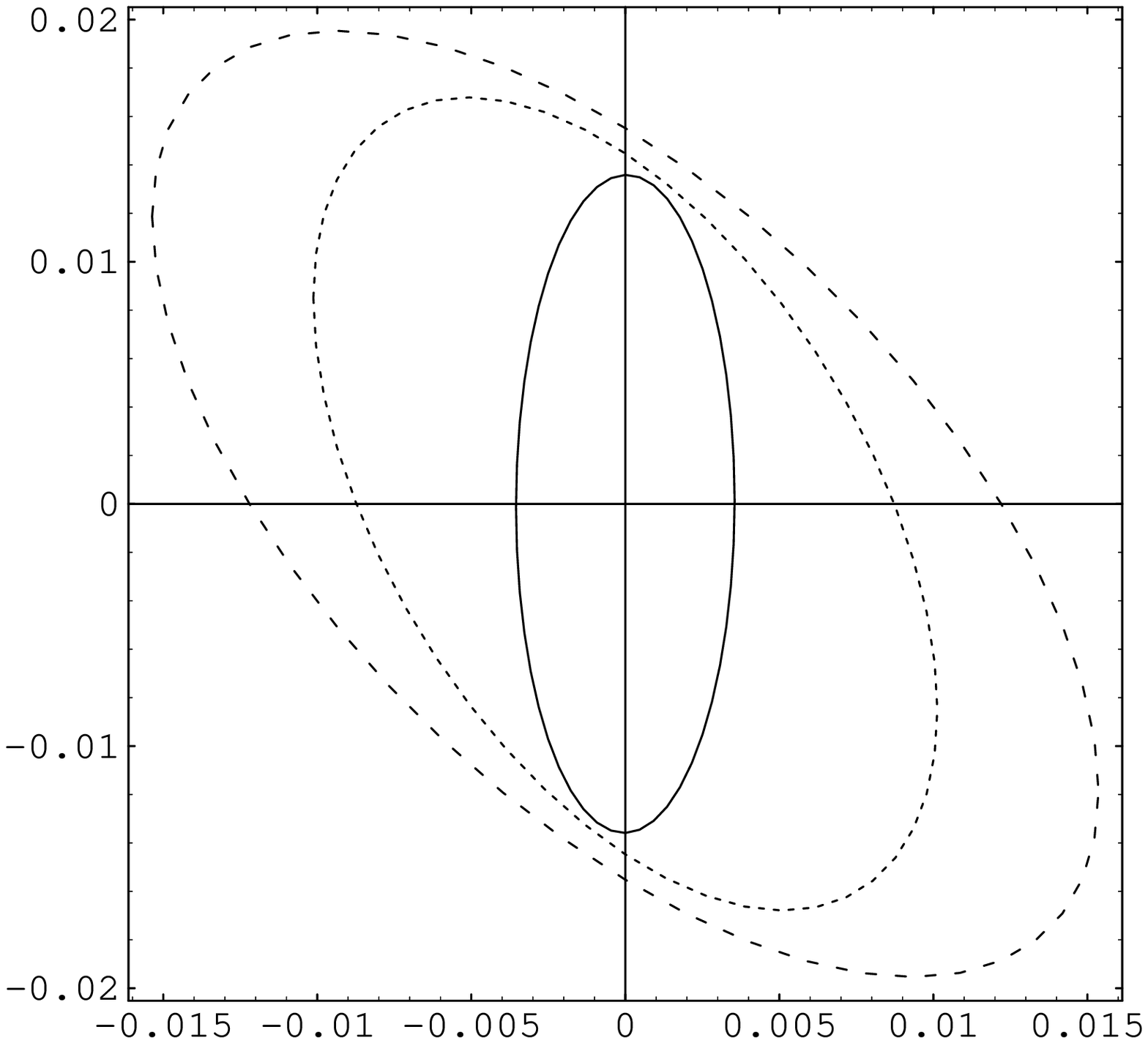,height=9cm}\hspace{2cm} 
\epsfig{file=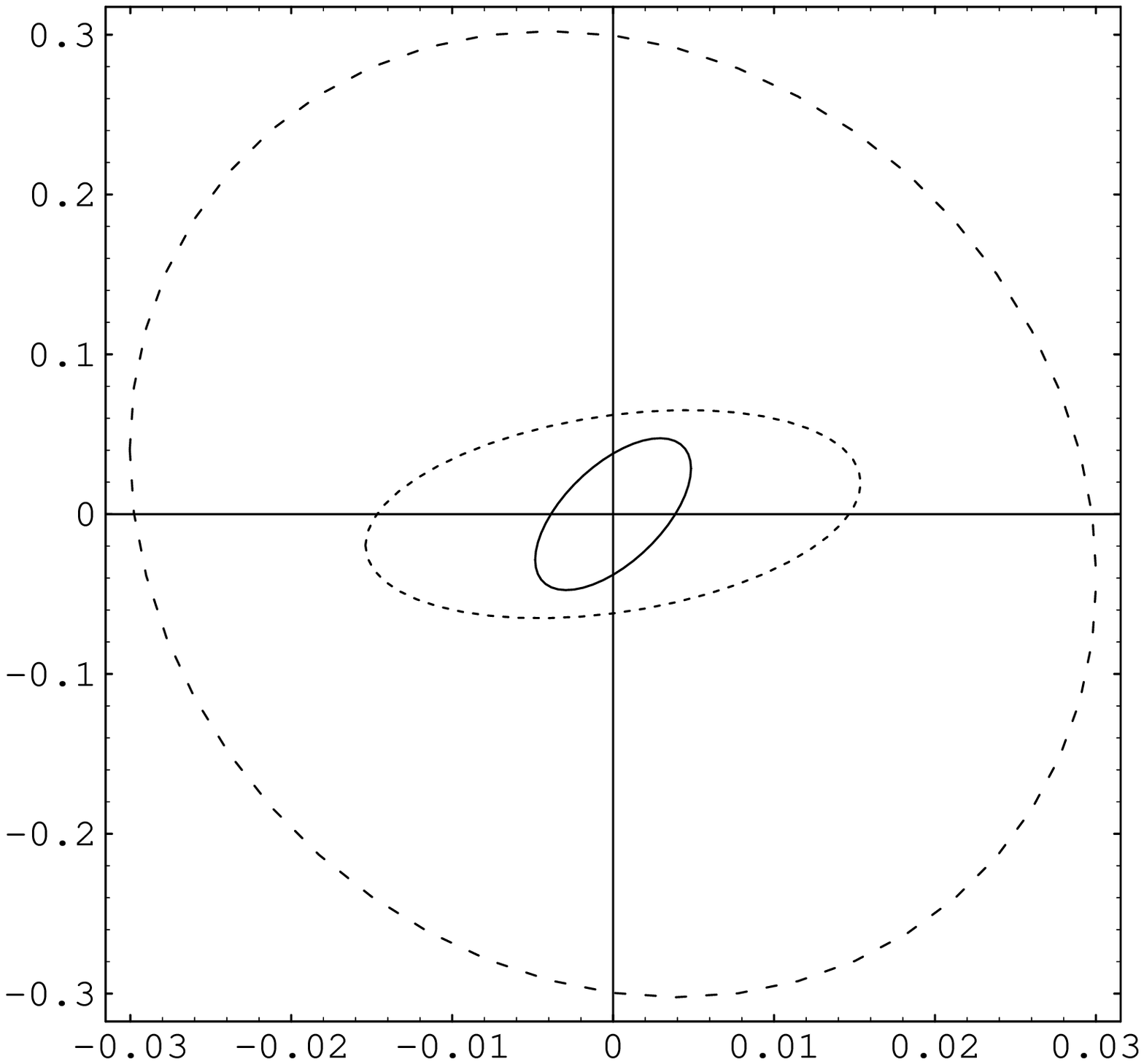,height=9cm}
\]
\vspace{-6.cm}\null\\
\hspace*{-1cm} \u \hspace{8cm} \u \\[3.2cm]
\hspace*{6cm} \z \hspace{8cm}  \z
\\
\hspace*{3.2cm} (a) \hspace{7.8cm}  (b)
%
\[
\epsfig{file=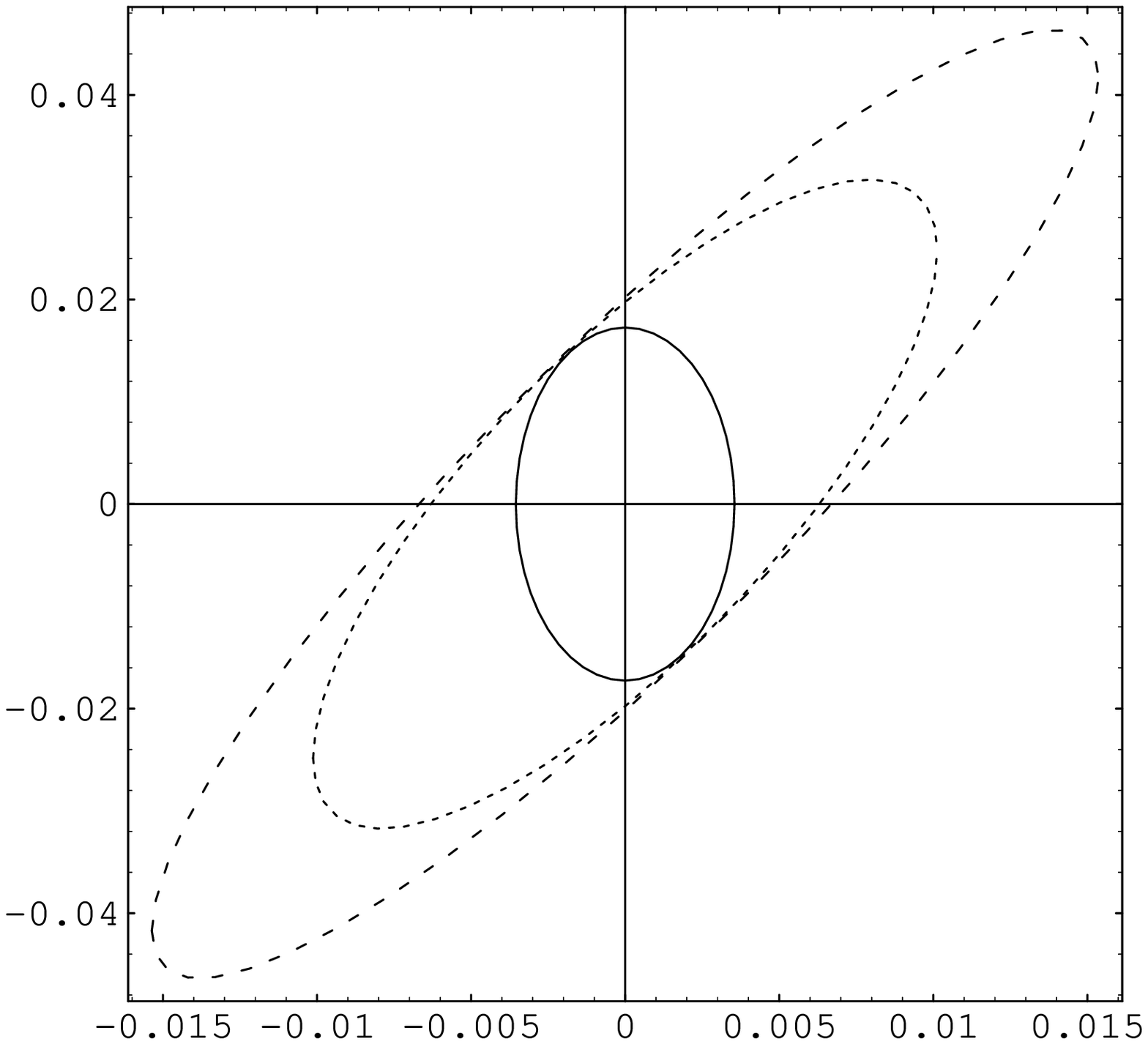,height=9cm}\hspace{2cm} 
\epsfig{file=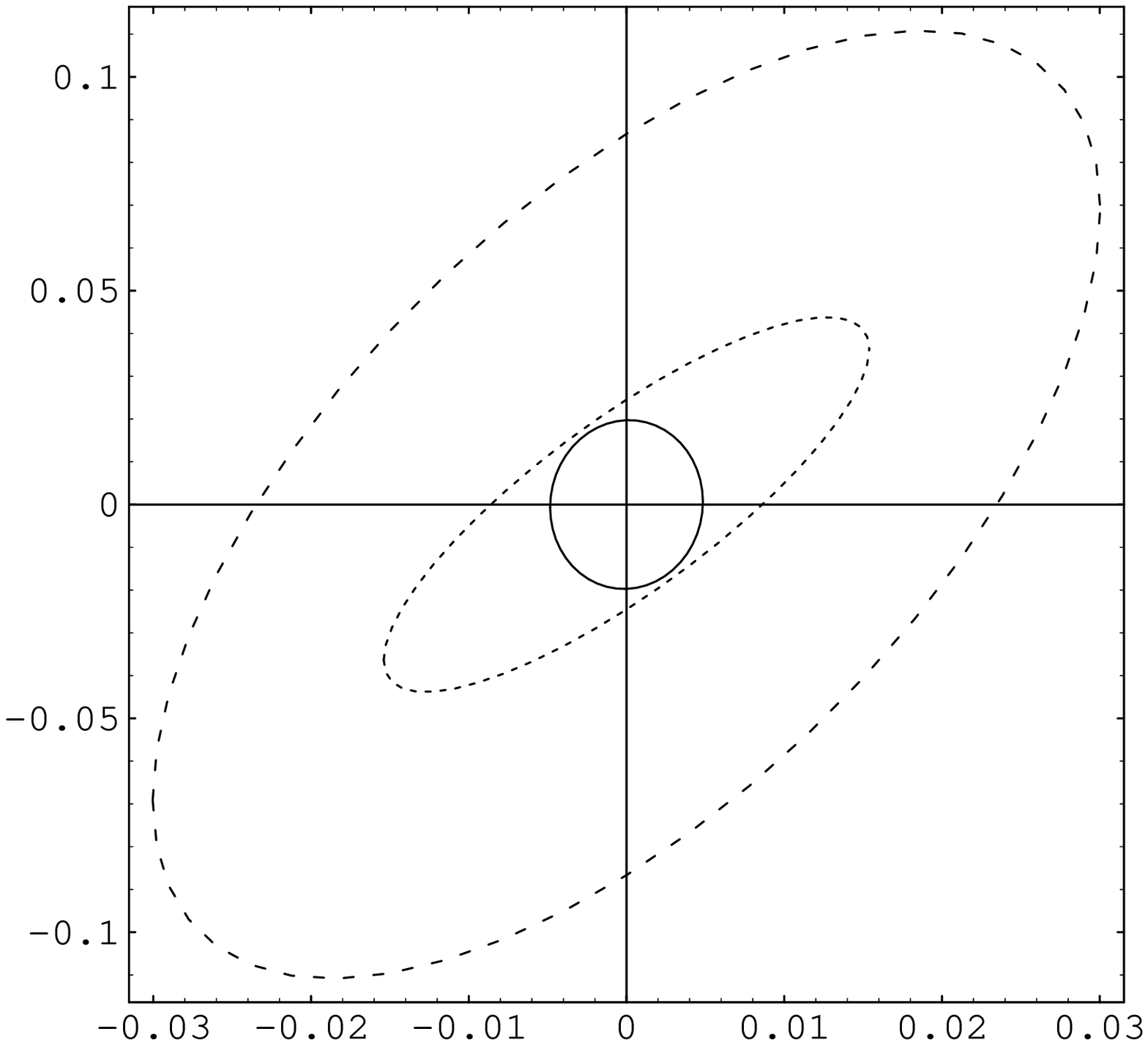,height=9cm}
\]
\vspace{-6.cm}\null\\
\hspace*{-1cm} \v \hspace{8cm} \v \\[3.2cm]
\hspace*{6cm} \z \hspace{8cm}  \z
\\
\hspace*{3.2cm} (c) \hspace{7.8cm}  (d)
\\[1.2cm]
\centerline{Fig 2}
\newpage
\vspace*{-2.cm}
\[
\epsfig{file=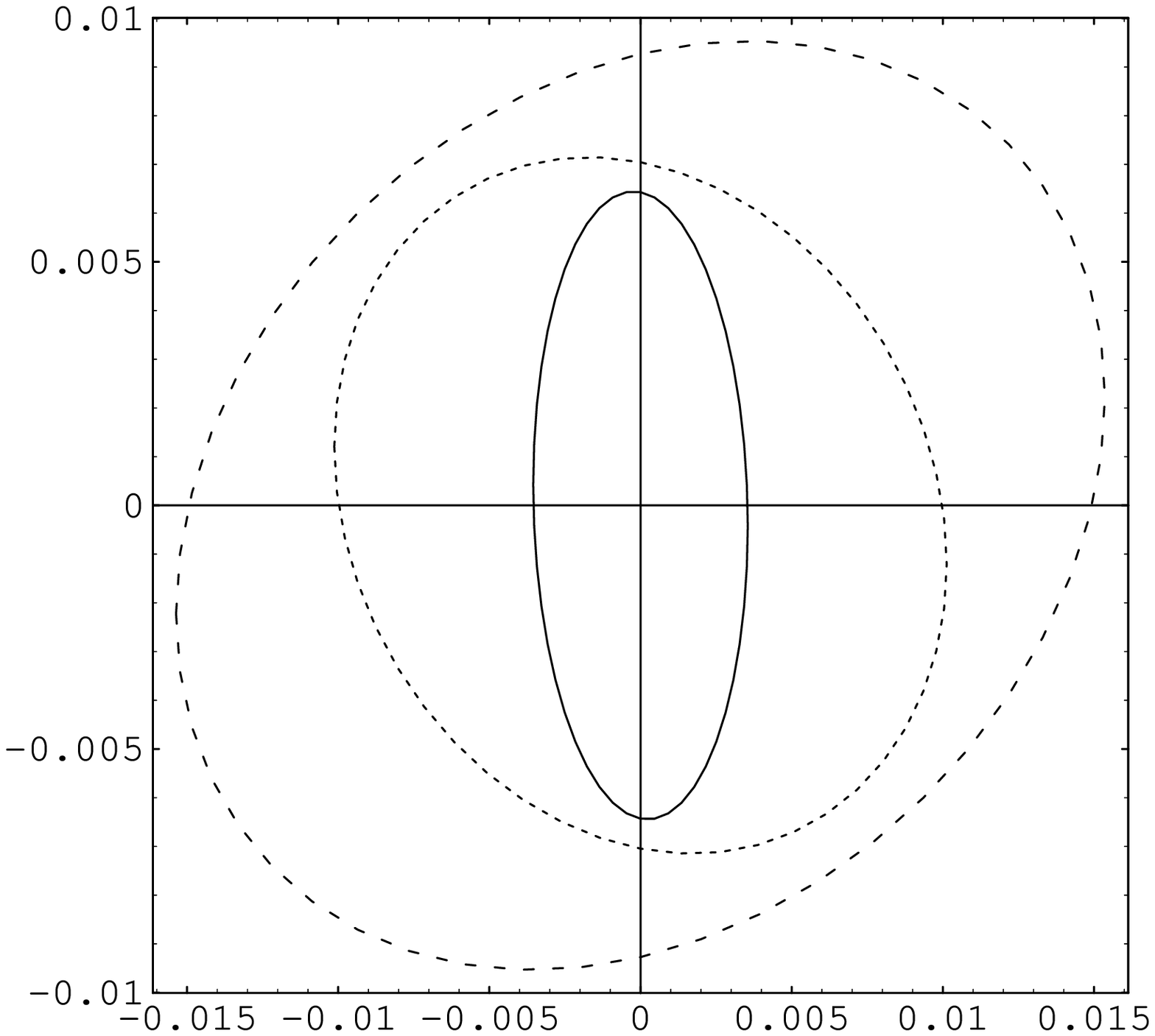,height=9cm}\hspace{2cm} 
\epsfig{file=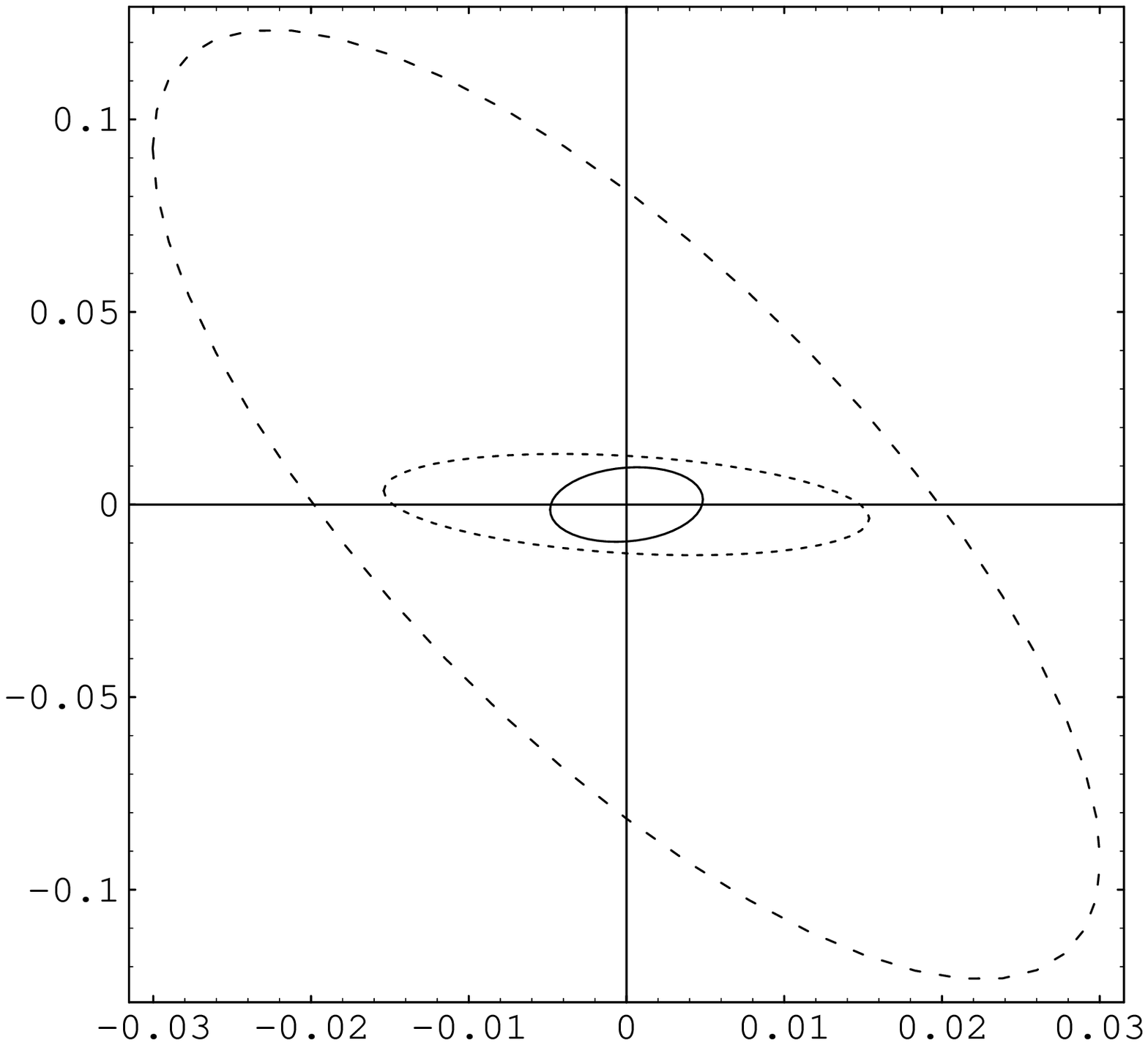,height=9cm}
\]
\vspace{-6.cm}\null\\
\hspace*{-1cm} \w \hspace{8cm} \w \\[3.2cm]
\hspace*{6cm} \z \hspace{8cm}  \z
\\
\hspace*{3.2cm} (e) \hspace{7.8cm}  (f)
%
\[
\epsfig{file=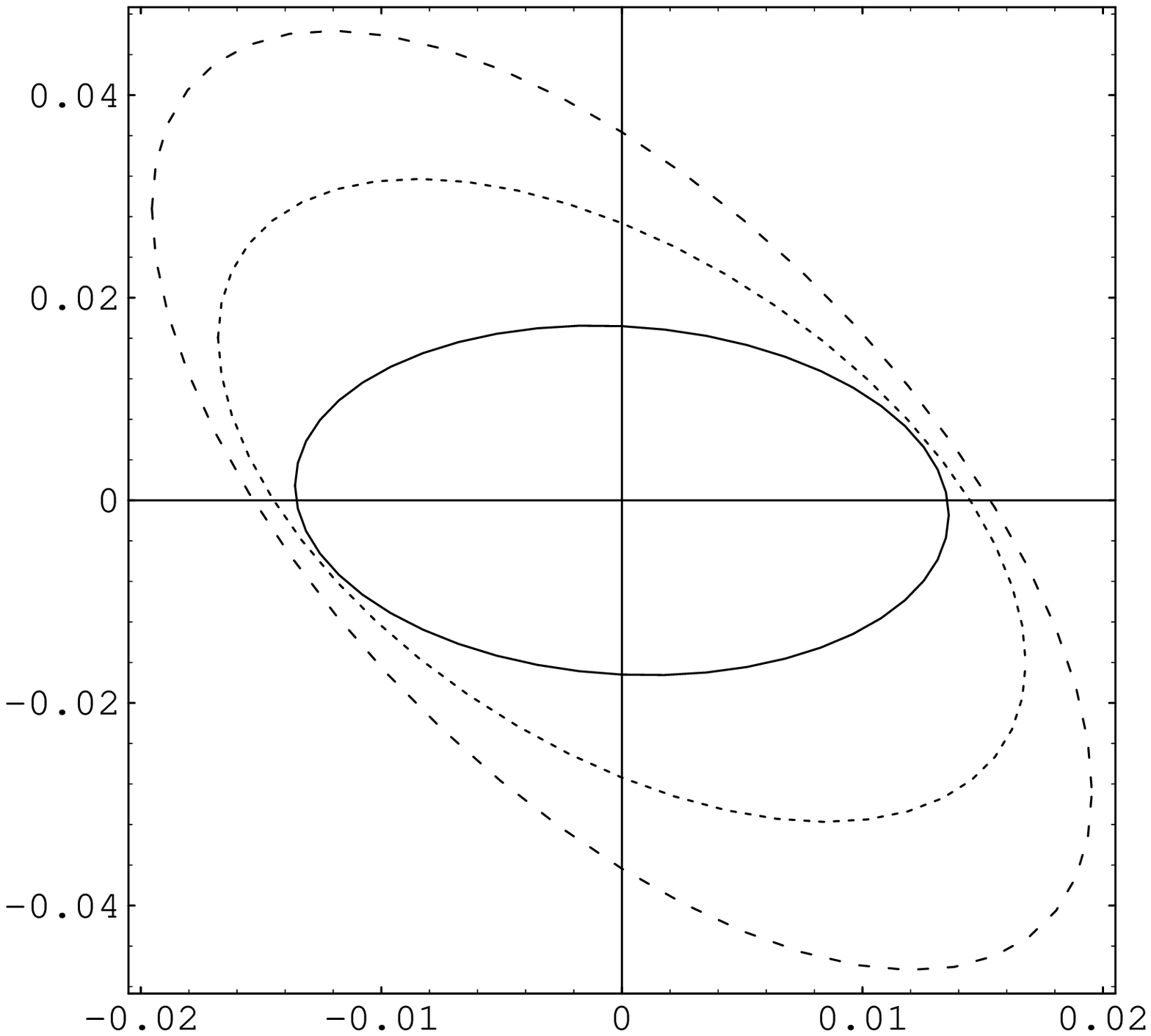,height=9cm}\hspace{2cm} 
\epsfig{file=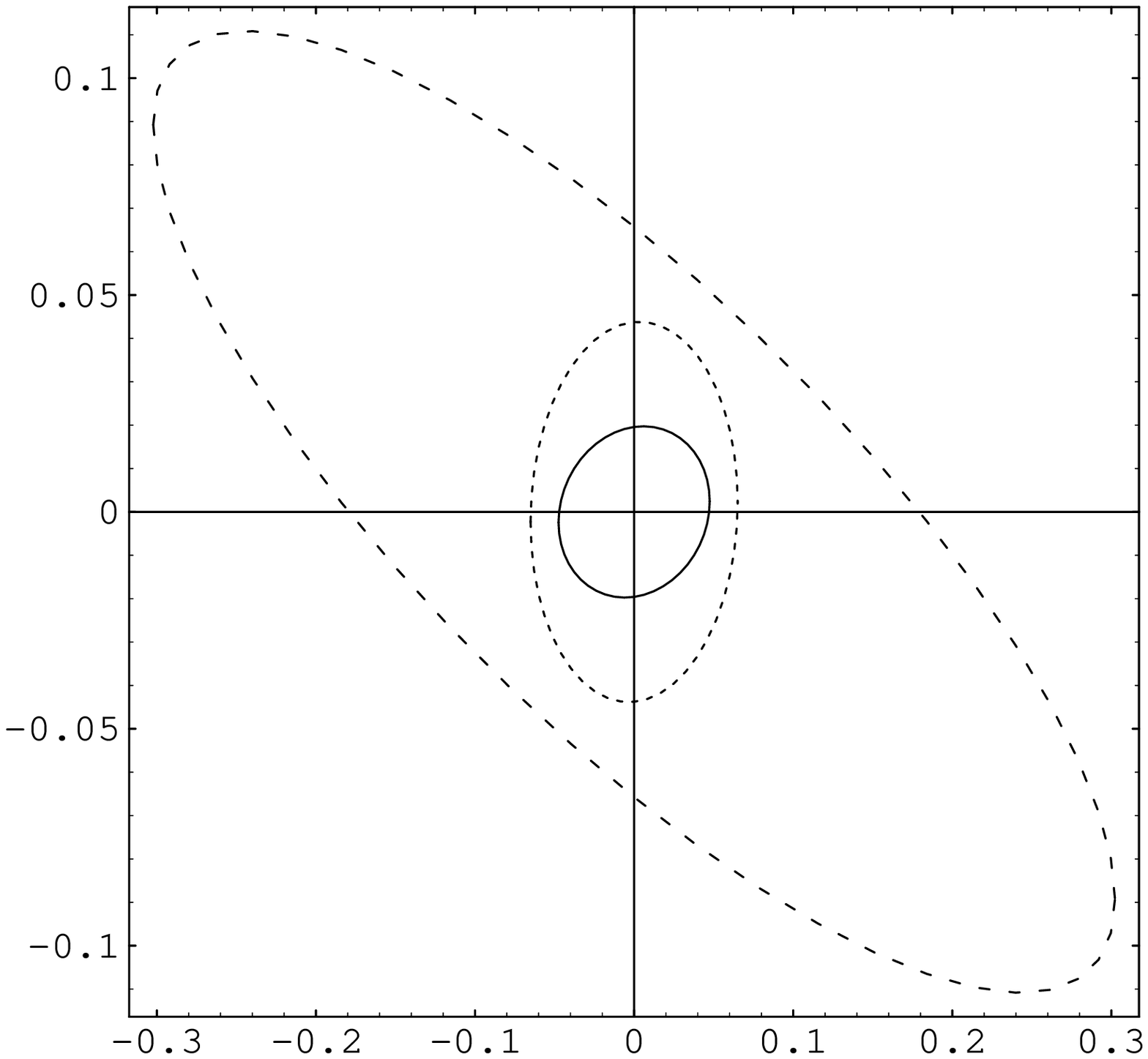,height=9cm}
\]
\vspace{-6.cm}\null\\
\hspace*{-1cm} \v \hspace{8cm} \v \\[3.2cm]
\hspace*{6cm} \u \hspace{8cm}  \u
\\
\hspace*{3.2cm} (g) \hspace{7.8cm}  (h)
\\[1.2cm]
\centerline{Fig 2}
\newpage
\vspace*{-2.cm}
\[
\epsfig{file=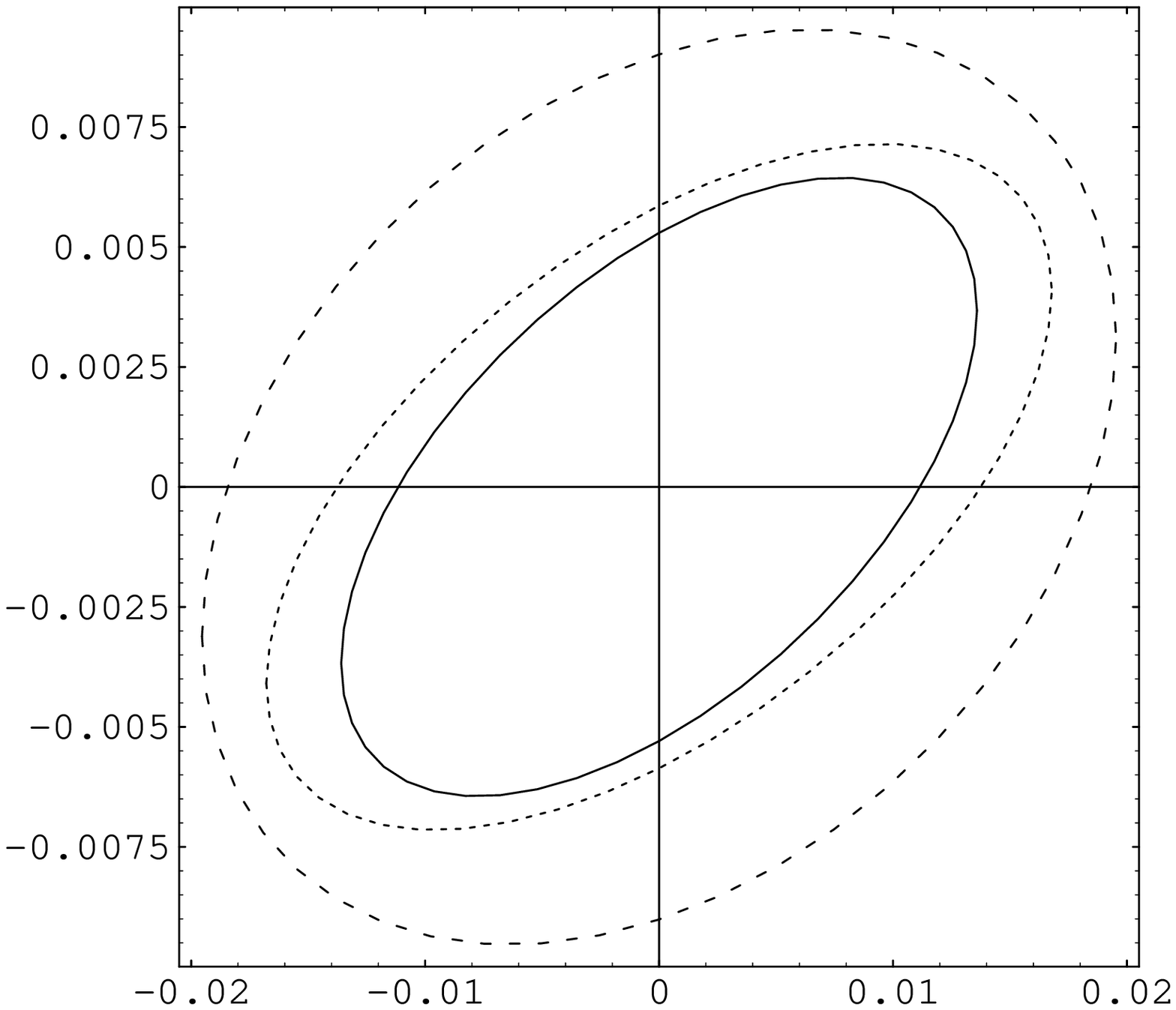,height=9cm}\hspace{2cm} 
\epsfig{file=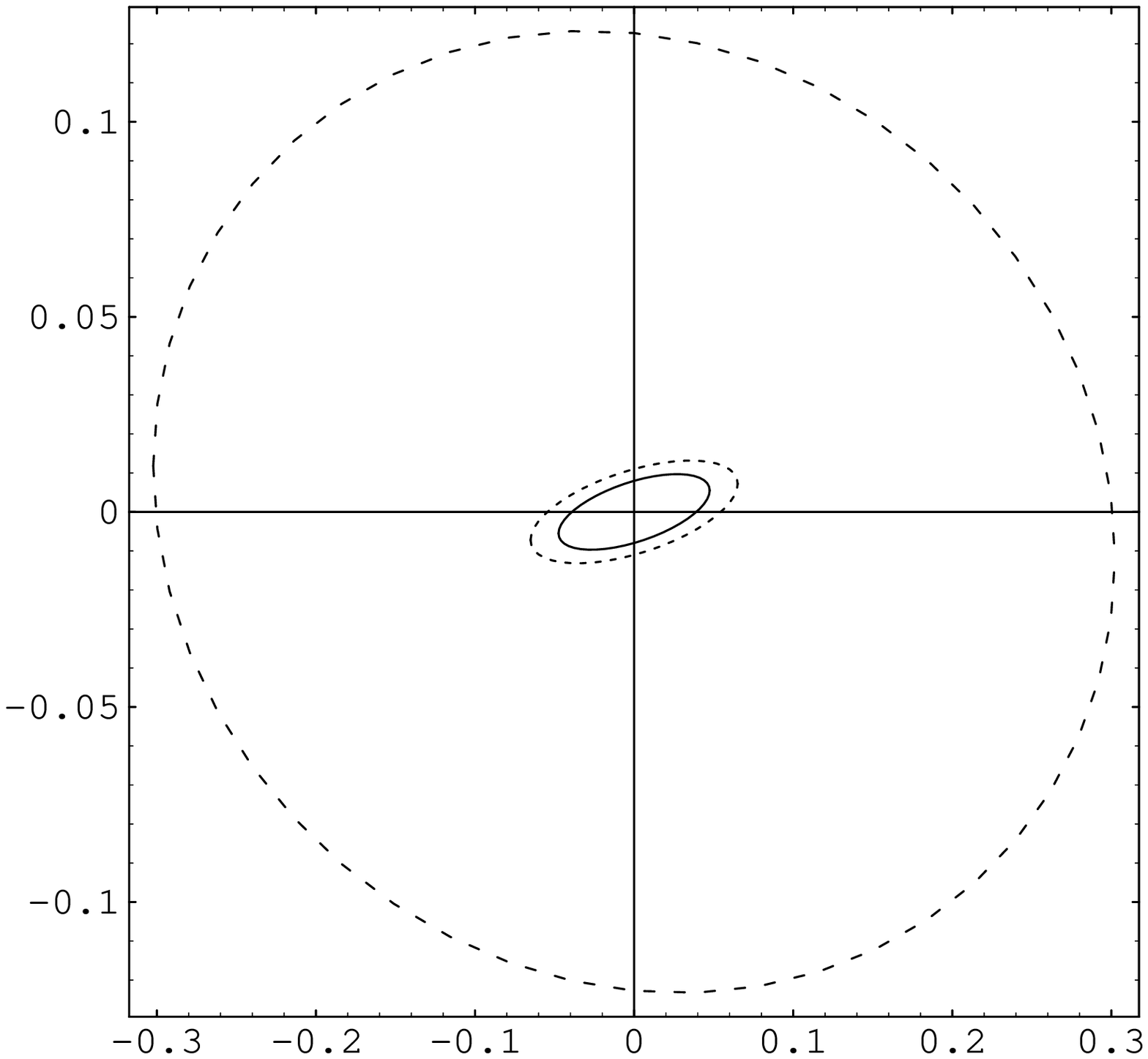,height=9cm}
\]
\vspace{-6.cm}\null\\
\hspace*{-1cm} \w \hspace{8cm} \w \\[3.2cm]
\hspace*{6cm} \u \hspace{8cm}  \u
\\
\hspace*{3.2cm} (i) \hspace{7.8cm}  (j)
%
\[
\epsfig{file=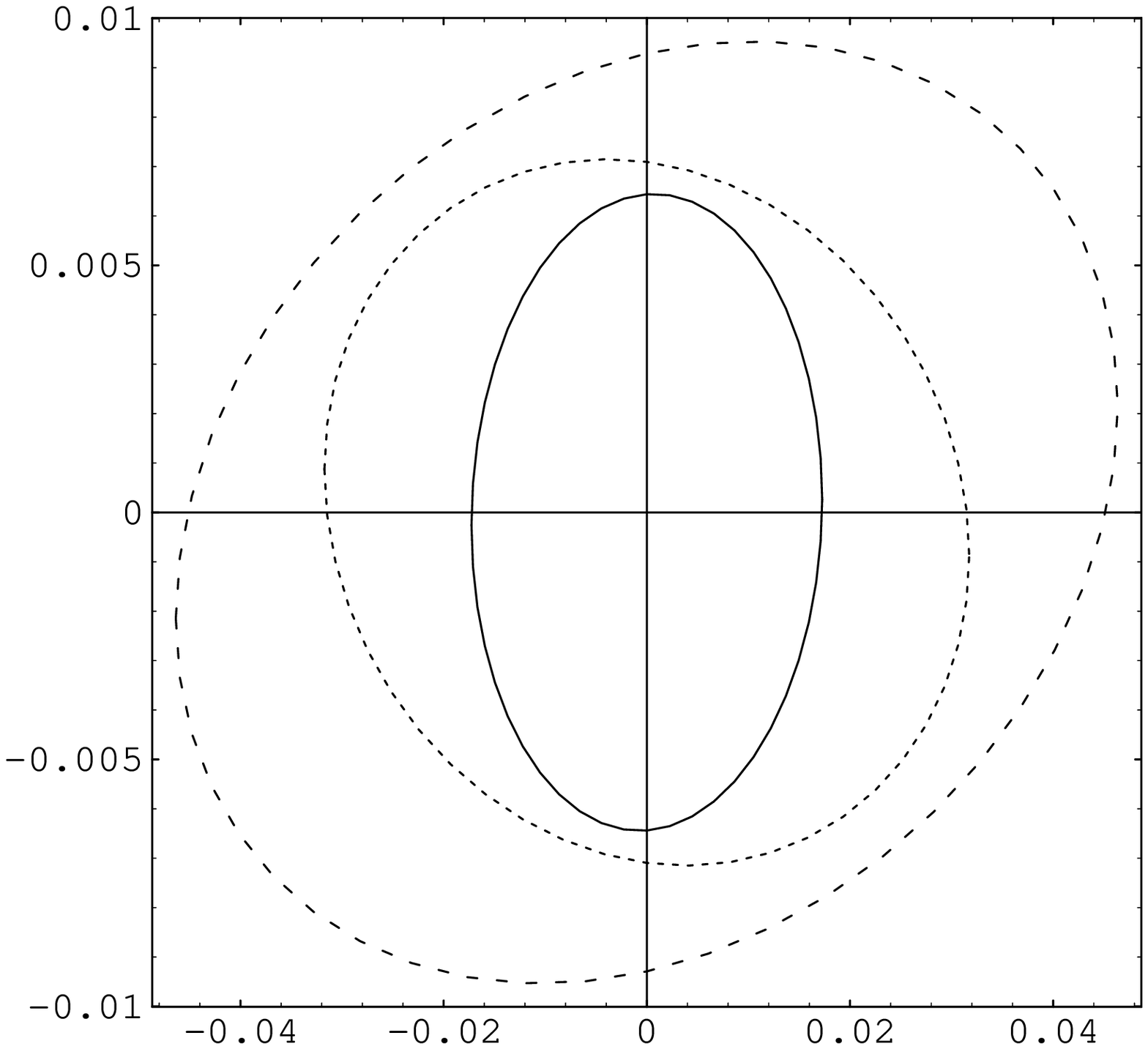,height=9cm}\hspace{2cm} 
\epsfig{file=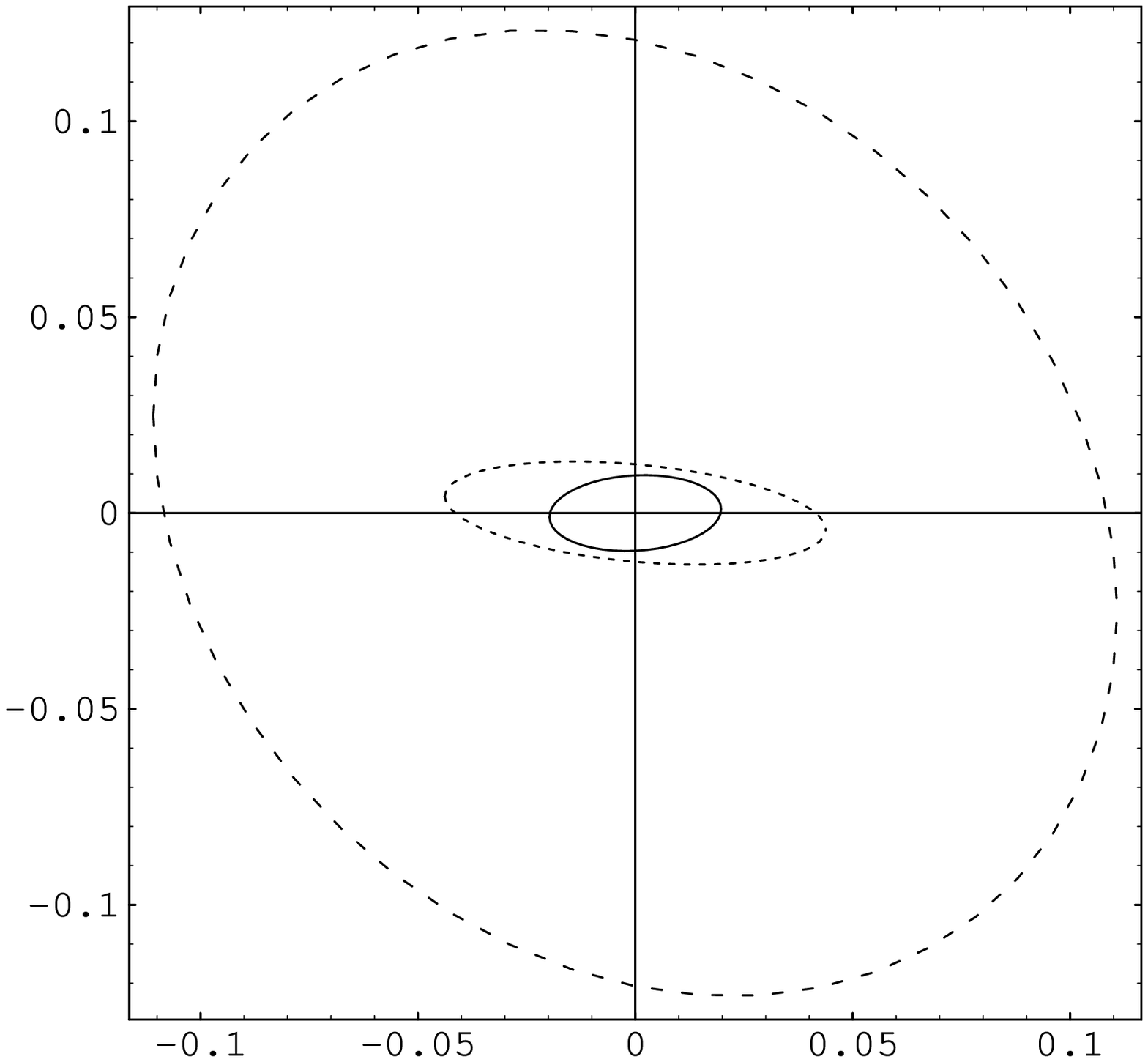,height=9cm}
\]
\vspace{-6.cm}\null\\
\hspace*{-1cm} \w \hspace{8cm} \w \\[3.2cm]
\hspace*{6cm} \v \hspace{8cm}  \v
\\
\hspace*{3.2cm} (k) \hspace{7.8cm}  (l)
\\[1.2cm]
\centerline{Fig 2}
\newpage
\vspace*{-2.cm}
\[
\epsfig{file=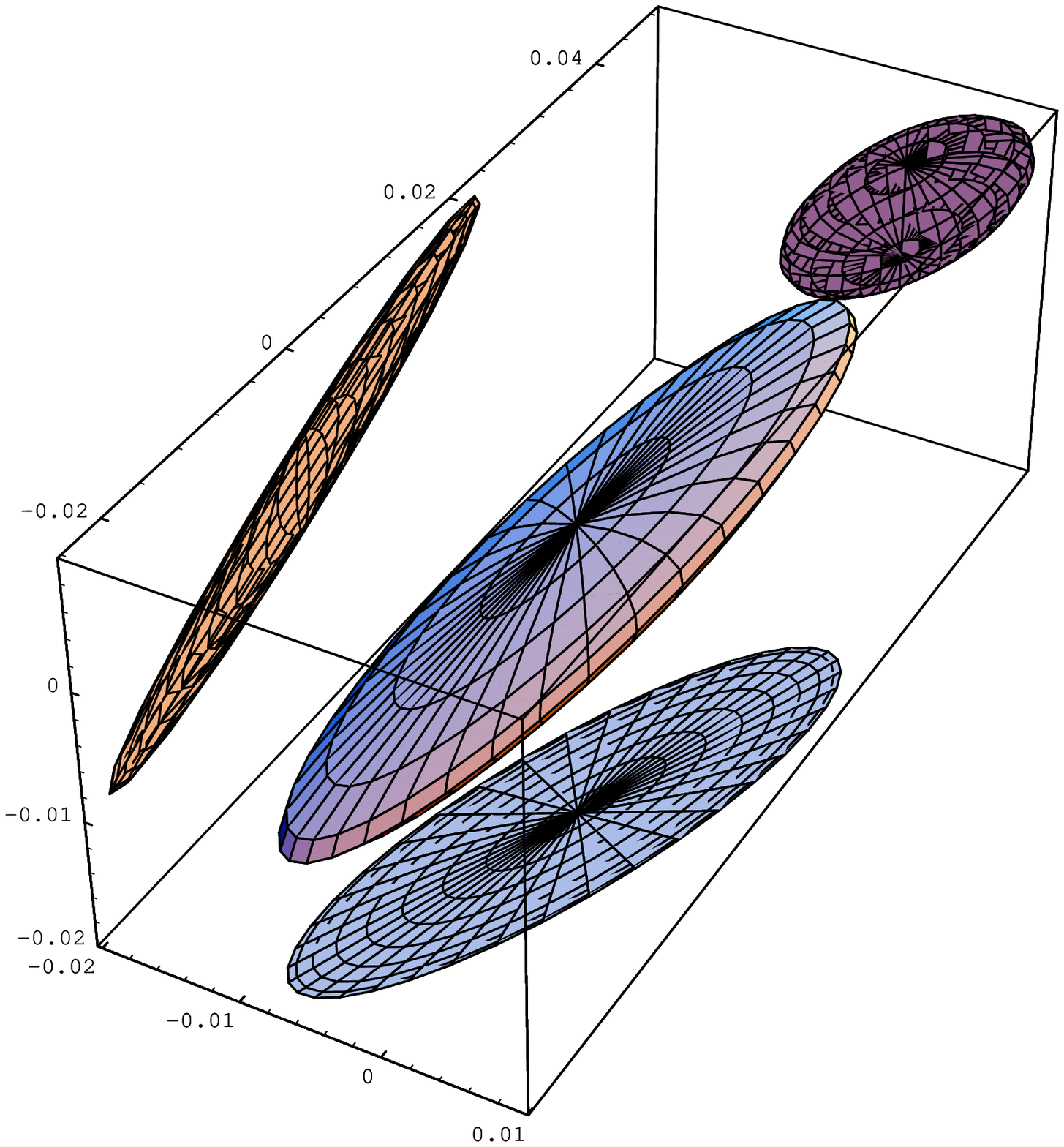,height=18cm}
\]
\vspace{-15cm}\null\\
\hspace*{3cm} 
{\Large
\v \\[5cm]
\hspace*{-0.5cm}\w \\[4.5cm]
\hspace*{3cm} \z \\[2cm]
\centerline{Fig 3} }
\newpage
\vspace*{-2.cm}
\[
\epsfig{file=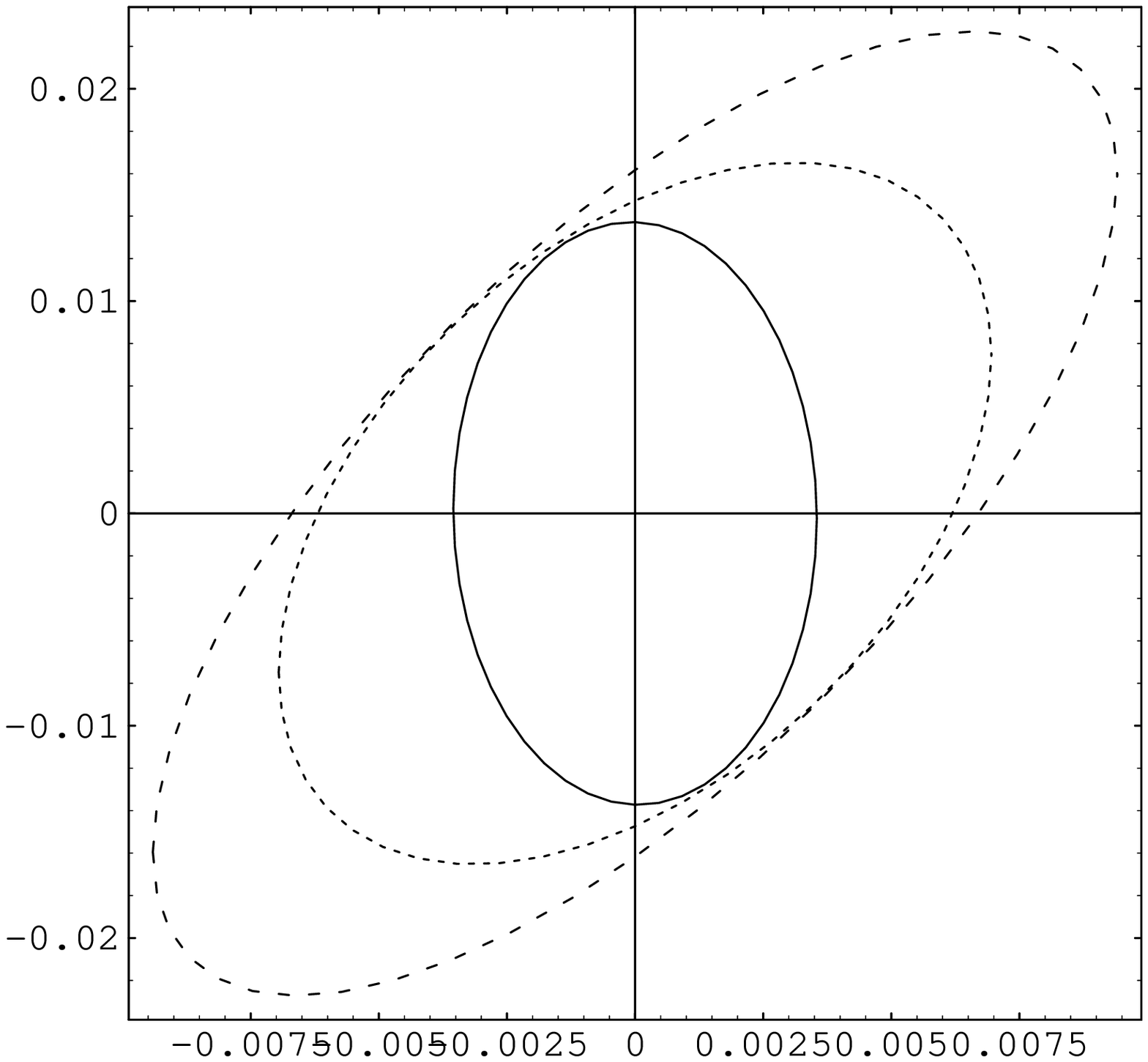,height=9cm}\hspace{2cm} 
\epsfig{file=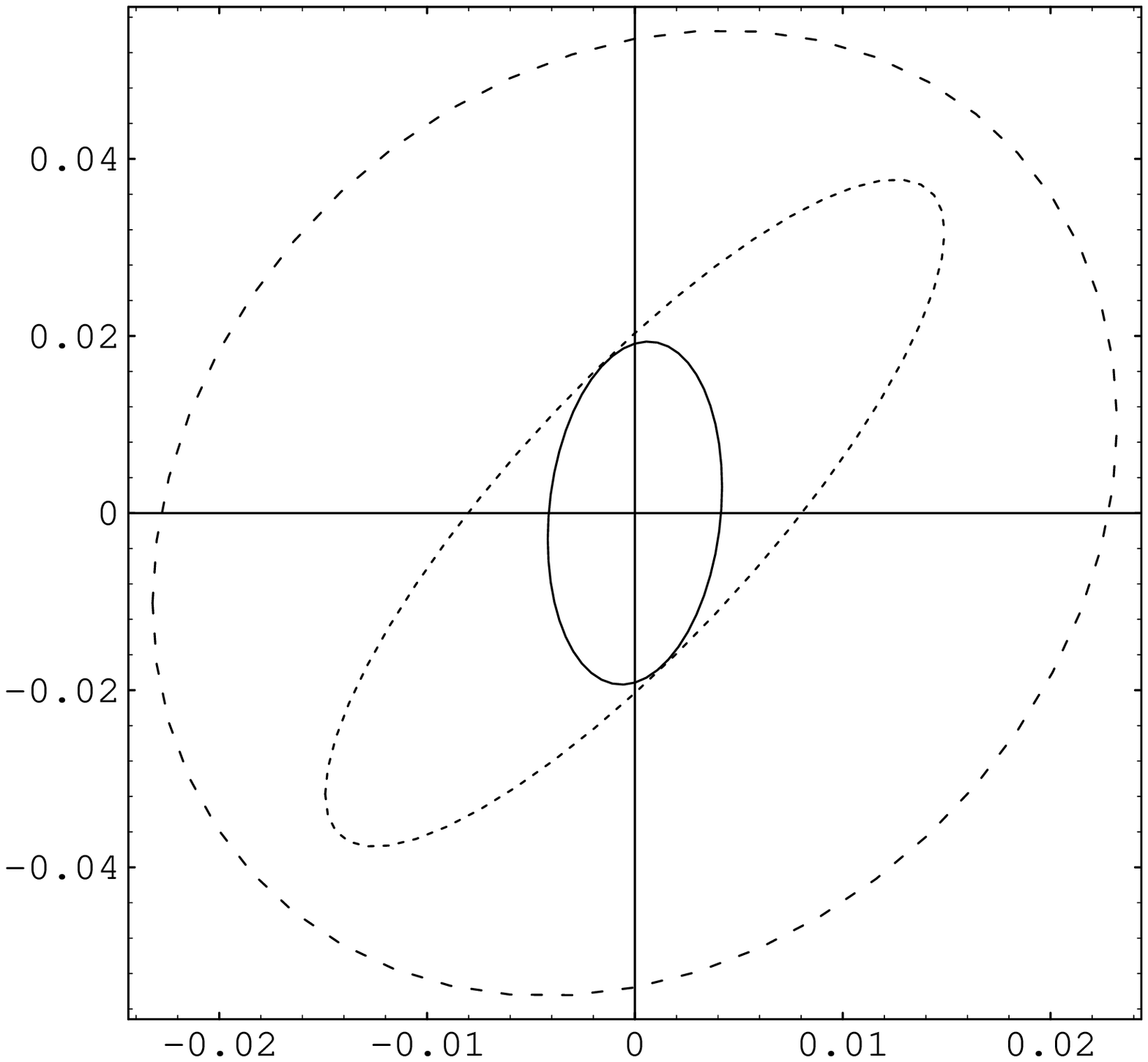,height=9cm}
\]
\vspace{-6.cm}\null\\
\hspace*{-1cm} \v \hspace{8cm} \v \\[3.2cm]
\hspace*{6cm} \z \hspace{8cm}  \z
\\
\hspace*{3.2cm} (a) \hspace{7.8cm}  (b)
\[
\epsfig{file=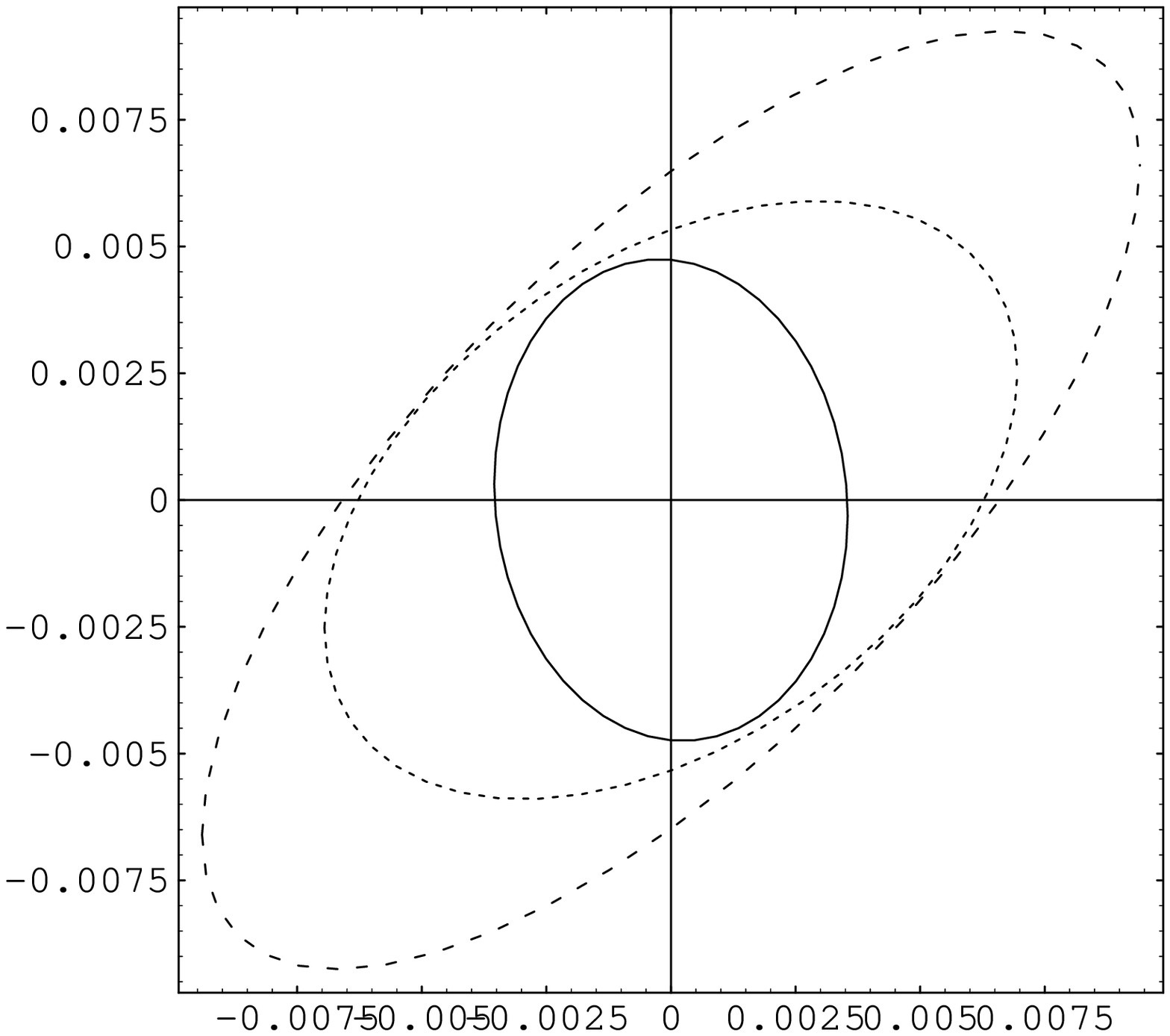,height=9cm}\hspace{2cm} 
\epsfig{file=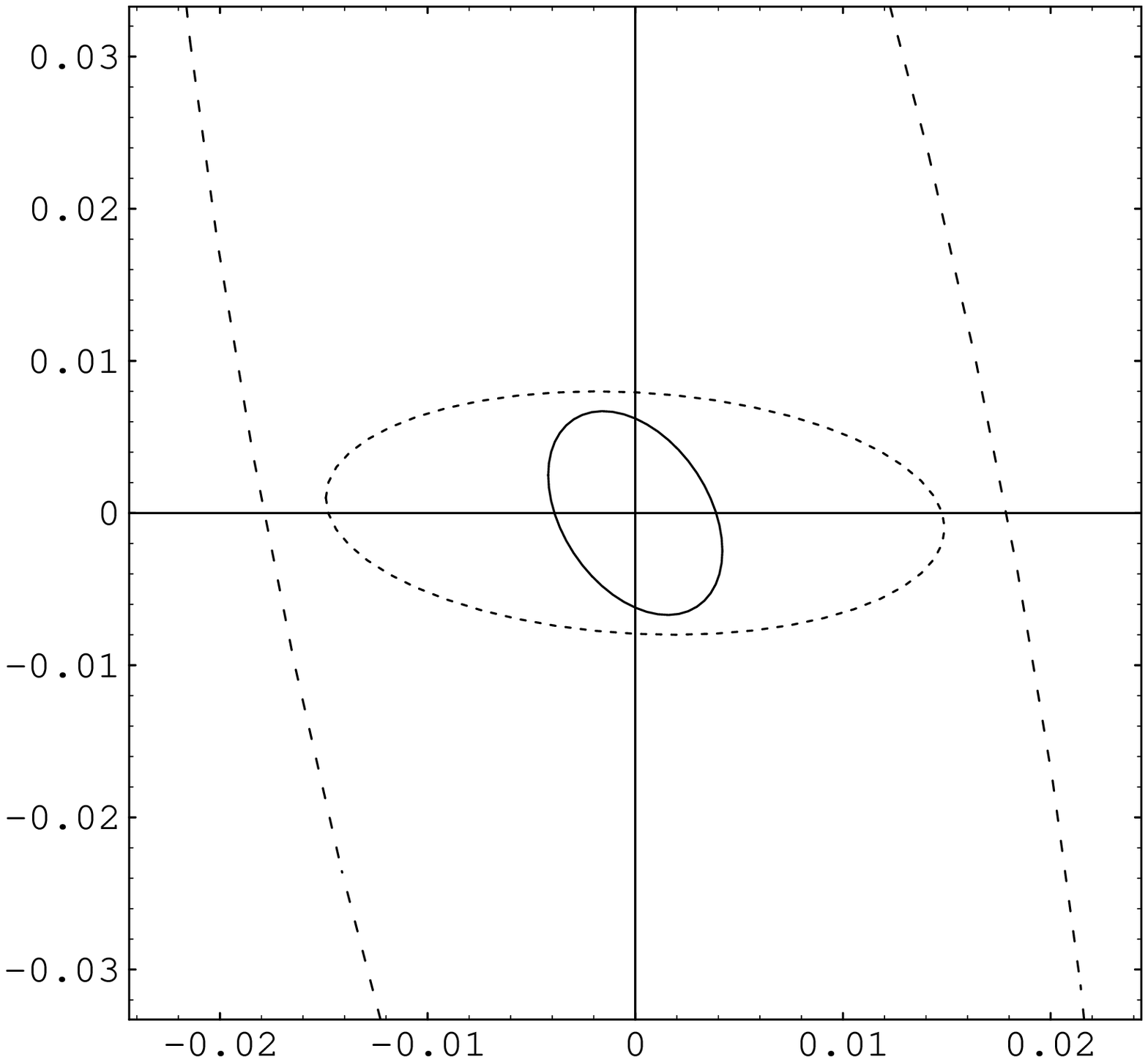,height=9cm}
\]
\vspace{-6.cm}\null\\
\hspace*{-1cm} \w \hspace{8cm} \w \\[3.2cm]
\hspace*{6cm} \z \hspace{8cm}  \z
\\
\hspace*{3.2cm} (c) \hspace{7.8cm}  (d)
\\[1.2cm]
\centerline{Fig 4}
\newpage
\vspace*{-2.cm}
\[
\epsfig{file=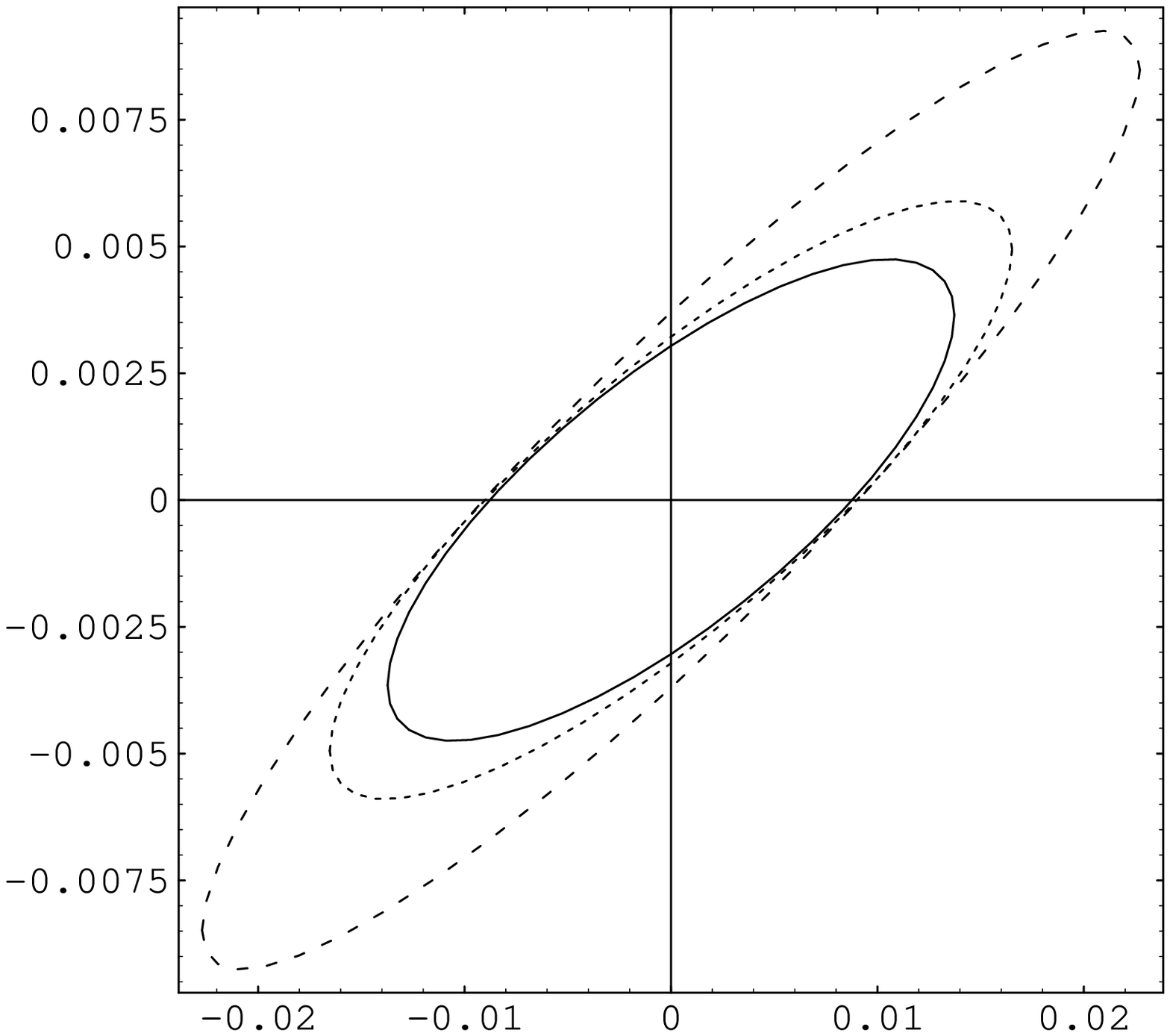,height=9cm}\hspace{2cm} 
\epsfig{file=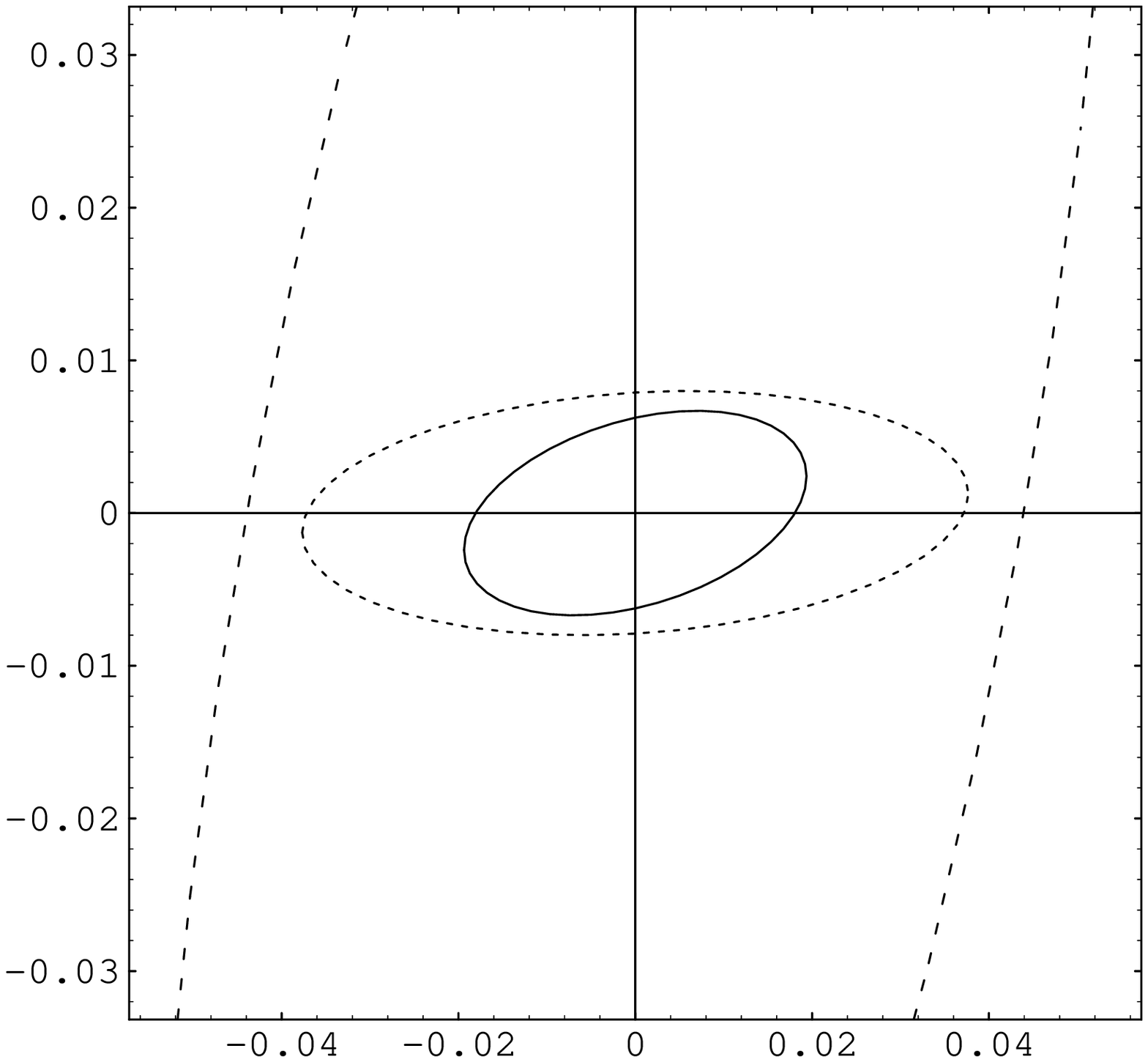,height=9cm}
\]
\vspace{-6.cm}\null\\
\hspace*{-1cm} \w \hspace{8cm} \w \\[3.2cm]
\hspace*{6cm} \v \hspace{8cm}  \v
\\
\hspace*{3.2cm} (e) \hspace{7.8cm}  (f)
\\[1.2cm]
\centerline{Fig 4}
\[
\epsfig{file=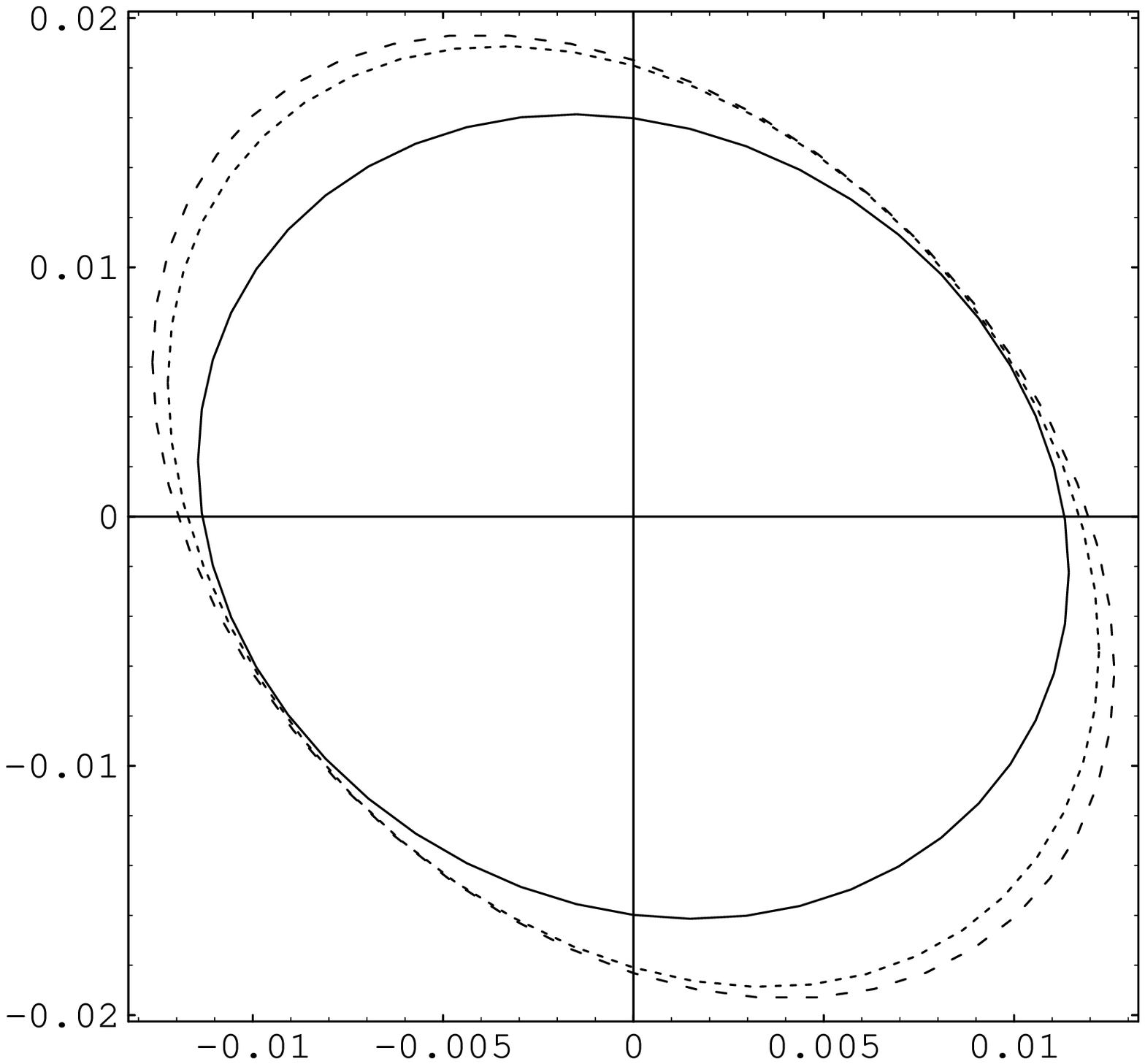,height=9cm}\hspace{2cm} 
\epsfig{file=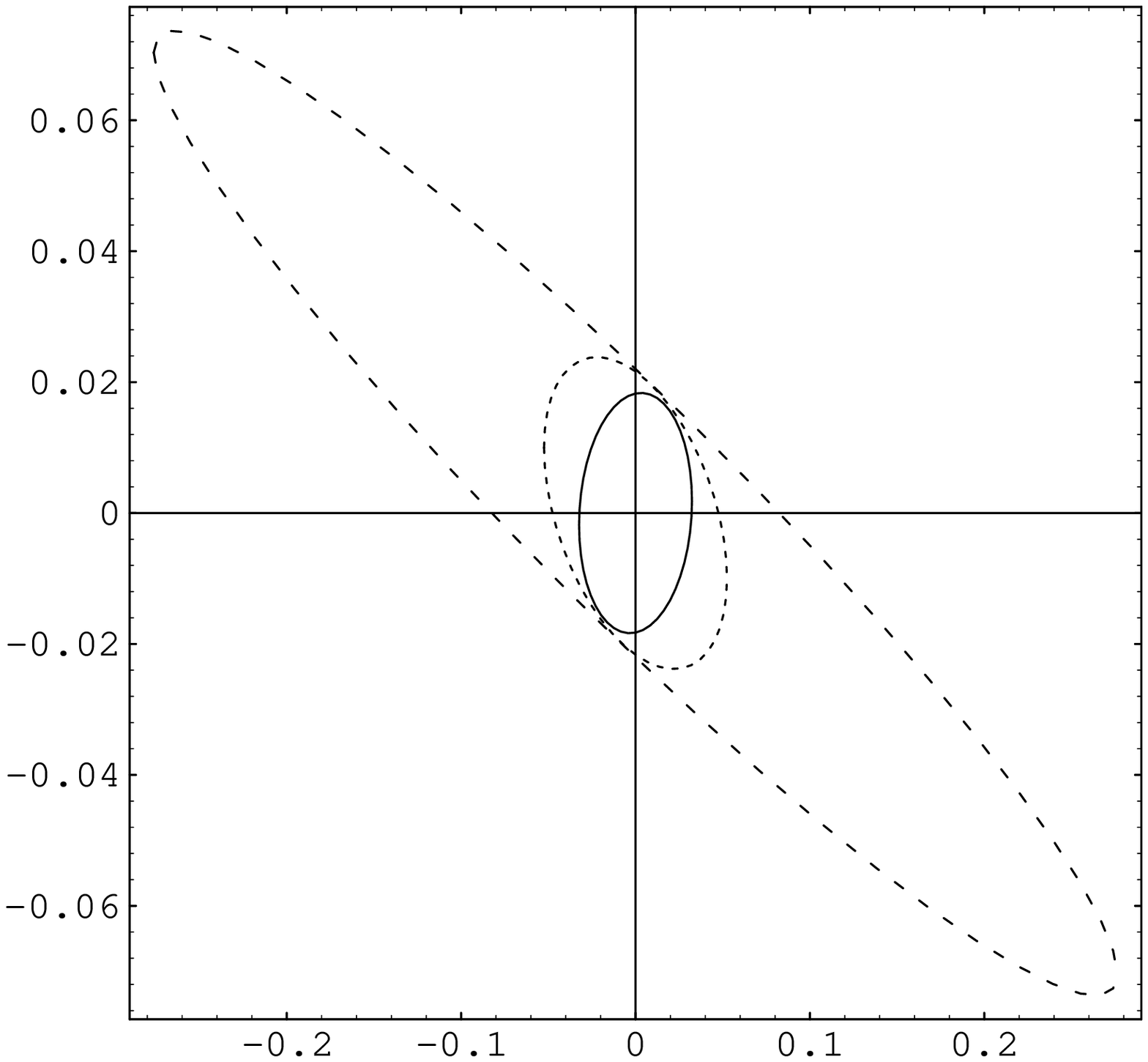,height=9cm}
\]
\vspace{-6.cm}\null\\
\hspace*{-1cm} \v \hspace{8cm} \v \\[3.2cm]
\hspace*{6cm} \u \hspace{8cm}  \u
\\
\hspace*{3.2cm} (a) \hspace{7.8cm}  (b)
\\[1.2cm]
\centerline{Fig 5}

\end{document}